\documentclass[aip,amsmath,amssymb,reprint]{revtex4-1}

\usepackage{graphicx}% Include figure files
\usepackage{dcolumn}% Align table columns on decimal point
\usepackage{bm}% bold math

\usepackage[utf8]{inputenc}
\usepackage[T1]{fontenc}
\usepackage{mathptmx}

\usepackage{subcaption}
\usepackage{ragged2e}
\DeclareCaptionJustification{justified}{\justifying}
\usepackage{color}
\definecolor{indigo}{RGB}{0,0,120}
\usepackage[colorlinks=true, linkcolor=indigo, citecolor=blue, urlcolor=indigo]{hyperref}

\usepackage{accents}
\newcommand\thickbar[1]{\accentset{\rule{.4em}{1.1pt}}{#1}}

\newcommand{\T}{{\cal T}}

\def\imply{\Rightarrow}
\newcommand{\pt}{\circ}

\newcommand{\tl}[1]{\tilde{#1}}
\newcommand{\dd}[2]{\frac {\partial #1}{\partial #2}}

\newcommand{\pdr}{\partial}
\newcommand{\DD}[2]{\frac {d #1}{d #2}}
\newcommand{\grad}{{\bf \nabla}}
\newcommand{\tr}{\, {\rm tr}\,}

\newcommand{\fl}{\noindent}

\newcommand{\beq}{\begin{equation}}
\newcommand{\eeq}{\end{equation}}
\newcommand{\beqs}{\begin{eqnarray}}
\newcommand{\eeqs}{\end{eqnarray}}

\newcommand{\half}{\frac{1}{2}}
\newcommand{\ov}[1]{\frac{1}{#1}}

\def\al{\alpha} 		
\def\del{\delta}
\def\D{\Delta}	
\def\g{\gamma}
\def\ka{\kappa} 
\def\eps{\epsilon} 
\def\la{\lambda}
		
\def\sig{\sigma}

\def\om{\omega}

\newcommand{\sech}{\,\text{sech}}
\newcommand{\cn}{\,\text{cn}}

\newcommand{\calF}{{\cal F}}

\newcommand{\bfv}{{\bf v}}
\newcommand{\bfw}{{\bf w}}
\newcommand{\bfx}{{\bf x}}
\newcommand{\bfy}{{\bf y}}

\newcommand{\bfr}{{\bf r}}

\newcommand{\bfF}{{\bf F}}

\newcount\colveccount  % column vec \colvec{nrows}{a11 & a12}{a21 & a22}
\newcommand*\colvec[1]{\global\colveccount#1  \begin{pmatrix} \colvecnext} \def\colvecnext#1{#1 \global\advance\colveccount-1
        \ifnum\colveccount>0 \\ \expandafter\colvecnext
        \else \end{pmatrix} \fi}
\begin{document}
%--------------------

\title{Nonlinear dispersive regularization of inviscid gas dynamics}

\author{Govind S. Krishnaswami} 

% \affiliation{Physics Department, Chennai Mathematical Institute,  SIPCOT IT Park, Siruseri 603103, India}

\email{govind,phatak,sonakshi@cmi.ac.in}

\author{Sachin S Phatak} 

% \affiliation{Chennai Mathematical Institute,  SIPCOT IT Park, Siruseri 603103, India}

\author{Sonakshi Sachdev} 

\affiliation{Physics Department, Chennai Mathematical Institute,  SIPCOT IT Park, Siruseri 603103, India}

\author{A. Thyagaraja}

\affiliation{Astrophysics Group, University of Bristol, Bristol, BS8 1TL, UK}

\email{athyagaraja@gmail.com}

Published in \href{https://doi.org/10.1063/1.5133720}{AIP Advances, 10, 025303 (2020)}, \qquad \href{https://arxiv.org/abs/1910.07836}{[arXiv:1910.07836]}

\date{February 6, 2020}

\begin{abstract}

Ideal gas dynamics can develop shock-like singularities with discontinuous density. Viscosity typically regularizes such singularities and leads to a shock structure. On the other hand, in 1d, singularities in the Hopf equation can be non-dissipatively smoothed via Korteweg-de Vries (KdV) dispersion. In this paper, we develop a minimal conservative regularization of 3d ideal adiabatic flow of a gas with polytropic exponent $\gamma$. It is achieved by augmenting the Hamiltonian by a capillarity energy $\beta(\rho) (\nabla \rho)^2$. The simplest capillarity coefficient leading to local conservation laws for mass, momentum, energy and entropy using the standard Poisson brackets is $\beta(\rho) = \beta_*/\rho$ for constant $\beta_*$. This leads to a Korteweg-like stress and nonlinear terms in the momentum equation with third derivatives of $\rho$, which are related to the Bohm potential and Gross quantum pressure. Just like KdV, our equations admit sound waves with a leading cubic dispersion relation, solitary waves and periodic traveling waves. As with KdV, there are no steady continuous shock-like solutions satisfying the Rankine-Hugoniot conditions. Nevertheless, in 1d, for $\gamma = 2$, numerical solutions show that the gradient catastrophe is averted through the formation of pairs of solitary waves which can display approximate phase-shift scattering. Numerics also indicate recurrent behavior in periodic domains. These observations are related to an equivalence between our regularized equations (in the special case of constant specific entropy potential flow in any dimension) and the defocussing nonlinear Schr\"odinger equation (cubically nonlinear for $\gamma = 2$), with $\beta_*$ playing the role of $\hbar^2$. Thus, our regularization of gas dynamics may be viewed as a generalization of both the single field KdV and nonlinear Schr\"odinger equations to include the adiabatic dynamics of density, velocity, pressure and entropy in any dimension.

\end{abstract}

\maketitle
\footnotesize
\tableofcontents
\normalsize
%--------------
\section{Introduction}
\label{s:intro}
%--------------

Gas dynamics has been an active area of research with applications to high-speed flows, aerodynamics and astrophysics. The equations of ideal compressible flow are known to encounter shock-like singularities with discontinuities in density, pressure or velocity \cite{whitham}. These singularities are often resolved by the inclusion of viscosity. However, as the KdV equation $(u_t + u u_x = \eps u_{xxx})$ illustrates, such singularities in the one-dimensional (1d) Hopf (or kinematic wave) equation $u_t + u u_x = 0$ can also be regularized conservatively via dispersion \cite{ablowitz}, as in dispersive shock wave theory (see \cite{whitham,biondini-etal,El-Hoefer,ali-kalisch-1} and references therein) with applications to undular bores in shallow water and blast waves in Bose-Einstein condensates. In this paper, we develop a minimal conservative regularization of ideal gas dynamics, which we refer to as R-gas dynamics. Somewhat analogous conservative `rheological' regularizations of vortical singularities in ideal Eulerian hydrodynamics, magnetohydrodynamics and two-fluid plasmas have been developed in \cite{thyagaraja,govind-sonakshi-thyagaraja-pop,govind-sonakshi-thyagaraja-2-fluid}. The current work may be regarded as a way of extending the single-field KdV equation to include the dynamics of density, velocity and pressure and also to dimensions higher than one. There is of course a well-known generalization of KdV to 2d, the Kadomtsev-Petviashvili (KP) equation \cite{kp}. However, unlike KP, our regularized equations are rotation-invariant and valid in any dimension. Now, recall\cite{gardner} that the dispersive regularization term in the KdV equation $u_t - 6uu_x + u_{3x} = 0$ arises from the gradient energy term in the Hamiltonian $H = \int(u^3 + (1/2)u_x^2)\, dx$, upon use of the Poisson brackets $\{u(x), u(y)\} = \pdr_x \del(x-y)$. In fact, KdV does not conserve mechanical and capillarity energies separately\cite{ali-kalisch-3, karczewska-rozmej-infeld}. By analogy with this, we obtain our regularized model by augmenting the Hamiltonian of ideal adiabatic flow of a gas with polytropic exponent $\gamma$, by a density gradient energy $\beta(\rho) (\grad\rho)^2$. Such a term arose in the work of van der Waals and Korteweg \cite{vdW,korteweg,dunn-serrin,gorban-karlin} in the context of capillarity, but can be important even away from interfaces in any region of rapid density variation, especially when dissipative effects are small, such as in weak shocks, cold atomic gases, superfluids and collisionless plasmas. It has also been used to model liquid-vapor phase transitions and in the thermomechanics of interstitial working \cite{dunn-serrin}. We argue that the simplest choice of capillarity coefficient that leads (using the standard Poisson brackets) to local conservation laws for mass, momentum, energy and entropy (with the standard mass, momentum and entropy densities) is $\beta(\rho) = \beta_*/\rho$ where $\beta_*$ is a constant. By contrast, the apparently simpler option of taking $\beta(\rho)$ constant leads, in 1d, to a KdV-like linear dispersive term $\rho_{xxx}$ in the velocity equation, but results in a momentum equation that, unlike KdV\cite{ali-kalisch-3}, is {\it not} in conservation form for the standard momentum density $\rho u$. A consequence of the constitutive law $\beta = \beta_*/\rho$ is that the ideal momentum flux $\rho u^2 + p$ is augmented by a stress $-\beta_* (\rho_{xx} - \rho_x^2/\rho)$ corresponding to a Kortweg-type \emph{grade} 3 elastic material \cite{dunn-serrin,gorban-karlin}. This leads to new nonlinear terms in the momentum equation with third derivatives of $\rho$, somewhat reminiscent of KdV. One of the effects of these nonlinear dispersive terms is to allow for `upstream influence'\cite{benjamin} which is forbidden by the hyperbolic equations of inviscid gas dynamics under supersonic conditions. Interestingly, our regularization term is also related to the quantum mechanical Bohm potential \cite{Bohm-1} and Gross quantum pressure (p.476 of \cite{gross}) encountered in superfluids. Moreover, unlike KdV, our equations extend in a natural way to any dimension. Remarkably, for potential flow $(\bfv = \grad \phi)$ in the isentropic case (globally constant entropy and $p \propto \rho^\gamma$), the R-gas dynamic equations may be transformed into the nonlinear Schr\"odinger equation (NLSE) via the Madelung transformation  \cite{madelung} $\psi = \sqrt{\rho} \exp\left( i\phi/2\sqrt{\beta_*} \right)$ with $\beta_*$ playing the role of $\hbar^2$. This equivalence, which may be regarded as a conservative analog of the Cole-Hopf transformation for Burgers, applies in any dimension, and results in a defocusing NLSE with $|\psi|^{2(\g - 1)} \psi$ nonlinearity, so that one obtains the celebrated cubic NLSE for $\gamma = 2$. The latter is known to admit an infinite number of conservation laws and display recurrence. It is noteworthy that the {\it quantum} version of the 1d cubic NLSE (Lieb-Liniger model) has recently been given a hydrodynamical description (generalized hydrodynamics \cite{bertini, castro-doyon-yoshimura}) with infinitely many local conservation laws and has been used to model 1d gases of ultracold Rubidium atoms which retain memory of their initial state \cite{schemmer}.

A brief summary of the paper and its organization follows. We begin in \S \ref{s:3d-hamil-form-R-gas-dyn} by giving the Lagrangian (in terms of Clebsch variables) and Hamiltonian formulations and equations of motion (EOM) of adiabatic R-gas dynamics in 3d. The mass, momentum, energy and entropy equations are all expressed in conservation form. In \S \ref{s:formulation-r-gas-dynm}, we specialize to 1d and discuss the special case of constant entropy (isentropic/barotropic) flow  in which case the velocity equation also acquires a conservation form. Sound waves are discussed in \S \ref{s:dispersive-sound-waves} and shown to be governed at long wavelengths by a cubic dispersion relation similar to that of the linearized KdV equation. In \S \ref{s:steady-trav-quadrature}, the local conservation laws are used to reduce the determination of steady and traveling wave solutions in 1d to a single quadrature of a generalization of the Ermakov-Pinney equation. A mechanical analogy and phase plane analysis is used to show that the only such non-constant bounded solutions are cavitons (in density) and periodic waves.  While these results hold for any value of $\g$, for $\g = 2$, closed-form $\text{sech}^2$ and cnoidal wave solutions are obtained, physically interpreted and compared with the corresponding KdV solutions. Aside from overall scales, steady solutions are parametrized by a pair of dimensionless shape parameters: a Mach number and a curvature. A parabolic embedding and a virial theorem for steady flows are given in Appendix \ref{a:parabolic-embed-LJ-id}. In \S \ref{s:weak-form}, the weak form of the R-gas dynamic equations is given, and in \S\ref{s:patched-shock-weak-sol} an attempt is made to find a steady shock-like profile by patching half a caviton with a constant solution. However, it is shown that there are no such continuous profiles that satisfy all the Rankine-Hugoniot conditions, though it may be possible to satisfy the mass flux condition alone. To study more general time-dependent solutions of R-gas dynamics and the evolution of initial conditions that could lead to shock-like discontinuities, we set up in \S \ref{s:IVP-numerical}, a semi-implicit spectral numerical scheme for the isentropic R-gas dynamic equations with periodic boundary conditions (BCs) in 1d. For $\g = 2$, our numerical solutions indicate that our regularization evades the gradient catastrophe through the formation of a pair of solitary waves at the top and bottom of a velocity profile with steep negative gradient. Though we do not observe a KdV-like solitary wave train, these solitary waves can suffer collisions and approximately re-emerge with a phase shift. We also observe a rapid decay of energy with mode number and recurrent behavior with Rayleigh quotient fluctuating between bounded limits, indicating an effectively finite number of active Fourier modes. In \S \ref{s:r-gas-to-nlse} we use a canonical transformation to reformulate 3d adiabatic R-gas dynamics in terms of a complex scalar field coupled to an entropy field and three Clebsch potentials. For isentropic potential flows, this formulation shows that R-gas dynamics for any $\g$ reduces to a defocusing 3d NLSE. In \S \ref{s:caviton-to-nlse}, the regularized Bernoulli equation is used to show that steady R-gas dynamic solutions map to solutions of NLSE with harmonic time dependence, with the $\g = 2$ caviton in 1d corresponding to the dark soliton of the cubic NLSE. In \S \ref{s:rayleigh-quotient-NLSE-conserved-quantities} we relate the conserved quantities and bounded Rayleigh quotient of NLSE to their R-gas dynamic analogues. This connection lends credence to our numerical observations, since the cubic NLSE with periodic BCs in 1d is known to possess an infinity of conserved quantities in involution \cite{faddeev-takhtajan}. We also note in \S \ref{s:neg-pressure-vortex-filament} that the {\it negative} pressure $\g = 2$ isentropic R-gas dynamic equations in 1d are equivalent to the vortex filament and Heisenberg magnetic chain equations. We conclude with a discussion in \S \ref{s:discussion}.

%--------------
\section{Hamiltonian and Lagrangian  formulations of 3d R-gas dynamics}
\label{s:3d-hamil-form-R-gas-dyn}
%--------------

It is well-known \cite{whitham} that adiabatic dynamics of an ideal gas with constant specific heat ratio $\g = c_p/c_v$ is governed by the continuity, momentum and internal energy equations 
	\beqs
	&& \rho_t + \grad \cdot (\rho \bfv) = 0, \quad
	(\rho v_i)_t + \pdr_j \left( p \del_{ij} + \rho v_i v_j \right) = 0 \quad \text{and} \cr
	&& \left( \frac{p}{\g - 1} \right)_t + p \grad \cdot \bfv  + \grad \cdot\left( \frac{p \bfv}{\g - 1}\right) = 0,
	\label{e:ideal-mass-mom-int-egy-3d}
	\eeqs
with the temperature in energy units given by $T = m p/\rho$ for a molecular mass $m$. In adiabatic flow, specific entropy (per unit mass) is advected ($D_t s \equiv \pdr_t s +  \bfv \cdot \grad s = 0$), while the entropy per unit volume is locally conserved, $\pdr_t (\rho s) + \grad \cdot (\rho s \bfv) = 0$. Though the terms `reversibly adiabatic' and `isentropic' are often used interchangeably, in this paper we use adiabatic for $D_t s = 0$ and isentropic for the special case where $s$ is a global constant. For adiabatic flow, $\rho$ and $p$ may be taken as independent variables with $s$ being a function of them. For a polytropic gas, $s = c_v \log\left( (p/\bar p)/ (\rho/ \bar \rho)^\g \right)$ where $\bar p, \bar \rho$ are reference values. These equations follow from the Hamiltonian
	\beq
    H_{\rm ideal} = \int \left[ \half \rho \bfv^2 + \frac{p}{\g-1} \right] d\bfr
    \label{e:3d-hamiltonian-ideal}
	\eeq
and Hamilton's equations $\dot f = \{ f, H \}$ using the (non-zero) non-canonical Poisson brackets (PB) \cite{morrison-greene}
	\beqs
	&& \{ \bfv(\bfx), \rho(\bfy) \} = \grad_y \del(\bfx - \bfy), \;\;\;
	\{ \bfv (\bfx), s(\bfy) \} = \frac{\grad s}{\rho} \del ( \bfx - \bfy) \cr
	&& \text{and} \quad \{ v_i(\bfx), v_j(\bfy) \} = \frac{\eps_{ijk} w_k}{\rho} \del (\bfx - \bfy).
	\label{e:PB-3d}
	\eeqs
where $\bfw = \grad \times \bfv$ is the vorticity. Our conservative regularization involves adding a density gradient term to the Hamiltonian while retaining the same PBs:
	\beq
    H = \int {\cal E} \; d\bfr \equiv  \int \left[ \half \rho \bfv^2 + \frac{p}{\g-1} + \frac{\beta_*}{2} \frac{(\grad \rho)^2}{\rho} \right] d\bfr.
    \label{e:3d-hamiltonian}
	\eeq
The density gradient energy, which could arise from capillarity \cite{vdW,korteweg}, has been chosen $\propto (\grad \rho)^2$ to ensure positivity, parity conservation and to prevent discontinuities in density, so as to conservatively regularize shock-like discontinuities. It involves the capillarity coefficient $\beta(\rho) = \beta_*/\rho$, where $\beta_*$ is a constant with dimensions $L^4 T^{-2}$. $\beta_*$ can be taken as $\la^2 c^2$ where $\la$ is a short-distance cut-off and $c$ a typical speed. This is the simplest form for $\beta(\rho)$ that ensures the mass, momentum and energy equations are all in conservation form for the ideal mass and momentum densities. It also leads to other nice properties such as a transformation to the NLSE for isentropic potential flow. 

The continuity and entropy equations following from (\ref{e:3d-hamiltonian}) and (\ref{e:PB-3d}) are as in the ideal model. The momentum and consequently the velocity equation however, differ due to the presence of a capillary force term $\beta_* \bfF$:
	\beqs
	&&\bfv_t + \bfv \cdot \grad \bfv + \frac{\grad p}{\rho} 
	= \beta_* \bfF
	= \beta_* \grad \left[ \half \frac{(\grad \rho)^2}{\rho^2} + \grad \cdot \left( \frac{\grad \rho}{\rho} \right) \right] \cr
	&& = \beta_* \grad \left[ \frac{\grad^2 \rho}{\rho} - \half \frac{(\grad \rho)^2}{\rho^2}\right] 
	= 2 \beta_* \grad \left( \frac{\grad^2 \sqrt{\rho}}{\sqrt{\rho}} \right)
	\label{e:3d-vel-eqn}
	\eeqs
Remarkably, $\beta_* \bfF = \beta_* \grad \Phi$ is a gradient, so that for barotropic flow ($\grad p/\rho = \grad h$), it augments the specific enthalpy $h \to h + \beta_* \Phi$. Thus, the vorticity evolves exactly as in ideal gas dynamics (in other words, we only regularize the `potential' part of the velocity and don't deal with vortical singularities as in \cite{thyagaraja, govind-sonakshi-thyagaraja-pop}). Thus, Kelvin's theorem would apply in R-gas dynamics, unchanged.
% For barotropic potential flow, this implies that each component of velocity is locally conserved.
The momentum and velocity equations may be expressed in terms of a regularized stress tensor: 
	\beqs
	&& \pdr_t (\rho v_i) + \pdr_j \left( \rho v_i v_j + \sig_{ij} \right) = 0
	\cr 
	\text{and} \quad && \pdr_t v_i + v_j \pdr_j v_i = - \ov{\rho} \pdr_j \sig_{ij} \qquad \text{where} \cr
	&& 
	\sig_{ij} = p \, \del_{ij} + \beta_* \left( \frac{(\pdr_i \rho) (\pdr_j \rho)}{\rho}  - \pdr_i \pdr_j \rho \right).
%	 = p_* \del_{ij} + \left( \sig_{ij} - \ov{d}\del_{ij} \tr \sig \right)
	 \label{e:stress-R-gas-dyn-3d}
	\eeqs
The scalar part of $\sig$ defines a regularized pressure $p_*$ which includes the Gross `quantum pressure' \cite{gross}:
	\beq 
	p_* = \ov{d} \tr \sig = p + \frac{\beta_*}{d} \left( \frac{(\grad \rho)^2}{\rho} - \grad^2 \rho \right) \;\; \text{where} \;\; d=3.
	\label{e:pstar-3d}
	\eeq
The energy equation for the energy density $\cal E$ defined in (\ref{e:3d-hamiltonian}) is given by:
	\beqs
	&& {\cal E}_t + \grad \cdot \left( \frac{\rho \bfv^2}{2} \bfv + \frac{\g}{\g - 1} p\bfv\right) + \cr
	&& \beta_* \grad \cdot\left[\frac{\grad \rho}{\rho} \grad \cdot (\rho \bfv) - \rho \bfv \grad \cdot \left( \frac{\grad \rho}{\rho} \right)  - \frac{\rho \bfv}{2}\frac{ (\grad \rho)^2}{\rho^2}  \right] = 0.
	\label{e:energy-eqn-3d}
	\eeqs
The fact that (\ref{e:energy-eqn-3d}) is in local conservation form follows from the PB formulation. Indeed, $\{ H, H \} = 0$ implies that ${\cal E}_t = \{ {\cal E}, H \}$ must be a divergence. The internal energy per unit volume is therefore
	\beq
	\rho \varepsilon_* = \rho \varepsilon + \frac{\beta_*}{2} \frac{(\grad \rho)^2}{\rho}
	\;\; \text{where} \;\:\; \varepsilon = \frac{p}{\rho(\g - 1)} = \frac{T}{(\g - 1) m}.
	\label{e:spec-int-energy}
	\eeq
These regularization terms in the pressure, enthalpy and internal energy depend upon density gradients and are therefore not strictly thermodynamic properties of the gas, any more than the regularized stress tensor. They are conservative analogues of the viscous stress tensor which depends on velocity gradients in dissipative gas dynamics.

Interestingly, the potential $\Phi$ in (\ref{e:3d-vel-eqn}) is also the Bohm potential $U$ \cite{Bohm-1} that arises as a correction to the classical potential $V$ in the quantum-corrected Hamilton-Jacobi equation for the Schr\"odinger wavefunction $\psi = \sqrt{\rho} e^{i S/\hbar}$: \small
	\beqs
	&& \rho_t + \grad \cdot \left( \rho \frac{\grad S}{m} \right) = 0 \quad \text{and} \quad 
	S_t + \frac{(\grad S)^2}{2m}+ V + U = 0 \cr
	&& \text{where} \quad U = - \frac{\hbar^2}{2m} \frac{\grad^2 \sqrt{\rho}}{\sqrt{\rho}} = - \frac{\hbar^2}{4 m} \left( \frac{\grad^2 \rho}{\rho} - \half \frac{(\grad \rho)^2}{\rho^2}  \right).
	\eeqs \normalsize
Our regularized stress $\sig$ also resembles the Korteweg stress $\sig^{\rm Kor}$ of Ref. \cite{korteweg,gorban-karlin}. Indeed, if $\beta = \beta_*/\rho$,
	\beqs
	\sig^{\rm Kor}_{ij} 
	&=&  p \del_{ij} - \rho \left[\pdr_k \left( \beta(\rho) \pdr_k \rho \right) \right] \del_{ij} + \beta(\rho) \pdr_i \rho \pdr_j \rho \cr
	&=& p \del_{ij} -\beta_* \left[ \left( \grad^2 \rho - \ov{\rho} (\grad \rho)^2 \right) \del_{ij} -  \frac{\pdr_i \rho \; \pdr_j \rho}{\rho} \right]. \quad
	\label{e:korteweg-stress}
	\eeqs
However, though $\sig^{\rm Kor}_{ij}$ has the term $({\beta_*}/{\rho}) (\pdr_i \rho) (\pdr_j \rho)$ in common with $\sig_{ij}$ (\ref{e:stress-R-gas-dyn-3d}), they are not quite equal. Thus, though the qualitative physical features of our equations may be similar to those of the Korteweg equations, ours additionally possess some remarkable mathematical properties facilitating the analysis in this paper.

Finally, if the flow domain is all of $\mathbb{R}^3$, then $\beta_*$ can be scaled out by defining ${\bf R} = {\bf r}/\sqrt{\beta_*}$ and $T = t/\sqrt{\beta_*}$, just as we may eliminate the dispersion coefficient in KdV on the whole real line. By contrast, in the presence of a characteristic length scale $l$, $\beta_*$ {\it cannot} be scaled out and $l/\la$ serves as a conservative analogue of the Reynolds number.

% the solutions to our equations for any positive $\beta_*$ can be obtained from those for $\beta_* = 1$ by an appropriate rescaling. 
%---------------
\subsection{Lagrangian formulation via Clebsch variables}
%---------------

To obtain a Lagrangian for R-gas dynamics, we use the Clebsch representation \cite{clebsch,zakharov-kuznetsov}  $\bfv = \grad \phi + (\la \grad \mu + \al \grad s)/\rho$. The PBs in (\ref{e:PB-3d}) are recovered by postulating canonical PBs among Clebsch variables:
    \beq
    \{\rho(\bfr), \phi(\bfr') \} = \{\alpha(\bfr), s(\bfr') \} = \{\la(\bfr), \mu(\bfr') \} = \del(\bfr - \bfr').
    \eeq
The Hamiltonian density in terms of Clebsch variables:
    \beq
    \mathcal{H} = \frac{\rho}{2} \left( \grad \phi + \frac{(\la \grad \mu + \al \grad s)}{\rho} \right)^2 + \rho \,\varepsilon(\rho, s) + \frac{\beta_*}{2} \frac{(\grad \rho)^2}{\rho}
    \eeq
where $\varepsilon(\rho, s)$ is the ideal specific internal energy (\ref{e:spec-int-energy}). The EOM (\ref{e:3d-vel-eqn}) follow as the Euler-Lagrange equations (EL) for the Bateman-Thellung \cite{bateman,thellung} Lagrangian density linear in velocities \cite{sudarshan-mukunda} augmented by the density gradient energy: 
    \beq
    \mathcal{L}_1 \; = \; \rho_t \phi + \la_t \mu + \alpha_t s - {\cal H}.
    \label{e:non-barotropic-lag}
    \eeq
The EL equations for $\al$ and $\la$ imply the advection of $s$ and $\mu$, while that for $\phi$ is the continuity equation and that for $s$ and $\mu$ are the evolution equations $\al_t + \grad \cdot (\al \bfv) = \rho T$ and $\la_t + \grad \cdot (\la \bfv) = 0$. The regularization only affects the EL equation for $\rho$. Upon using $p = \rho^2 \partial \varepsilon /\partial \rho$, it becomes the time-dependent Bernoulli equation for adiabatic R-gas dynamics:
    \beq
    \phi_t - \frac{\bfv^2}{2} + \bfv \cdot \grad \phi + \varepsilon(\rho, s) + \frac{p}{\rho} -\beta_* \left( \frac{\grad^2 \rho}{\rho} - \half \frac{(\grad \rho)^2}{\rho^2}\right) = 0.
    \eeq
Using these, one obtains (\ref{e:3d-vel-eqn}) for $\bfv$. There are of course related Lagrangians for the same EOM, e.g.,
	\beqs
	&& \mathcal{L}_2 = -\rho \phi_t - \la \mu_t - \alpha s_t - \mathcal{H} \quad \text{and} \quad \mathcal{L}_3 = \rho \left(\frac{\bfv^2}{2} - \varepsilon \right) \cr
	&& - \frac{\beta_*}{2} \frac{(\grad \rho)^2}{\rho} + \phi \left( \rho_t + \grad \cdot (\rho \bfv) \right)  - \la \frac{D \mu}{Dt} - \al \frac{D s}{Dt}.
	\label{e:alternate-lagrangians}
	\eeqs
Thus, we may interpret $\phi,\la$ and $\al$ as Lagrange multipliers enforcing the EOM for $\rho, \mu$ and $s$.

%--------------
\section{Formulation of one-dimensional regularized gas dynamics}
\label{s:formulation-r-gas-dynm}
%--------------

%--------------
\subsection{Hamiltonian and equations of motion}
%--------------

In what follows, we will primarily be interested in 1d adiabatic R-gas dynamics where $\rho, s$ and $p$ are independent of two of the Cartesian coordinates and $\bfv = (u(x,t),0,0)$. The non-zero PBs (\ref{e:PB-3d}) simplify as $\bfw = 0$: $\{u , u \} = 0$ and 
	\beq
	\{ u(x), s(y) \} = \frac{s'}{\rho} \del(x-y) \;\; \text{and} \;\;  \{ \rho(x), u(y) \} = \pdr_y \del(x-y).
	\label{e:PB-u-rho-s}
	\eeq
The total mass $(\int \rho \: dx)$, entropy  $(\int \rho \, s \: dx)$ and more generally $\int \rho \Sigma(s) dx$ for any $\Sigma(s)$ are Casimirs of this algebra. As before, the dynamics is generated by a Hamiltonian that involves a capillary energy
	\beq
	H = \int \left[\half \rho u^2 + \frac{p}{\g - 1} + \half \beta(\rho) \: \rho_x^2 \right] dx,
	\label{e:beta-reg-hamiltonian}
	\eeq
where $\beta(\rho)$ will be chosen by requiring that the momentum equation be in conservation form. The continuity and entropy equations are as in the ideal model:
	\beq
	\rho_t + (\rho u )_x = 0, \;\;
	s_t + u s_x = 0 \;\; \text{with} \;\;
	s = c_v \log\left( \frac{p \bar \rho^\g}{\bar p \rho^\g} \right).
	\label{e:cont-entropy-eqn} 
	\eeq
Thus, even with our regularization we continue to have $D_t p = c_s^2 D_t \rho$ where $c_s^2 = (\pdr p/\pdr \rho)_s = \g p/\rho$.
The regularized momentum  and velocity equations are
	\beqs
	&& (\rho u)_t + (\rho u^2 + p)_x = \rho \left[ \left( \beta \rho_x  \right)_x - \half \beta' \rho_x^2 \right]_x
	\quad
	\text{and} \cr
	&& u_t + u u_x = - \frac{p_x}{\rho} + \left[ \left( \beta \rho_x  \right)_x - \half \beta' \rho_x^2 \right]_x.
	\label{e:vel-mom-r-gas-dynm}
	\eeqs
The simplest way for the former to be in conservation form is for the momentum density to equal $\rho u$ and for the regularization term to be a divergence. $\beta = \beta_*/\rho$ is the simplest capillarity coefficient that ensures this, giving 
	\beqs
	&& (\rho u )_t + \left[ \rho u^2 + p - \beta_* \left(\rho_{xx} - \frac{\rho_x^2}{\rho} \right)\right]_x = 0
	\quad \text{and} \cr
	&& \frac{Du}{Dt} + \frac{p_x}{\rho} 
	= \beta_* f 
	=  \beta_* \left[ \frac{\rho_{2x}}{\rho} - \frac{\rho_x^2}{2 \rho^2} \right]_x 
	= 2 \beta_* \left[ \frac{(\sqrt{\rho})_{xx}}{\sqrt{\rho}}\right]_x. \quad
	\label{e:reg-vel-eqn-gas}
	\eeqs 
We note that the apparently simpler choice of constant $\beta$ leads to a KdV-like $\rho_{xxx}$ term in the velocity equation but prevents the momentum equation from being in conservation form. Our regularization amounts to modifying the  pressure $p \to p_*$ in the momentum and velocity equations:
	\beqs
	&& (\rho u)_t + (\rho u^2 + p_*)_x = 0
	\quad \text{and} \quad 
	u_t + u u_x = - \frac{{p_*}_x}{\rho}
	\cr
	&& \text{where} \quad p_* = p - \beta_* (\rho_{xx} - \rho_x^2/\rho).
	\label{e:reg-vel-mom-eqn-p_*}
	\eeqs
It is instructive to compare our velocity equation with Korteweg's. For capillarity coefficient $\beta = \beta_*/\rho$, the 1d Korteweg velocity equation following from (\ref{e:korteweg-stress}) is
	\beq
	u_t + u u_x + \frac{p_x}{\rho} = \beta_* f^{\rm Kor} = \frac{\beta_*}{\rho} \left( \rho_{xx} -  \frac{2 \rho_x^2}{\rho}\right)_x.
	\eeq	
Unlike our force per unit mass $\beta_* f$ which is a gradient, $\beta_* f^{\rm Kor}$ is not. Thus, in the barotropic case of \S\ref{s:barotropic-r-gas-dyn} where $p_x/\rho = h_x$, our velocity equation (\ref{e:reg-vel-eqn-gas}) (but not Korteweg's) comes into conservation form as in ideal gas dynamics. Finally, our energy equation is also in local conservation form,
	\beqs
	&& \left( \half \rho u^2 + \frac{p}{\g -1} +  \frac{\beta_*}{2} \frac{\rho_x^2}{\rho} \right)_t + \left( \half \rho u^2 u + \frac{\g}{\g - 1} pu\right)_x \cr
	 && + \: \beta_* \left( \frac{\rho_x}{\rho} (\rho u)_x - \rho u \left( \frac{\rho_x}{\rho} \right)_x  - \half \frac{u \rho_x^2}{\rho} \right)_x = 0.
	\label{e:reg-total-energy-PB}
	\eeqs
%	\beq
%	\text{or} \quad \left( \half \rho u^2 + \frac{p}{\g -1} +  \frac{\beta_*}{2} \frac{\rho_x^2}{\rho} \right)_t + \left( \half \rho u^2 u + \frac{\g}{\g - 1} pu - \beta_* \left( u \rho_{xx} - \frac{3}{2} \frac{u \rho_x^2}{\rho} - u_x \rho_x \right) \right)_x = 0
%	\eeq
It takes a compact form in terms of a regularized specific internal energy $\varepsilon_*$ and enthalpy $h_*$: 
	\beqs
	&& \left( \half \rho u^2 + \rho \varepsilon_* \right)_t + \left( \rho \left(\half u^2 +  h_* \right) u + \beta_* u_x \rho_x  \right)_x = 0
	\cr
	&& \text{where} \quad \rho \varepsilon_* = \frac{p}{\g -1} + \frac{\beta_*}{2} \frac{\rho_x^2}{\rho} 
	\;\; \text{and} \;\;
	\rho h_* = \rho \varepsilon_* + p_*.
	\label{e:reg-total-energy-eps_*-enthalpy-PB}
	\eeqs 
%\beq
%	\rho \varepsilon_* = \frac{p_*}{\g - 1} + \frac{\beta_* \rho_x^2}{\rho}\left[ \half - \ov{\g - 1} \right] - \frac{\beta_* \rho_{xx}}{\g - 1}
%\eeq
The internal energy equation may be interpreted as the $1^{\rm st}$ law of thermodynamics for adiabatic flow:
	\beq
	D_t \varepsilon_* + p_*  D_t \left( \frac{1}{\rho} \right) + \frac{\beta_*}{\rho} (u_x \rho_x)_x = T D_t s = 0.
	\label{e:reg-1st-law-adiab-specific}
	\eeq
Evidently, the gas does work against the pressure $p_*$ as well as a new type of reversible, non-dissipative work due to the regularization while ensuring that the specific entropy $s$ is constant along the flow.

%---------------
%\subsection{\Blue Symmetries of 1d R-gas dynamics}
%---------------

The action corresponding to ({\ref{e:non-barotropic-lag}}) possesses three obvious symmetries: (a) constant shift in $\phi$, (b) space translation and (c) time translation, leading to the local conservation laws for mass $(\rho)$, momentum $(\rho u)$ and energy $(\half \rho u^2 + \rho \varepsilon_*)$ densities (\ref{e:cont-entropy-eqn}, \ref{e:reg-vel-eqn-gas}, \ref{e:reg-total-energy-PB}). In addition, under an infinitesimal Galilean boost ($t \to t$, $x \to x-ct$), the fields transform as
    \beqs
    \del\phi &=& c(t\phi_x - x),\quad \del u = c(t u_x -1) \quad \text{and} \cr 
    \del \Upsilon &=& ct\Upsilon_x \quad \text{for} \quad \Upsilon = \rho, \al, s, \la \quad \text{and} \quad \mu,
    \eeqs 
leading to a change in the Lagrangian (\ref{e:alternate-lagrangians}) by a spatial derivative $\del \mathcal{L}_2 = ct \left( p - \beta_* \rho_{xx} \right)_x$. The corresponding Noether charge and flux densities are
    \beq
    j^t = \sum_{\chi} \dd{\mathcal{L}_2}{\dot \chi} \del\chi \; \; \; \text{and} \;\; \; j^x = \sum_\chi \dd{\mathcal{L}_2}{\chi_x}\del\chi - ct (p-\beta_* \rho_{xx}).
    \eeq
Here, we sum over $\chi = \rho, \phi, \al, s, \la$ and $\mu$. The resulting conservation law $\pdr_t j^t + \pdr_x j^x = 0$ or $(\rho (x - t u) )_t + ( x \rho u - t F^{\rm p})_x = 0$ involves an explicitly time-dependent Galilean charge and flux, where $F^{\rm p} = \rho u^2 + p - \beta_* \rho \left(\log \rho \right)_{xx}$ is the regularized momentum flux (\ref{e:reg-vel-eqn-gas}). Thus, $G = \int (x - tu) \, \rho \, dx$ is conserved even though $\{ G, H \} = P$ where $P = \int \rho u \, dx$ is the total momentum. $P$, $G$ and $H$ satisfy a 1d Galilei algebra with the total mass $M$ furnishing a central extension: $\{ G, P \} = M$.

%---------------
\subsection{Isentropic R-gas dynamics of a polytropic gas}
\label{s:barotropic-r-gas-dyn}
%---------------

Sans entropy sources/sinks and boundaries, one is mainly interested in cases where $s = \bar s$ is initially constant and by (\ref{e:cont-entropy-eqn}), independent of time. Thus, we consider isentropic flow where $p$ and $\rho$ satisfy the barotropic relation
	\beq
	p = (\g - 1) K \rho^\g \quad \text{where} \quad
	K = \frac{e^{\bar s/c_v}}{\g - 1} \frac{\bar p}{\bar \rho^\g} \: > \:  0
	\label{e:entropy-barotropic}
	\eeq
is a constant that encodes the constant value of entropy and labels isentopes. A feature of isentropic flow is that in addition to the continuity, momentum and energy equations, the velocity equation is {\it also} in conservation form:
	\beqs
	&& \rho_t + F^{\rm m}_x = 0, \quad
	(\rho u)_t + F^{\rm p}_x = 0, \cr
	&&\left[ \half \rho u^2 + \rho \eps + \frac{\beta_*}{2} \frac{\rho_x^2}{\rho} \right]_t + F^{\rm e}_x = 0 \;\;\;\text{and} \;\;\; u_t + F^{\rm u}_x = 0.
	\label{e:unsteady-barotropic-eqns}
	\eeqs
Here $\varepsilon = K \rho^{\g-1}$ and $h = \g K \rho^{\g-1}$ are specific internal energy and enthalpy. The corresponding fluxes are 
% {\Blue (more precisely flux or current densities)}
	\beqs
	&& F^{\rm m} = \rho u, \quad F^{\rm p} = \rho u^2 + p - \beta_* \left( \rho_{xx} - \frac{\rho_x^2}{\rho} \right), \cr
	 && F^{\rm e} =\left(\frac{u^2}{2} + h \right) \rho u + \; \beta_* \left( \frac{\rho_x}{\rho} (\rho u)_x - \rho u \left( \frac{\rho_x}{\rho} \right)_x  - \frac{u \rho_x^2}{2\rho} \right) \cr
	&& \text{and} \quad F^{\rm u} = \half u^2 + h - \beta_* \left( \frac{\rho_{xx}}{\rho} - \half \frac{\rho_x^2}{\rho^2} \right).
	\label{e:barotropic-curr}
	\eeqs
In ideal gas dynamics $F^{\rm u} = F^{\rm e}/F^{\rm m}$, but no such algebraic relation holds when $\beta_* \ne 0$. The emergence of a $4^{\rm th}$ conservation law in the isentropic case is tied to the global constancy of entropy. These equations follow from the degenerate Landau PBs $\{ \rho, \rho \} = \{ u, u \} = 0$ and $\{ \rho(x), u(y) \} = \pdr_y \del(x-y)$ whose Casimirs include $M = \int \rho \: dx$ and $\int u \: dx$. These PBs become canonical $[\{ \rho(x), \phi(y) \} =\del (x - y)]$ upon introducing a velocity potential $u(y) = \phi_y$. The corresponding EOM follow the Lagrangian ${\cal L}_1 = \rho_t \phi - {\cal H}$ where ${\cal H} = \half \rho \phi_x^2 + \rho \varepsilon(\rho) + \half \beta_* \rho_x^2/\rho$. As above, the local conservation laws for mass, momentum, energy and Galilei charge follow from Noether's theorem. However, the conservation law $u_t + F^{\rm u}_x = 0$ (\ref{e:unsteady-barotropic-eqns}) does not arise from a symmetry via Noether's theorem. This is because $C = \int u \: dx$ is a Casimir, it acts trivially on all observables: $\del \phi = \{ C, \phi \} = 0$, etc.

% M is not a Casimir of $(\rho,\phi)$ PB:  $\del \phi = \{ M , \phi \} = 1$.

%--------------
\section{Dispersive sound, steady and traveling waves}
\label{s:dispersive-sound-steady-trav}
%--------------

%------------
\subsection{Dispersive sound waves}
\label{s:dispersive-sound-waves}
%-----------

To discuss sound waves it is convenient to nondimensionalize the variables in (\ref{e:cont-entropy-eqn}) and (\ref{e:reg-vel-mom-eqn-p_*}):
	\beqs
	&& x = l \hat x,  \quad
	t = \frac{l}{\bar c} \hat t, \quad
	\rho = \bar \rho \hat \rho, \quad p = \bar p \hat p, \quad {\bar c}^2 = \frac{\g \bar p}{\bar \rho}, \cr
	&& u = \bar c \hat u, \quad
	\hat s = \frac{s}{c_v} = \log\left(\frac{\hat p}{\hat \rho^\g} \right).
	\label{e:non-dim-var}
	\eeqs
Here, $l$ is a macroscopic length. The nondimensional (hatted) variables satisfy
	\beqs
	&&\hat \rho_{\hat t} + (\hat \rho \hat u)_{\hat x} = 0, \quad
	\hat s_{\hat t} + \hat u \hat s_{\hat x} = 0 \quad \text{and}\cr
	&&\hat u_{\hat t} + \hat u \hat u_{\hat x} = - \ov{\g} \frac{\hat p_{\hat x}}{\hat \rho} + \frac{\eps^2}{\hat \rho} \left( \hat \rho_{\hat x \hat x} - \frac{\hat \rho_{\hat x}^2}{\hat \rho} \right)_{\hat x}.
	\label{e:non-dim-eoms}
	\eeqs
Here $\beta_* = {\bar c}^2 \la^2$ where $\la$ is a regularization length and $\eps = \la/l$ its nondimensional version.

\begin{figure}
	\begin{center} 
 \includegraphics[width = 8.7cm]{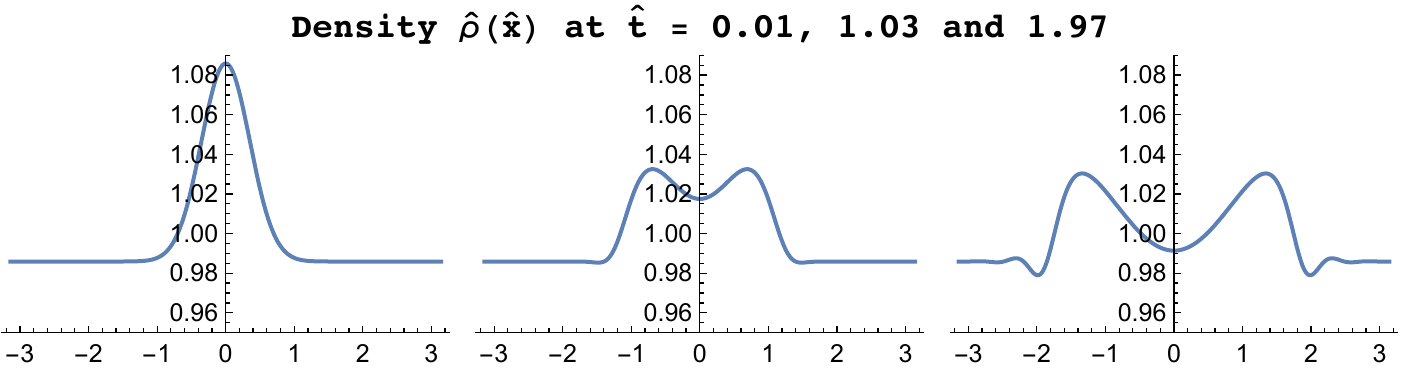} 
 \caption{R-gas dynamic evolution of a pulse ($\hat \rho(\hat x,0) = 1 + 0.1 e^{-4 \hat x^2}$ and $\hat u(\hat x, 0) = 0$) showing d'Alembert-like splitting of the pulse. Dispersion and nonlinearity modify the shape and produce `forerunners' and `backrunners'. The evolution is for $\g = 2$ and $\eps= 0.1$ with periodic BCs using the scheme of \S\ref{s:IVP-numerical} with $n_{\rm max} = 20$ Fourier modes, a time step $\D = 0.01$ and `nonlinearity strength' $\delta = 0.1$.}
 \label{f:gaussian-rho-compare-dalembert}
	\end{center}
	\end{figure}

A homogenous, stationary fluid [$\hat \rho = 1$, $\hat p = 1$, $\hat u = 0$ and $\hat s = 0$] is a solution of (\ref{e:non-dim-eoms}). To study sound, we consider linear perturbations $\hat p = 1 + \del \tl p$, $\hat \rho = 1 + \del \tl \rho$, $\hat s = \del \tl s$ and $\hat u = \del \tl u$ around this solution where $\del \ll 1$. The entropy equation upon linearization gives $\tl s_t = 0$. If we initially choose $\tl s(x,0) \equiv 0$ then $\tl s(x,t) \equiv 0$ and the entropy $\hat s(x,t) = \log\left({\hat p}/{\hat \rho^\g} \right) = 0$. Linearizing this we get $\tl p = \g \tl \rho$. The linearized continuity and velocity equations are 
	\beq
	\tl \rho_{\hat t} + \tl u_{\hat x} = 0 \quad \text{and} \quad
	\tl u_{\hat t} = - \tl \rho_{\hat x} + \eps^2 \tl \rho_{\hat x \hat x \hat x}.
	\label{e:sound-wave-eqn-dispersive}
	\eeq
Thus, we arrive at an equation for dispersive sound $\tl \rho_{\hat t \hat t} = \tl \rho_{\hat x \hat x} - \eps^2 \tl \rho_{4\hat x}$. The $4^{\rm th}$ derivative is reminiscent of elasticity, so our regularization force is like a tension. Fig.~\ref{f:gaussian-rho-compare-dalembert} shows the splitting of a pulse in density into two smaller pulses including effects of dispersion and weak nonlinearity.
Eqns. (\ref{e:sound-wave-eqn-dispersive}) have several conserved quantities including  
	\beqs
	 && M = \int \tl\rho \, d\hat x,\;\; P = \int \tl u \: d\hat x, \;\; H_1 = \half \int [\tl u^2 + \tl \rho^2 + \eps^2 \tl \rho_{\hat x}^2] d \hat x \cr
	&&\text{and} \quad
	H_2 = \half \int \left[ \tl \rho_t^2 + \tl \rho_{\hat x}^2 + \eps^2 \rho_{\hat x \hat x}^2  \right] d \hat x.
	\label{e:sound-waves-cons-qtys}
	\eeqs
Putting $\tl \rho \propto e^{i(k \hat x - \om \hat t)}$ we get a dispersion relation akin to that of linearized KdV ($u_t + uu_x = \epsilon u_{3x}$), $\om_{\rm KdV} = k + \eps k^3$:
	\beq
	\om^2 = k^2( 1 + \eps^2 k^2) \; \text{or} \; \om = \pm \left(k + \half \eps^2 k^3 + \cdots \right).
	\eeq
The phase velocity is $v_p = \om/k = \pm (1 + \eps^2 k^2)^{1/2} \approx \pm( 1 + \half \eps^2 k^2)$, while the group velocity is
	\beqs
	v_g &=& \dd{\om}{k} = \pm \frac{1 + 2 \eps^2 k^2}{\sqrt{(1 + \eps^2 k^2)}}
	\approx \pm (1 + 2 \eps^2 k^2)(1 - \half \eps^2 k^2 + ..) \cr
	&=& \pm \left( 1 + \frac{3}{2} \eps^2 k^2 + \cdots \right).
	\eeqs
Note that the regularization increases the phase speed while $|v_g|$ always exceeds $|v_p|$.

%-----------------
\subsection{Steady and traveling waves in one-dimension}
\label{s:steady-trav-quadrature}
%-----------------

Traveling waves are those where $\rho, u, p$ and $s$ are functions only of $(x-ct)$, where $c$ is the velocity of the wave. The entropy equation $s_t + u s_x = 0$ becomes $(u-c) \,s' = 0$. Thus, either $s = \bar s$ is a constant in space and time or $u = c$. In the former case, we have isentropic flow. In the latter, $s$ can be an arbitrary function of $(x - ct)$, but the fluid is at rest (`aerostatic') in a frame moving at velocity $c$. We will focus on the first possibility and look at steady solutions, subsequently `boosting' them to get traveling waves.

% {\Blue While we will consider aerostatic solutions, those with non-constant $s$ do not arise as a limit of non-aerostatic traveling wave solutions (with $u \not \equiv c$), so we do not study them here.}

%----------------
\subsubsection{Isentropic steady solutions}
\label{s:set-up-steady-barotropic}
%----------------

For steady flow $(c=0)$ the mass, momentum and velocity fluxes (\ref{e:barotropic-curr}) are constant:
	\beqs
	&& F^{\rm m} = \rho u, \quad 
	F^{\rm p} = \rho u^2 + (\g -1 ) K \rho^\g - \beta_* \left( \rho_{xx} - \frac{\rho_x^2}{\rho} \right)\cr
	&& \text{and} \quad F^{\rm u} = \half u^2 + \g K \rho^{\g-1} - \beta_* \left( \frac{\rho_{xx}}{\rho} - \half \frac{\rho_x^2}{\rho^2} \right).
	\label{e:fluxes-j-Pi-B}
	\eeqs
Moreover, the steady continuity equation $u \rho_x + u_x \rho = 0$ implies that the constant energy flux of Eqn. (\ref{e:barotropic-curr}) is not independent: $F^{\rm e} = F^{\rm m} F^{\rm u}$. Eliminating $u = F^{\rm m}/\rho$ we get two expressions for $\rho_{xx}$:
	\beqs
	\beta_* \rho_{xx} &=& - F^{\rm p} + \frac{{(F^{\rm m})}^2}{\rho} + (\g - 1) K \rho^\g + \beta_* \frac{\rho_x^2}{\rho} \quad \text{and}
	\cr
	\beta_* \rho_{xx} &=& - F^{\rm u} \rho + \frac{{(F^{\rm m})}^2}{2 \rho} + \g K \rho^{\g} + \frac{\beta_*}{2} \frac{\rho_x^2}{\rho}.
	\label{e:rhoxx-steady-eqn-two-versions}
	\eeqs
Taking a linear  combination allows us to eliminate the $\rho^\g$ term and arrive at the second order equation
	\beqs
	\beta_* \rho_{xx} &=& - V'(\rho) + \frac{(\g + 1) \beta_*}{2} \frac{\rho_x^2}{\rho}, \quad  \text{where} \cr 
	V'(\rho) &=& F^{\rm p} \g - F^{\rm u}(\g -1) \rho - \frac{(\g + 1) {(F^{\rm m})}^2}{2\rho}.
	\label{e:steady-reg-gas-eqn-rho}
	\eeqs
In Appendix \ref{a:parabolic-embed-LJ-id}, a different linear combination that eliminates the $\rho_x^2/\rho$ term is considered, leading to additional results. The current choice makes it easier to treat all values of $\g$ in a uniform manner. Interpreting $x$ and $\rho$ as time and position, this describes a Newtonian particle of mass $\beta_*$ moving in a (linear + harmonic + logarithmic) potential $V$ on the positive half-line subject also to a `velocity-dependent' force $\propto \rho_x^2/\rho$. This ensures that the motion is `time-reversal' $(x \to - x)$ invariant. The qualitative nature of trajectories is elucidated via a $\rho$-$\rho_x$ phase plane analysis in Appendix \ref{a:vect-fld-phase-portrait}. There are only two types of non-constant bounded solutions for $\rho(x)$: solitary waves of depression (cavitons) and periodic waves. The latter correspond to closed trajectories around an elliptic fixed point (O-point) in the phase portrait while cavitons correspond to the homoclinic separatrix orbit that encircles an O-point and begins and ends at a hyperbolic X-point to its right. The location of these fixed points are determined (for any $\gamma$) by the roots of the quadratic $V'(\rho) = 0$ whose discriminant $\D = \g^2 (F^{\rm p})^2 - 2(\g^2 -1) F^{\rm u}{(F^{\rm m})}^2$ must therefore be positive. In the generic non-aerostatic situation (i.e. $u \not \equiv 0$ or equivalently $F^{\rm m} \neq 0$), the only cases when we get non-constant bounded solutions for $\rho$ are (a) $F^{\rm p}, F^{\rm u} > 0$: both periodic solutions and cavitons and (b) $F^{\rm u} < 0$: only periodic solutions. 

\vspace{.1cm}

{\fl \bf Remark:} Eqn.~(\ref{e:steady-reg-gas-eqn-rho}) is a generalization of the Ermakov-Pinney equation \cite{ermakov, pinney} which corresponds to $V(\rho) = \rho^4/2$, $\g = 2$ and $\beta_* = 1$. This leads to an alternate approach to understanding (\ref{e:steady-reg-gas-eqn-rho}), since the transformation $z^2 = 1/\rho$ converts it into a Newton equation with sextic potential and no velocity dependent force for any $\g$:
	\beq
	z_{xx} =  \half (\g - 1) F^{\rm u} z - \g F^{\rm p} \, z^3 + \frac{(\g +1)}{4(F^{\rm m})^2} z^5.
	\eeq
% Moreover, the transformation $z^2 = 1/\rho$ reduces the Ermakov-Pinney equation to $z_{xx} = 1/{z^3}$.
{\fl \bf Reduction to quadrature:} Subtracting the two equations in (\ref{e:rhoxx-steady-eqn-two-versions}), we get a first order ordinary differential equation (ODE) for $\rho$: 
	\beqs
	&& \frac{\beta_*}{2} \rho_x^2 = -\frac{(F^{\rm m})^2}{2} + F^{\rm p} \rho - F^{\rm u} \rho^2 + K  \rho^{\g + 1} \equiv  \rho^{\g + 1} (K - U) \cr
	&& \text{where} \quad K = \half \left( \frac{\beta_*}{\rho^{\g + 1}} \right) \rho_x^2 + U \cr
	&& \text{and} \quad U (\rho)= \ov{\rho^{\g + 1}}\left[ \frac{(F^{\rm m})^2}{2} - F^{\rm p} \rho + F^{\rm u} \rho^2 \right].
	\label{e:rhox-vs-rho-diff-eqn}
	\eeqs
The `potential energy' $U$ is related to the potential $V$ via $\rho^{\g + 1} U'(\rho) = V' (\rho)$. This allows us to reduce the determination of the steady density profile $\rho(x)$ to quadrature:
	\beqs
	&& dx = \frac{d \rho}{\sqrt{(2/\beta_*)\rho^{\g + 1}(K - U(\rho))}} \cr 
	&& = \frac{d \rho}{\sqrt{(2/\beta_*) ( K  \rho^{\g + 1} - F^{\rm u} \rho^2 + F^{\rm p} \rho  -((F^{\rm m})^2/2))}}.
	\label{e:quadrature-steady-rho}
	\eeqs
For integer $\g \geq 1$ (\ref{e:quadrature-steady-rho}) is a hyperelliptic integral though it reduces to an elliptic integral when $\g = 2$ (see \S \ref{s:exact-cavitons-cnoidal-g-2}). For other values of $\g$, steady solutions may be found via the parabolic embedding of Appendix \ref{a:parabolic-embed-LJ-id}.

%-----------
\subsubsection{Nondimensionalizing the steady equation}
\label{s:non-dim-moduli}
%--------------

\begin{figure}	
\begin{center}
		\includegraphics[width=8cm]{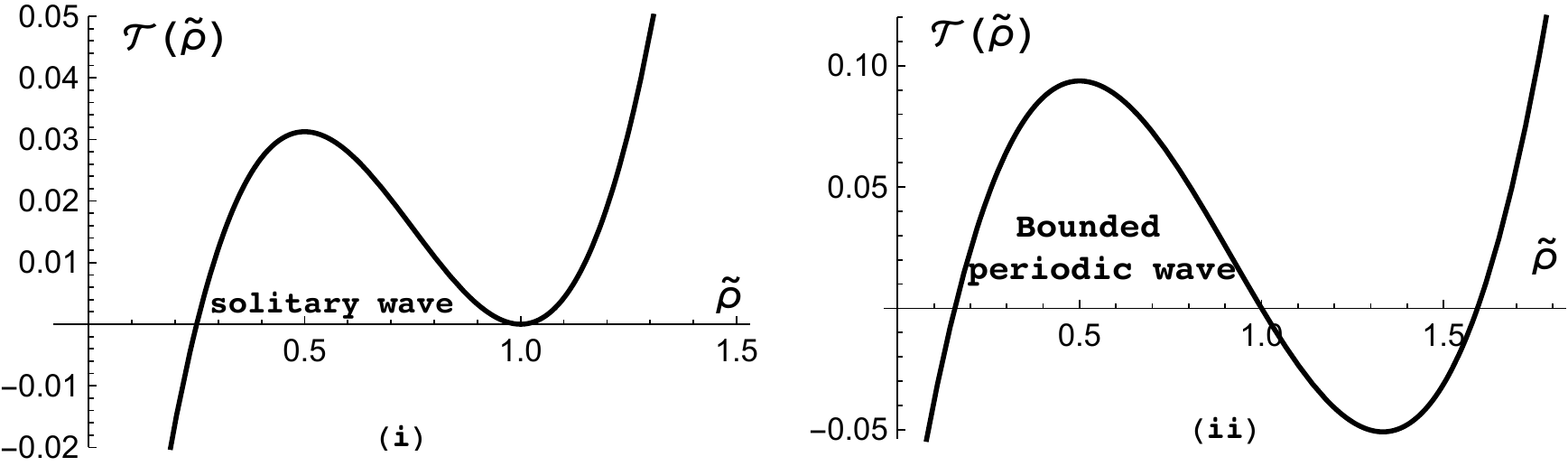}
		\end{center}
	\caption{(i) Density of caviton lies between simple and double root of ${\mathcal T}(\tl \rho)$ (\ref{e:non-linear-ode-for-rho-xi-M1-param}). (ii) Periodic wave density lies between simple roots where $\T > 0$. Here $\g = 2$ and in (i) $\ka_\circ = 0$, $M_\circ = 0.5$ while in (ii) $\ka_\circ = -0.25$ and $M_\circ = 0.5$.}
	\label{f:T-rho-vs-rho}
\end{figure}

To integrate (\ref{e:rhox-vs-rho-diff-eqn}), it is convenient to replace the four constants $(F^{\rm m}, F^{\rm p}, F^{\rm u}, K)$ with two dimensionless shape parameters $(\ka_\pt, M_\pt)$ and two dimensional ones $(\rho_\pt, c_\pt)$ that set scales. These parameters are adapted to the solutions one seeks to find: $\rho_\pt$ is the density at a point $x_\pt$ where $\rho_x = 0$. For a caviton, $\rho_\pt$ can be the asymptotic or trough density while for a periodic wave, it can be the trough or crest density (or the trough density of an unbounded solution with the same $K$). This choice will simplify the expressions for the constant fluxes (\ref{e:fluxes-j-Pi-B}) and $K$ when evaluated at $x_\pt$. For example, $K = U(\rho_\pt)$ gives 
	\beq
	K = \ov{\rho_\pt^{\g + 1}}\left[ \frac{(F^{\rm m}_\pt)^2}{2} - F^{\rm p}_\pt \rho_\pt + F^{\rm u}_\pt \rho^2_\pt \right] = \frac{p_\pt \rho_\pt^{-\g}}{\g - 1}	= \frac{c_\pt^2 \rho_\pt^{1 -\gamma}}{\gamma (\gamma - 1)}.
\label{e:K-at-rhostar}
	\eeq
where $p_\pt$ and $c_\pt$ are the pressure and sound speed at $x_\pt$. We may use $c_\pt^2$ to trade $\beta_*$ for a regularization length $\la_\pt = \sqrt{\beta_*}/c_\pt$ which is used to define the nondimensional position $\xi = x/\la_\pt$. Next, let $M_\pt^2 = {u_\pt^2}/{c_\pt^2}$ be the square of the Mach number at $x_\pt$. Positive $M_\pt$ corresponds to rightward flow at $x_\pt$ and vice-versa. We will take $M_\pt \geq 0$ with the remaining steady solutions obtained by taking $M_\pt \to - M_\pt$. Using these definitions, we rewrite (\ref{e:rhox-vs-rho-diff-eqn}) as a 1st order ODE for the nondimensional density $\tl \rho(\xi) = \rho(\la_\pt \xi)/\rho_\pt$:
	\beqs
	&& \half \left( \frac{d \tl \rho}{d\xi}\right)^2 = {\cal T}(\tl \rho) \equiv \frac{\tl \rho^{\g + 1}}{\g (\g - 1)} + \left(\ka_\pt - \frac{M_\pt^2}{2} - \ov{\g - 1} \right) \tl \rho^2 \cr
	&& \qquad \qquad \qquad \qquad \quad + \left(M_\pt^2 + \ov{\g} - \ka_\pt \right) \tl \rho - \half M_\pt^2 \cr
	&& = \frac{\tl \rho^{\g +1}}{\g(\g - 1)} - \half M_\pt^2 ( \tl \rho - 1)^2 - \frac{\tl \rho^2}{\g - 1} + \frac{\tl \rho}{\g} - \ka_\pt \, \tl \rho \left( 1 - \tl \rho \right).
	\label{e:non-linear-ode-for-rho-xi-M1-param}
	\eeqs
Here, $\ka_\pt = \la_\pt^2 \rho''(x_\pt)/\rho_\pt = \tl \rho''(\xi_\pt)$ measures the curvature of the density profile at $x_\pt = \la_\pt \xi_\pt$. A virtue of this nondimensionalization is that unlike the four parameters in (\ref{e:rhox-vs-rho-diff-eqn}), only two parameters $\ka_\pt$ and $M_\pt$ appear in (\ref{e:non-linear-ode-for-rho-xi-M1-param}). Eqn. (\ref{e:non-linear-ode-for-rho-xi-M1-param}) describes zero energy trajectories $\tl \rho(\xi)$ of a unit mass particle in the potential $- {\cal T}(\tl \rho)$. Evidently, the allowed values of $\tl \rho$ must lie between adjacent positive zeros of ${\cal T}$ with ${\cal T} > 0$ in between (Fig.~\ref{f:T-rho-vs-rho}). To obtain (\ref{e:non-linear-ode-for-rho-xi-M1-param}), we used the following expressions for the constant fluxes and entropy: 
	\beqs
	&& (F^{\rm m})^2 = \rho_\pt^2 c_\pt^2 M_\pt^2, \quad
	F^{\rm p} = \rho_\pt c_\pt^2 \left( M_\pt^2 + \ov{\gamma} - \ka_\pt \right), \cr
	&& F^{\rm u} = c_\pt^2 \left( \frac{M_\pt^2}{2} + \ov{\gamma - 1} - \ka_\pt \right) \quad \text{and} \quad 
	K = \frac{c_\pt^2 \rho_\pt^{1 -\gamma}}{\gamma (\gamma - 1)}.
	\label{e:steady-sol-currents}
	\eeqs 
Conversely, we may invert (\ref{e:steady-sol-currents}) by first determining $\rho_\pt$ by solving the algebraic equation
	\beq
	K = U(\rho_\pt) = \rho_\pt^{-(1 + \g)}\left[ \frac{(F^{\rm m})^2}{2} - F^{\rm p} \rho_\pt + F^{\rm u} \rho_\pt^2 \right]
	\label{e:determine-rho-star-from-K}
	\eeq
following from (\ref{e:rhox-vs-rho-diff-eqn}). The remaining new parameters follow from (\ref{e:steady-sol-currents}) and (\ref{e:fluxes-j-Pi-B}):
	\beq
	c_\pt^2 =  \frac{K \g(\g -1)}{\rho_\pt^{1 - \g}}, \;\; M_\pt^2 = \frac{(F^{\rm m})^2}{\rho_\pt^2 c_\pt^2} 
	\;\; \text{and} \;\; \ka_\pt = \frac{(F^{\rm m})^2 -  \rho_\pt F^{\rm p}}{\rho_\pt^2 c_\pt^2}  + \frac{1}{\g}.
	\eeq

%---------------
\subsubsection{Exact cavitons and cnoidal waves for $\g = 2$}
\label{s:exact-cavitons-cnoidal-g-2}
%---------------

\begin{figure*}	
	\centering
	\begin{subfigure}[t]{5cm}
	\centering
		\includegraphics[width=5cm]{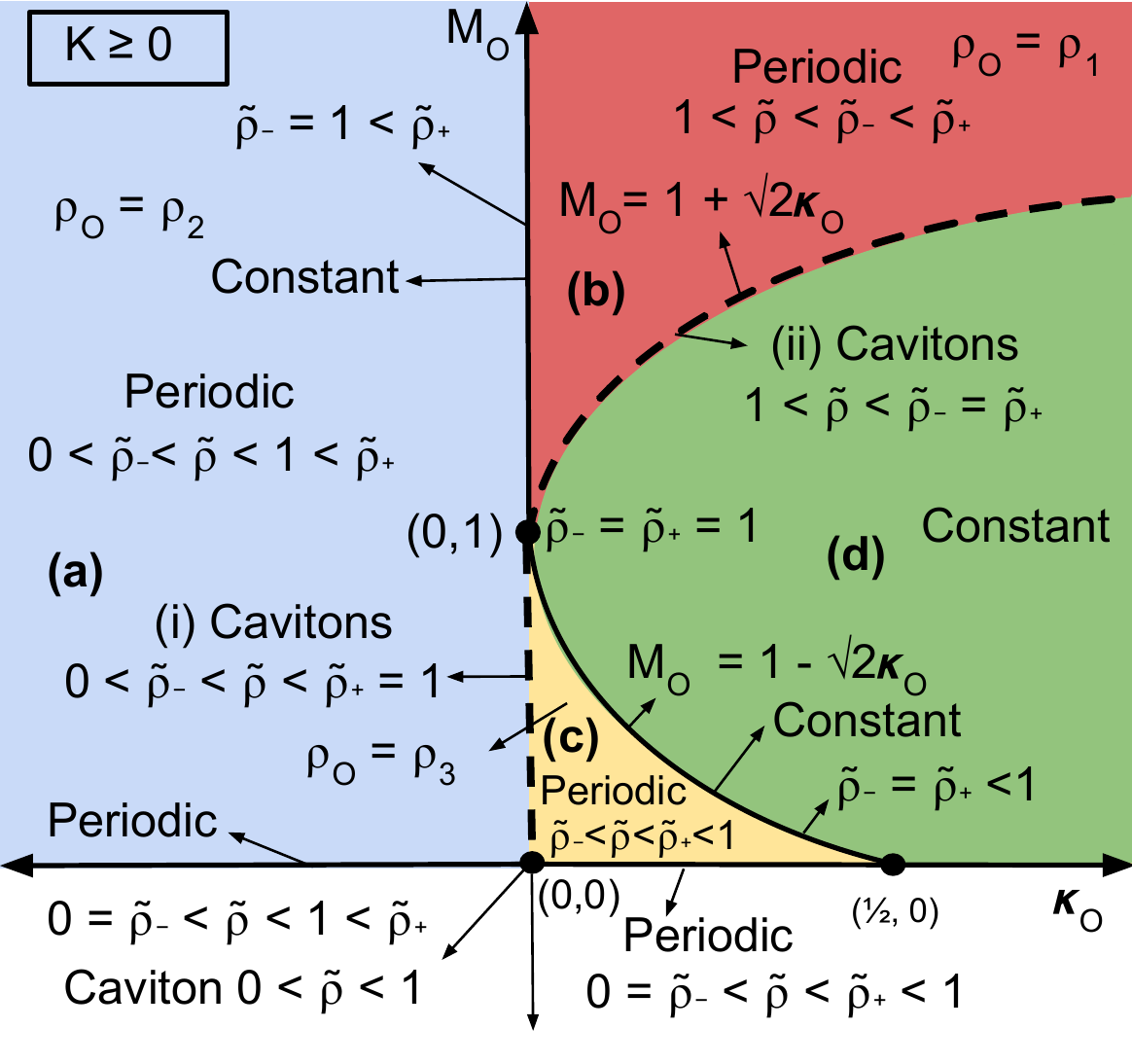}
	\caption{}
	\label{f:kappa-M-K>0}	
	\end{subfigure} \qquad 
	\begin{subfigure}[t]{5cm}
		\includegraphics[width=5cm]{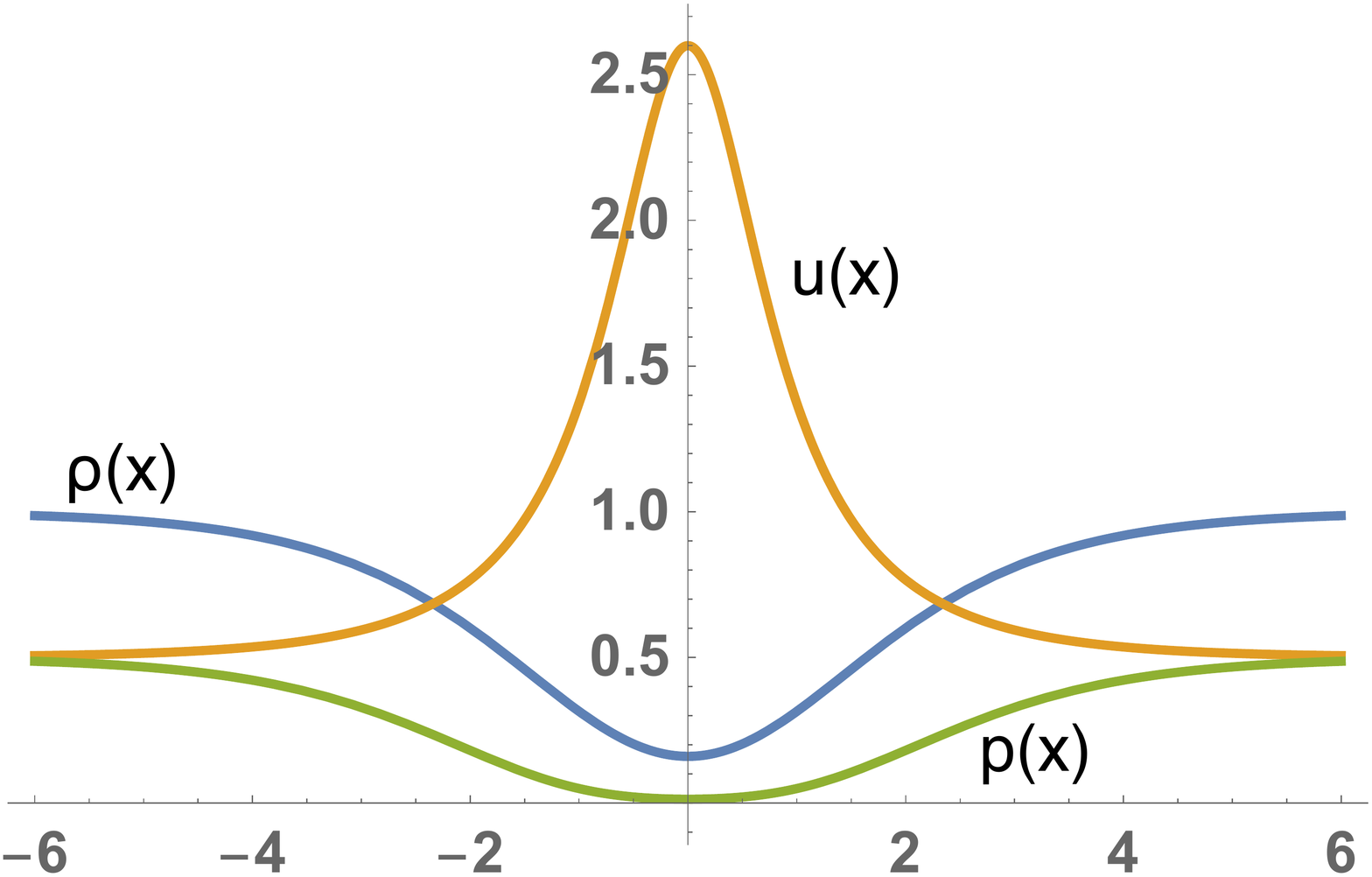}
		\caption{}	
	\end{subfigure}
	\qquad 
	\begin{subfigure}[t]{5cm}
	\centering
		\includegraphics[width=5cm]{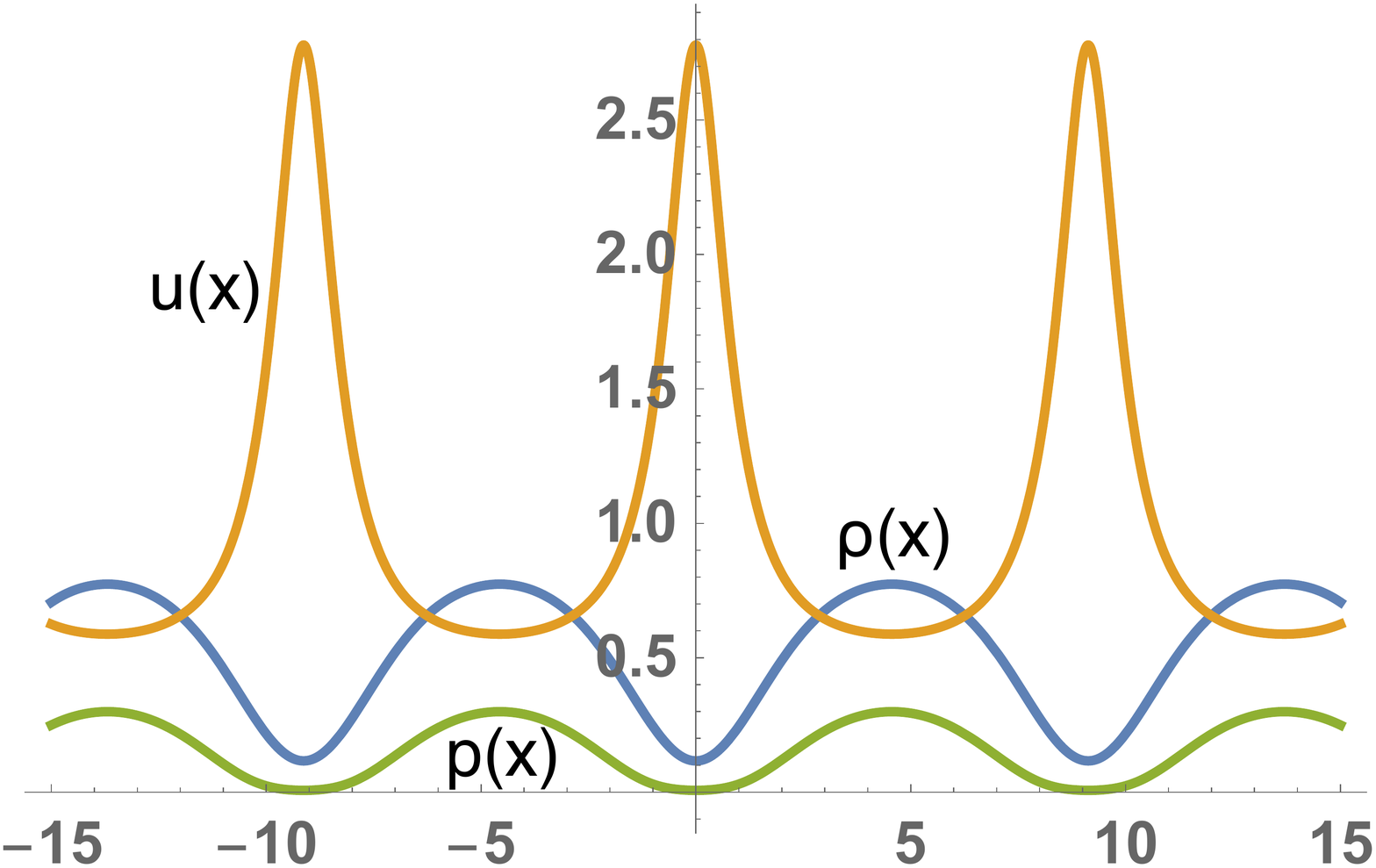}
		\caption{}
	\end{subfigure}
	\caption{\label{f:crho-u-p-caviton-traveling-cnoidal} (a) Nature of steady bounded solutions (\ref{e:non-linear-ode-for-rho-xi-M1-param}) for $\gamma = 2$ in $\kappa_\pt$-$M_\pt$ parameter space. (b)  and (c) Density, velocity and pressure for $K = 1/2$ for (b) traveling cavitons ($\rho_\pt =c_\pt =1, M_\pt = 0.5, \ka = 0$ and $\tl c = 0.1$) and (c) traveling periodic cnoidal waves $\rho_\pt = c_\pt = 1, M_\pt = 0.5, \ka = 0.1$ and $\tl c = 0.2$ (lower triangular region in $|M_\pt - \tl c|$-$\ka_\pt$ plane). Cavitons are solutions where the temperature drops isentropically in a small region of size $\la_\circ$.}
\end{figure*}

For $\g = 2$, ${\cal T}(\tl \rho)$ (\ref{e:non-linear-ode-for-rho-xi-M1-param}) becomes a cubic with roots $\tl \rho = 1 $ and 
	\beq
	\tl \rho_\pm  = \half \left(1 + M_\pt^2 - 2 \ka_\pt \pm \sqrt{ (1 + M_\pt^2 - 2 \ka_\pt)^2 -4 M_\pt^2} \right).
	\label{e:steady-rho-plus-minus}
	\eeq
The density of periodic and solitary waves must lie between adjacent positive roots with ${\cal T} > 0$ in between. Interestingly, it can be shown that if for $\g = 2$, all three roots of $\T$ are positive, then the same holds for any $1 < \g < 2$. So some qualitative features of solutions for $\g = 2$ are valid more generally. For $\g = 2$, the nature of solutions on the $\ka_\pt$-$M_\pt$ plane (Fig. \ref{f:kappa-M-K>0}) changes when the two relevant roots coalesce, i.e., when the discriminant of the cubic ${\cal T}$ vanishes:
	\beqs
	\Delta(\ka_\pt, M_\pt) &=& \left[(\tl\rho_+ - \tl\rho_-)(\tl\rho_- - 1)(1 - \tl\rho_+)\right]^2 \cr
	&=& 4\ka_\pt^2 \left[M_\pt^4 - 2 M_\pt^2 (1 + 2 \ka_\pt) + (1 - 2 \ka_\pt)^2\right] = 0.\qquad 
	\eeqs
$\D$ vanishes only along the vertical axis $\ka_\pt = 0$, the two parabolic curves $M_\pt = 1 \pm \sqrt{2\ka_\pt}$ and their reflections in the $\ka_\pt$-axis. In what follows, we restrict to rightward flow by taking $M_\pt \geq 0$. There are three regions in the upper half $\ka_\pt$-$M_\pt$ plane (pictured in Fig.~\ref{f:kappa-M-K>0}, colors online) admitting {\it periodic solutions}: (a) the blue second quadrant $\ka_\pt < 0$, (b) the red north-east region above $M_\pt = 1 + \sqrt{2 \ka_\pt}$ and (c) the yellow triangular region below the curve $M_\pt = 1 - \sqrt{2 \ka_\pt}$. In the green wedge (d) lying within the parabola but above the horizontal axis, non-constant solutions are unbounded since either $\tl \rho_\pm$ are negative (when $M_\pt < -1 + \sqrt{2 \ka_\pt}$) or not real (when $M_\pt > -1 + \sqrt{2 \ka_\pt}$). 

{\it Solitary waves} (cavitons) occur only on the dashed boundaries (i) between (a) and (c) ($0 < M_\pt < 1$, $\ka_\pt = 0$) and (ii) between (b) and (d) ($M_\pt = 1 + \sqrt{2 \ka_\pt}$). When $\ka_\pt = M_\pt = 0$, ${\cal T} =  \tl \rho (\tl \rho - 1)^2/2$ so that we have an aerostatic ($u \equiv 0$) caviton with $0 \leq \tl \rho \leq 1$. {\it Constant solutions} occur when $\tl\rho \equiv$ any zero of ${\cal T}$. ${\cal T}$ has a double zero $(\tl \rho = 1)$ along the vertical axis, the double zero $\tl\rho_+ = \tl\rho_-$ along the curve $M_\pt = 1- \sqrt{2\ka_\pt}$ and the triple zero $\tl \rho = 1$ at $(\ka_\pt = 0, M_\pt = 1)$. At all other points, ${\cal T}$ has either one or three positive simple zeros.

Interestingly, when we reinstate dimensions, the periodic solutions from regions (a), (b) and (c) of the $\ka_\pt$-$M_\pt$ plane (Fig \ref{f:kappa-M-K>0}) are physically identical. They differ by the choice of nondimensionalizing density $\rho_\pt$ which could be any one of the roots of the cubic in (\ref{e:determine-rho-star-from-K}). Thus, the parameters $(\ka_\pt, M_\pt, c_\pt, \rho_\pt)$ generically furnish a 3-fold cover (redundancy) of the original space of constants $((F^{\rm m})^2, F^{\rm u}, F^{\rm p}, K)$. For solitary waves, it degenerates into a double cover: the two families of cavitons ((i) and (ii)) in Fig.~\ref{f:kappa-M-K>0} differ via the choice of $\rho_\pt$ as trough or asymptotic density. Moreover, $M_\pt$ is the Mach number at the trough in (i) while it is the asymptotic Mach number in (ii). In a caviton, the flow goes from asymptotically subsonic to supersonic at $x = 0$. A caviton is like a pair of normal shocks joined at the trough. Finally, the map between the two sets of parameters  becomes a 1-fold cover for the aerostatic cavitons at $\ka_\pt = M_\pt = 0$, since their trough densities vanish and $\rho_\pt$ can only be chosen as the asymptotic density. 

In light of the above remarks, we now restrict attention to the yellow triangular region (c) where $\ka_\pt > 0$ and $0 < M_\pt < 1 - \sqrt{2 \ka_\pt}$. Here, the roots of ${\cal T}$ are $0 < \tl \rho_- < \tl \rho_+ < 1$ and (\ref{e:non-linear-ode-for-rho-xi-M1-param}) is reduced to quadrature:
	\beqs
	&& \xi(\tl\rho) - \xi(\tl\rho_-) = \int_{\tl\rho_-}^{\tl\rho} \frac{d\rho'}{\sqrt{(\rho' - \tl\rho_-)(\rho' - \tl\rho_+)(\rho' - 1)}} \cr
	&& = \frac{2}{\sqrt{1 - \tl\rho_-}} F\left( \arcsin \sqrt{\frac{\tl\rho - \tl\rho_-}{\tl\rho_+ - \tl\rho_-}} \;\vert\;  {\frac{\tl\rho_+ - \tl\rho_-}{1 - \tl\rho_-}} \right).
	\eeqs
Here, $F(\phi\vert m) = \int_0^\phi {(1 - m \sin^2\theta)^{-1/2}} d\theta$ is the incomplete elliptic integral of the first kind with $m$ the square of the elliptic modulus $k$ (see 17.4.62 of \cite{abramowitz-stegun}). Inverting, we write $\tl \rho (\xi)$ in terms of the Jacobi $\cn(u,m)$ function:
	\beq
	\tl \rho (\xi) = \tl \rho_+ - (\tl \rho_+ - \tl \rho_-) \: \cn^2 \left( \frac{\sqrt{1 - \tl \rho_-}}{2}(\xi - \xi_-), \frac{\tl \rho_+ - \tl \rho_-}{1 - \tl \rho_-}\right).
	\label{e:cnoidal-wave-rho-til-triangle}
	\eeq
Here $\xi_- = \xi(\tl \rho_-)$. The periodic wave extends from a trough at $\tl \rho_-$ to a crest at $\tl \rho_+$ with amplitude and wavelength 
	\beqs
	&& {\cal A}_{\ka_\pt > 0} = \sqrt{ (1 + M_\pt^2 - 2 \ka_\pt)^2 -4 M_\pt^2} \quad \text{and} \cr 
	&& {\Lambda}_{\ka_\pt > 0} = \int_{\tl \rho_-}^{\tl \rho_+} \frac{2 \: d \tl \rho}{\sqrt{2 {\cal T}(\tl \rho)}} = \frac{4}{\sqrt{1 - \tl \rho_-}} K \left( \frac{\tl \rho_+ - \tl \rho_-}{1 - \tl \rho_-}\right). \quad
	\label{e:amplitude-waveleng-steady-cnoidal}
	\eeqs
Here $K(m) = F(\frac{\pi}{2}|m)$ is the complete elliptic integral of the $1^{\rm st}$ kind. When we approach the left boundary $\ka_\pt \to 0^+$ with $0 < M_\pt < 1$, the wavelength diverges ($K(1/2 \ka_\pt) \sim \sqrt{\ka_\pt} \log \ka_\pt$ for small $\ka_\pt$) and the periodic waves turn into cavitons which extend from a trough density $\tl \rho = \tl \rho_- = M_\pt^2 < 1$ to an asymptotic density $\tl \rho = \tl \rho_+ = 1$:
	\beq
	\tl \rho(\xi) = 1 - (1 - M_\pt^2) \sech^2 \left( \sqrt{\frac{1-M_\pt^2}{4}} \xi \right) \;\; \text{for} \;\; 0 \leq M_\pt \leq 1.
%	= M_\pt^2 - (M_\pt^2 -1) \tanh^2 \left( \sqrt{\frac{1-M_\pt^2}{4}} \xi \right).
	\label{e:caviton-nondim-gamma=2}
	\eeq
On the other hand, when we approach the upper boundary $M_\pt = 1 - \sqrt{2 \ka_\pt}$, ${\cal A} \to 0$ and we get constant solutions. By contrast, on the lower boundary $(M_\pt = 0, 0< \ka_\pt < \half)$ we continue to have periodic solutions except that $\tl \rho$ vanishes at the trough ($\tl \rho_- = 0$) while the crest density $\tl \rho_+ = 1 - 2 \ka_\pt$:
	\beqs
	 \tl \rho (\xi) &=& (1 - 2 \ka_\pt) \,\text{sn}^2 \left( \frac{\xi}{2}, 1 - 2\ka_\pt \right) \quad \text{and} \cr
	{\Lambda} &=&  2 i \left[ \sqrt{\frac{2}{\ka_\pt}} K \left( \ov{2\ka_\pt} \right) - 2 K(2\ka_\pt) \right].
	\label{e:cnoidal-wave-gamma=2}
	\eeqs
When $\ka_\pt = M_\pt = 0$ in (\ref{e:caviton-nondim-gamma=2})/(\ref{e:cnoidal-wave-gamma=2}), we get an {\it aerostatic caviton} which reaches {\it vacuum conditions} at $\xi = 0$:
	\beq
	\tl \rho(\xi) =  \tanh^2(\xi/2) \quad \text{with} \quad u \equiv 0.
	\label{e:aerostatic-caviton-Mpt-0}
	\eeq
The dimensional $\rho, u$ and $p$, are got by reinstating the constants ($\la_\pt$, $\rho_\pt$, $c_\pt$). Writing $x = \la_\pt \xi$ we have
	\beq
	\rho(x) = \rho_\pt \tl \rho\left( \xi \right), \; 
	u(x) =  \frac{c_\pt M_\pt}{\tl \rho(\xi)} \quad \text{and} \quad
	p(x) = \frac{c_\pt^2 \rho_\pt}{\g} {\tl \rho(\xi)}^\g
	\eeq
for isentropic flow. Reversing the sign of $M_\pt$ reverses the flow direction leaving $p$ and $\rho$ unaltered. Moreover, since $\tl \rho \geq 0$ and $u = F^{\rm m}/\rho$, the flow is unidirectional with positive velocity solitary waves being waves of elevation in $u$ and vice-versa. A caviton is superficially a bifurcation of the constant solution $\rho(x) \equiv \rho({\pm\infty})$. However, though the caviton and constant solutions have the same constant specific entropy, they have different values of mass and energy (per unit length). For instance, the energy density (\ref{e:beta-reg-hamiltonian}) of an aerostatic caviton is less than that of the constant state:
    \beq
	{\cal E}_{\rm aerostatic \; caviton} = p_\infty \left[ 1 - \frac{2 \rho(x)}{\rho_\infty} \left( 1 - \frac{\rho(x)}{\rho_\infty}  \right) \right] < {\cal E}_{\rm const.} = p_\infty.
    \eeq
As a consequence, the uniform state cannot, by any isentropic time-dependent motion with fixed BCs at $\pm \infty$, in the absence of sources and sinks, evolve via R-gas dynamics to the caviton, or vice versa, since the two states have different invariants. However, one could imagine creating, say an aerostatic caviton, by starting with a uniform state and introducing a point sink at the origin, to suck fluid out. A symmetrical pair of expansion waves would travel to infinity on both sides, without affecting the conditions at infinity. When the density reaches $0$ at the origin, we stop the sink. The pressure gradient will then be balanced by the regularizing density gradient force, and the solution should tend to the aerostatic caviton as $t\to \infty$. Since temperature $T = K m (\g-1) \rho^{\g-1}$, the caviton corresponds to a region of width $\lambda$ where the temperature drops. Loosely, the regularizing force is like Pauli's exchange repulsion, capable of maintaining a depression in density with variable but isentropic temperature and pressure distributions.

%---------------
\subsubsection{Traveling waves of isentropic R-gas dynamics}
\label{s:traveling-waves}
%-----------

Here we generalize the above steady solutions to waves traveling at speed $c$: $(\rho,u,p)(x,t) = (\rho,u,p)(x-ct)$. The continuity, velocity and momentum equations (\ref{e:unsteady-barotropic-eqns}) are readily integrated, giving the constant fluxes: 
	\beqs
	&& F^{\rm m} = \rho({u - c}),\;\; 
	F^{\rm u} = \frac{u^2}{2} - c u - \beta_* \left[ \frac{\rho''}{\rho} -  \frac{\rho'^2}{2 \rho^2} \right] + \g K \rho^{\g-1}\cr 
	&&  \text{and} \quad 
	F^{\rm p} = \rho u( u - c) + (\g - 1) K \rho^\g - \beta_* \left(\rho'' - \frac{\rho'^2}{\rho} \right).
	\eeqs 
Eliminating $u = c + F^{\rm m}/\rho$ and taking a linear combination leads us to a $2^{\rm nd}$ order nonlinear ODE for $\rho$: 
	\beqs
	\beta_* \rho'' &=& - V'(\rho) + \frac{(\g + 1) \beta_*}{2} \frac{\rho'^2}{\rho} \quad \text{where} \cr
	 V'(\rho) &=&   \g (c F^{\rm m}  - F^{\rm p}) - (\g -1) \left(\frac{c F^{\rm p}}{F^{\rm m}} - F^{\rm u}  - \frac{c^2}{2} \right) \rho \cr
	&& + \frac{(\g +1) (F^{\rm m})^2}{2 \rho}.
	\label{e:traveling-reg-gas-eqn-rho}
	\eeqs 
Proceeding as in \S \ref{s:set-up-steady-barotropic} and \S \ref{s:non-dim-moduli}, this may be reduced to the nonlinear first order equation
	\beqs
	&& \half \left( \DD{\tl \rho}{\xi} \right)^2 = {\cal T}( \tl \rho) \equiv\frac{\tl \rho^{\g + 1}}{\g(\g-1)} -  \half (\tl \rho - 1)^2 (\tl c - M_\pt)^2 \cr
	&& \qquad \qquad \qquad \qquad \quad - \frac{\tl \rho^2}{\g - 1} + \frac{\tl \rho}{\g} - \ka_\pt \tl \rho ( 1 - \tl \rho)  \quad \text{where} \cr
	&&  \beta_* = \la_\pt^2 c_\pt^2, \quad  \tl \rho = \frac{\rho}{\rho_\pt}, \quad \tl c = \frac{c}{c_\pt}, \quad M_\pt = \frac{u_\pt}{c_\pt}, \quad \ka_\pt = \tl \rho''(\xi_\pt)  \cr
	&& \text{and} \quad \xi = \frac{x - ct}{\la_\pt} \quad \text{with} \quad \tl \rho'(\xi_0) = 0.
	\eeqs
Comparing with (\ref{e:non-linear-ode-for-rho-xi-M1-param}), we see that the passage from steady to traveling waves involves only a shift in the square of the Mach number $M_\pt^2 \to (M_\pt - \tl c)^2$. Here, $\tl c$ is the speed of the traveling wave in units of the sound speed $c_\pt$ at the point where $\rho'(x - ct) = 0$. Thus, for each value of $\beta_*$, we have a 5-parameter ($\ka_\pt, M_\pt, \rho_\pt,c_\pt, \tl c$) space of traveling cavitons and periodic waves. The dimensionful $\rho, u$ and $p$ for any value of $(M_\pt - \tl c)$ are given by
	\beqs
	&& \rho(x,t) = \rho_\pt \tl \rho\left( \xi \right), \quad 
	u(x,t) = c_\pt \left( \frac{M_\pt - \tl c}{\tl \rho(\xi)} + \tl c \right) \quad \text{and} \cr
	&& p(x,t) = \frac{c_\pt^2 \rho_\pt}{\g} {\tl \rho(\xi)}^\g
\quad \text{where} \quad \xi = \frac{x- \tl c \,  c_\pt t}{\la_\pt}
	\eeqs

%----------------

\vspace{.1cm}

{\fl \bf Traveling cavitons for $\g = 2$:} In these nondimensional variables, traveling cavitons have the profile
	\beq
	\tl \rho(\xi) = 	
	 1 - (1 - (\tl c - M_\pt)^2) \sech^2 \left( \sqrt{\frac{1-(\tl c - M_\pt)^2}{4}} \xi \right),
	\eeq
for $(\tl c - M_\pt )^2 < 1$. These cavitons are reminiscent of the solitary waves of depression/elevation of the KdV equation $u_t \mp 6 u u_x + u_{xxx}=0$ that move rightward with speed $c > 0$: $u(x,t)= \mp (c/2) \sech^{2}\,\left((\sqrt{c}/2)(x - ct) \right)$. Just as in KdV, narrower cavitons (width $\propto 1/\sqrt{|\tl c - M_\pt|-1}$) are taller (height $\propto |\tl c - M_\pt|-1$) and have a higher speed relative to the speed at the reference location where the density has an extremum ($\tl c - M_\pt \propto c - u_\pt$ is the speed of the traveling wave in the rest frame of the fluid at the reference location).

%------------------

\vspace{.1cm}

{\fl \bf Traveling cnoidal waves for $\g = 2$:} The nondimensional density profile of traveling cnoidal waves is
	\beqs
	&& \tl \rho(\xi) = 
\tl \rho_+ - (\tl \rho_+ - \tl \rho_-) \;\text{cn}^2 \left( \frac{\sqrt{1 - \tl \rho_-}}{2}(\xi - \xi_-), \frac{\tl \rho_+ - \tl \rho_-}{1 - \tl \rho_-}\right) \cr
	&& \quad \text{for} \quad 0 < \ka_\pt < \half 
	\quad \text{and} \quad |M_\pt -\tl c| < 1 - \sqrt{2 \ka_\pt}.
	\eeqs
Here, $\tl \rho_\pm$ are got by replacing $M_\pt^2 \to (M_\pt - \tl c)^2$ in (\ref{e:steady-rho-plus-minus}).
Their wavelength is given by the same formula (\ref{e:amplitude-waveleng-steady-cnoidal}). Our cnoidal waves are very similar to those of KdV $(u_t - 6 u u_x + u_{xxx} = 0)$:
	\beq
	u= f(\xi) = f_2 \left[1 - \:\text{cn}^2 \,\left[ \sqrt{\frac{f_1 - f_2}{2}} (\xi - \xi_3), \: \frac{f_2 - f_3}{f_1 - f_3} \right]\right] + f_3.
	\eeq
Here $ \xi = x - ct$ and $f(\xi_3) = f_3$ and $f_{1,2,3}$ are the roots of $f^3 + \half c f^2 + A f + B$ with $A$ and $B$ constants of integration.

%--------------
\section{Weak form and shock-like profiles}
\label{s:weak-and-shock-like-sol}
%--------------

%--------------
\subsection{Weak form of R-gas dynamic equations}
\label{s:weak-form}
%--------------

The R-gas dynamic equations ((\ref{e:cont-entropy-eqn}) and (\ref{e:reg-vel-eqn-gas})) involve $u_x, p_x, \rho_x, \rho_{xx}$ and $\rho_{xxx}$. Thus, classical solutions need to be $C^1$ in $u$ and $p$ and $C^3$ in $\rho$. However, by multiplying the conservation equations by $C^\infty$ test functions $\phi, \psi$ and $\zeta$ and integrating by parts, we arrive at a weak form of the equations:
	\beqs
	&& \int [\rho_t \phi - (\rho u) \phi_x] dx = 0,\cr
	&& \int \left[ (\rho u)_t \psi - \left(p + \rho u^2 + \beta_* \frac{\rho_x^2}{\rho} \right) \psi_x + \beta_* \rho \psi_{xxx} \right] dx = 0 \cr
	&& \text{and} \quad \int \left[\frac{\rho u^2}{2} + \frac{p}{\g-1} + 
	\frac{\beta_*  \rho_x^2}{2 \rho} \right]_t \zeta dx \cr
	&& = \int \left[ \left[\frac{\rho u^3}{2} + \frac{\g p u}{\g - 1} +  \frac{3\: \beta_* \:u }{2} \frac{\rho_x^2}{\rho} \right] \zeta_x - \beta_* u \rho_x \zeta_{xx} \right] dx.
	\label{e:weak-form-of-Rgas-dyn}
	\eeqs
Thus, for weak solutions, it suffices that $\rho$ be $C^1$ and $u$ and $p$ be merely continuous.

%--------------
\subsection{Steady shock-like profile from half a caviton}
\label{s:patched-shock-weak-sol}
%--------------

\begin{figure}	
\begin{center}
	\begin{subfigure}[t]{5cm}
		\includegraphics[width=5cm]{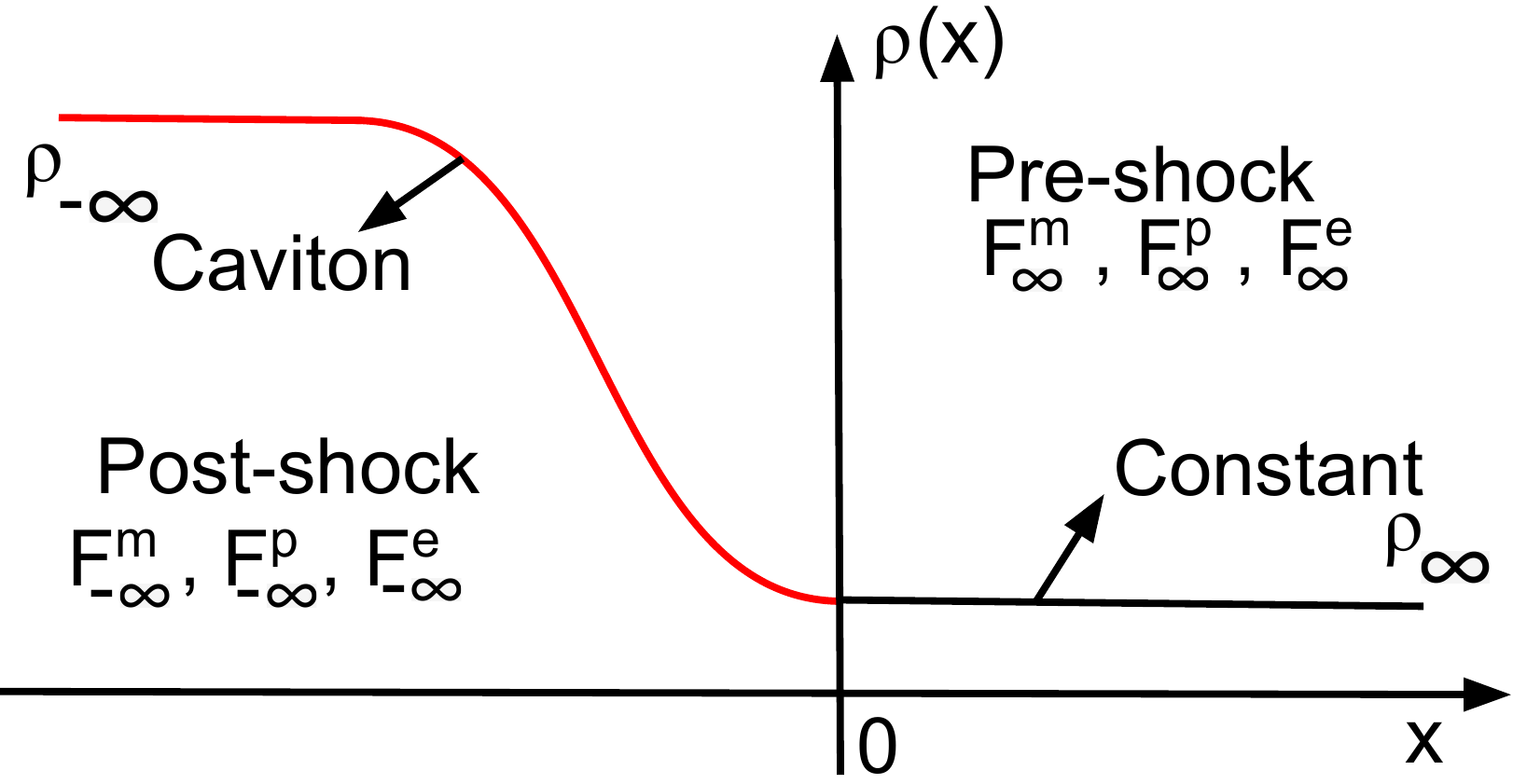}
	\end{subfigure}
		\end{center}
	\caption{Steady shock-like profiles from patching half a caviton with a constant state violate one or more Rankine-Hugoniot conditions.}
	\label{f:patched-shock}
\end{figure}

Here, we try to use the steady solutions of \S \ref{s:steady-trav-quadrature} to model the structure of a normal shock propagating to the right in the lab frame. As in Fig.~\ref{f:patched-shock}, in the shock frame, the shock is assumed to be located around $x = 0$. The undisturbed pre-shock medium is to the right $(x > 0)$ while the `disturbed' post-shock medium is to the left ($x<0$) \cite{whitham}. As $x \to \pm \infty$ the variables approach the asymptotic values $\rho_{\pm \infty}, u_{\pm \infty}$ and $p_{\pm \infty}$ with $\rho_{+ \infty} < \rho_{-\infty}$. The Rankine-Hugoniot (RH) conditions are obtained by equating the conserved fluxes $F^{\rm m}$ (\ref{e:cont-entropy-eqn}), $F^{\rm p}$ (\ref{e:vel-mom-r-gas-dynm}) and $F^{\rm e}$ (\ref{e:reg-total-energy-PB}) at $x = \pm \infty$:
	\beqs
	&& (\rho u )_{-\infty} = (\rho u )_{\infty}, \quad (\rho u^2 + p)_{-\infty} = (\rho u^2 + p)_{\infty} \quad \text{and} \cr
	&& \left( \half \rho u^2 u + \frac{\g}{\g - 1}pu \right)_{-\infty} = \left( \half \rho u^2 u + \frac{\g}{\g - 1}pu \right)_{\infty}.
	\eeqs 
In our $\g = 2$ cavitons (\ref{e:caviton-nondim-gamma=2}), the flow is subsonic at $x = \pm \infty$ and supersonic at $x = 0$. We exploit this in trying to find a shock-like steady solution by patching half a caviton with a constant solution. Thus, we seek a steady solution where $\rho(x) \equiv \rho_{\infty}$ in the pre-shock medium and is the left half of a caviton for $x < 0$. The half-caviton profile is got from (\ref{e:caviton-nondim-gamma=2}) by taking the reference location $x_\pt = -\infty$ so that $\rho_\pt = \rho_{-\infty}$ and $\ka_\pt = 0$:
	\beqs
    && \rho(x) = \rho_{-\infty} \left[1 - (1 - M_{-\infty}^2) \sech^2 \left[ \sqrt{\frac{1-M_{-\infty}^2}{4}} \frac{x}{\la_{-\infty}} \right] \right]\cr
    && \qquad \qquad \text{for} \quad x < 0 \quad \text{with} \quad 0 \leq M_{-\infty}^2 \leq 1.
	\eeqs
Here $\la_{-\infty}^2 = \beta_*/c_{-\infty}^2$. On the other hand, $\rho_{+\infty}$ must correspond to a constant solution with fluxes $F^{\rm m, p, e}_{\infty}$. It can take one of two density values corresponding to the X/O point (\ref{e:rho-pm-fixedpts-steady}):
	\beq
	 \rho_{\infty}^{\rm X,O} = \frac{F^{\rm m}_{\infty} (2 F^{\rm p}_{\infty} \pm \sqrt{\D_{\infty}})}{2 F^{\rm e}_{\infty}} \;\; \text{with} \;\; \D = 4 (F^{\rm p}_{\infty})^2 - 6 F^{\rm e}_{\infty} F^{\rm m}_{\infty}.
	\eeq
We now attempt to patch these two solutions requiring $\rho$ and $\rho_x$ to be continuous at $x = 0$. Since $\rho$ has a local minimum at the trough of a caviton, $\rho_x(0^-) = 0 = \rho_x(0^+)$ of the undisturbed medium. However, a difficulty arises in trying to ensure that $\rho$ is continuous at $x=0$. Indeed, suppose we impose the RH conditions, $F^{\rm m}_\infty = F^{\rm m}_{- \infty}$, $F^{\rm p}_{-\infty} = F^{\rm p}_{\infty}$ and $F^{\rm e}_{-\infty} = F^{\rm e}_{\infty}$, then both the pre- and post-shock regions correspond to a common phase portrait. We observe (see also Fig. \ref{f:phase-por-caviton}) that the caviton trough density $\rho(0^-) = \rho_{- \infty} M_{-\infty}^2$ is strictly less than both $\rho_{\infty}^{\rm X,O}$. Thus, the post-shock semi-caviton solution cannot be continuously extended into the pre-shock region.

Stated differently, in the above patched shock construction, if $\rho_{+\infty}$ is chosen to be equal to $\rho(0^-)$ in order to make $\rho$ continuous, then the RH conditions are violated. Let us illustrate this with a $\g = 2$ aerostatic example. We take the pre-shock region to be vacuum ($\rho = u = p \equiv 0$ for $x > 0$) and try to patch this at $x=0$ with the left half of the aerostatic caviton [$\rho(x) = \rho_{-\infty} \tanh^2(x/2 \la_{-\infty})$ and $u \equiv 0$] of (\ref{e:aerostatic-caviton-Mpt-0}). This caviton corresponds to the values $F^{\rm m}_{-\infty} = F^{\rm e}_{-\infty} = \ka_{-\infty} = M_{-\infty} = 0$ and has trough density $\rho(0^-) = 0$ with trough density gradient $\rho_x(0^-) = 0$ as well. Since the caviton is isentropic, its trough pressure $p(0^-) = K \rho(0^-)^2 = 0$. Thus, $\rho$ and $p$ are both $C^1$ at $x=0$, while $u \equiv 0$ so that this is a weak solution in the sense of \S\ref{s:weak-form}. However, it violates the RH conditions: while $F^{\rm m} \equiv 0$ is continuous, $F^{\rm p}$ and $F^{\rm u}$ are not. In fact, in the pre-shock vacuum region $F^{\rm p} = F^{\rm u} = 0$ while in the post-shock region they are non-zero, as evaluating them at  $x = - \infty$ shows:
	\beq
	F^{\rm p}_{-\infty} = p(-\infty) = K_- \rho_{-\infty}^2 \neq 0 
	\quad \text{and} \quad
	F^{\rm u}_{- \infty} = 2 K_- \rho_{-\infty} \neq 0.
	\eeq
In the post-shock region $K_- = F^{\rm p}_{- \infty}/\rho_{-\infty}^2 \neq 0$ whereas in the pre-shock vacuum $K$ is arbitrary since $p = \rho = 0$ for $x > 0$. In conclusion, there are no continuous steady shock-like solutions in the shock frame that satisfy the RH conditions. To see how initial conditions (ICs) that  would lead to shocks in the ideal model are regularized, we turn to a numerical approach.

%--------------
\section{Numerical solutions to the initial value problem}
\label{s:IVP-numerical}
%--------------

%------------
\subsection{Spectral method with nonlinear terms isolated}
\label{s:numerical-scheme}
%------------

In this section, we discuss the numerical solution of the isentropic R-gas dynamic initial value problem (IVP). It is convenient to work with the nondimensional variables $\hat \rho$, $\hat u$, $\hat s$, etc. of \S \ref{s:dispersive-sound-waves}. The continuity equation is $\hat \rho_{\hat t} + (\hat \rho \hat u)_{\hat x} = 0$ while for isentropic flow, $\hat s$ is a global constant which may be taken to vanish by adding a constant to entropy. Thus, we can eliminate $\hat p = \hat \rho^\g$ in the velocity and momentum equations, both of which are in conservation form (\ref{e:unsteady-barotropic-eqns}): 
	\beqs
	&& \hat u_{\hat t} + \left[\half \hat u^2 + \ov{\g-1} \hat \rho^{\g - 1} - \eps^2 \left( \frac{\hat \rho_{\hat x \hat x}}{\hat \rho} - \half \frac{\hat \rho_{\hat x}^2}{\hat \rho^2} \right) \right]_{\hat x} = 0 \quad \text{or} \cr
	&& (\hat \rho \hat u)_{\hat t} + \left[ \hat \rho \hat u^2 + \ov{\g} \hat \rho^\g - \eps^2 \left( \hat \rho_{\hat x \hat x} - \frac{\hat \rho_{\hat x}^2}{\hat \rho} \right) \right]_{\hat x} = 0.
	\label{e:non-dim-barotropic-vel-momentum}
	\eeqs
The energy equation is $\pdr_{\hat t} \hat{\cal E} + \hat F^{\rm e}_{\hat x} = 0$ where the energy density and flux are
	\beqs
	&& \hat {\cal E} = \half \hat \rho \hat u^2 + \frac{\hat \rho^\g}{\g(\g - 1)} + \frac{\eps^2}{2} \frac{\hat \rho_{\hat x}^2}{\hat \rho}
	\quad \text{and} \cr 
	&& \hat F^{\rm e} = \left[\frac{\hat \rho \hat u^2}{2} + \frac{\hat \rho^\g}{\g-1} \right] \hat u - \eps^2 \left[ \hat u \hat \rho_{\hat x \hat x} - \frac{3}{2}  \frac{\hat u\hat \rho_{\hat x}^2}{\hat \rho} - \hat u_{\hat x} \hat \rho_{\hat x} \right].
	\eeqs
These equations follow from the PB $ \{ \hat \rho(\hat x), \hat u(\hat y) \} = \pdr_{\hat y} \del(\hat x - \hat y)$ and the Hamiltonian
	\beq
	\hat H = \int \left[\half \hat \rho \hat u^2 + \frac{\hat \rho^\g}{\g(\g - 1)} + \frac{\eps^2}{2} \frac{\hat \rho_{\hat x}^2}{\hat \rho} \right] d\hat x.
	\eeq
% In these variables, the dimensions of the PB $1/{\bar c}^2 \bar \rho l^2$ (see (\ref{e:non-dim-var})) cancels out. Moreover, if we introduce the velocity potential $\hat u = \hat \phi_{\hat x}$, then $\{ \hat \rho(\hat x), \hat \phi(\hat y) \} = \del(\hat x - \hat y)$ and the Bernoulli and continuity equations take the canonical forms $\hat \phi_{\hat t} = - \deldel{\hat H}{\hat \rho}$ and $\hat \rho_{\hat t} = \deldel{\hat H}{\hat \phi}$.
We will consider ICs that are fluctuations around a uniform state. For stability of the numerics, we separate the linear and nonlinear terms in the equations and treat the former implicitly and the latter explicitly. Introducing the book-keeping parameter $\del$ (which  will also enter through the ICs and may eventually be set to 1), we write
	\beq
	\hat \rho(\hat x, \hat t) = 1 + \del\, \tl \rho(\hat x, \hat t) \quad \text{and} \quad
	\hat u(\hat x, \hat t) = \thickbar u(\hat x, \hat t) + \del\, \tl u(\hat x, \hat t).
	\eeq
We will consider flow in the domain $-\pi \leq \hat x \leq \pi$ with periodic BCs and thus expand $\tl \rho$ and $\tl u$ as	\beq
 	\tl \rho = \sum_{ -\infty}^{\infty} \rho_n(\hat t) e^{in \hat x}, \quad
	\tl u = \sum_{ -\infty}^{\infty}  u_n(\hat t) e^{in \hat x} \;\; \text{with} \;\; {(\rho,u)}_{-n} = (\rho,u)_n^*.
	\eeq
Since $\int \hat \rho \: d \hat x$ is conserved, $\rho_0$ can be taken time-independent. Furthermore, we choose the constant $\bar \rho$ used to nondimensionalize $\rho$ as the (conserved) average density, so that $\rho_0 = 0$. We also suppose that the `background' flow velocity $\thickbar u$ is independent of position $\hat x$. Since $\int \hat u \: d\hat x$ is conserved and $\int \hat u \: d\hat x = 2\pi (\thickbar u(\hat t) + \del \: u_0(\hat t))$, we may absorb $\del \: u_0(\hat t)$ into $\thickbar u(\hat t)$ and thereby take $u_0 = 0$. 
Next, we write the continuity equation with nonlinear terms isolated:
	\beq
	\tl \rho_{\hat t} = - (\thickbar u \tl \rho + \tl u)_{\hat x} - \del\, {\calF^{\rm m}} \quad \text{where} \quad {\calF^{\rm m}} = (\tl \rho \tl u)_{\hat x}.
	\label{e:cont-linear-nonlinear-split}
	\eeq
The velocity equation (\ref{e:non-dim-barotropic-vel-momentum}) in conservation form is
	\beqs
	&& \del \tl u_{\hat t} + \left( \frac{(\thickbar u + \del \tl u)^2}{2} + \ov{\g-1} (1 + \del \tl \rho)^{\g-1} \right)_x \cr
	&& - \eps^2 \del \left( \frac{\tl \rho_{\hat x \hat x}}{1 + \del \tl \rho} - \frac{\del}{2} \frac{\tl \rho_{\hat x}^2}{(1 + \del \tl \rho)^2} \right)_{\hat x} = 0.
	\eeqs
Separating out the linear part we get
	\beqs
	\tl u_{\hat t} &=& - \left(\thickbar u \tl u + \tl \rho - \eps^2 \tl \rho_{\hat x \hat x} \right)_{\hat x} - \del \calF^{\rm u}
	\qquad \text{\rm where} \cr
	\calF^{\rm u} 
%	&=& \left[ \frac{\tl u^2}{2} 
%	+ \ov{\del^2} \ov{\g-1} \left\{ (1 + \del \tl \rho)^{\g-1} - \del (\g-1) \tl \rho \right\} 
%	- \frac{\eps^2}{\del} \left( \frac{\tl \rho_{\hat x \hat x}}{1 + \del \tl \rho} - \tl \rho_{\hat x \hat x} - \frac{\del}{2} \frac{\tl \rho_{\hat x}^2}{(1 + \del \tl \rho)^2} \right) \right]_{\hat x} \cr
	&=& \left( \frac{\tl u^2}{2} 
	+ \ov{\del^2}\ov{\g-1} \left\{ (1 + \del \tl \rho)^{\g-1} - \del (\g-1) \tl \rho \right\}\right)_x\cr
	&& + \left(\frac{\eps^2}{(1 + \del \tl \rho)^2} \left(  \tl \rho \tl \rho_{\hat x \hat x} ( 1 +  \del \tl \rho)  + \half \tl \rho_{\hat x}^2  \right)\right)_{\hat x}.
	\label{e:vel-linear-nonlinear-split} 
	\eeqs
In (\ref{e:cont-linear-nonlinear-split}) and (\ref{e:vel-linear-nonlinear-split}) the linear terms are at ${\cal O}(\del^0)$ while the nonlinearities involve $\del$ to higher powers, depending on the value of $\g$. Interestingly, for $\g = 2$, the pressure gradient doesn't contribute to the nonlinear part of the acceleration:
	\beq
	\calF^{\rm u}_{\g=2} = \left[ \frac{\tl u^2}{2} 
	+ \frac{\eps^2 }{(1 + \del \tl \rho)^2}\left( \tl \rho \tl \rho_{\hat x \hat x}( 1+ \del \tl \rho)  + \half \tl \rho_{\hat x}^2  \right) \right]_{\hat x}.
	 \eeq
Expanding $\calF^{\rm m} = \sum_n \calF^{\rm m}_n e^{i n \hat x}$ and $\calF^{\rm u} = \sum_n \calF^{\rm u}_n e^{i n \hat x}$, the EOM in Fourier space become
	\beqs
	&& \dot \rho_n = - i n (\thickbar u \rho_n + u_n) - \del \calF^{\rm m}_n \quad
	\text{and} \cr
	&& \dot u_n = - i n (\thickbar u u_n + (1 + \eps^2 n^2) \rho_n) - \del \calF^{\rm u}_n.
	\label{e:cont-vel-baro-fourier}
	\eeqs
When nonlinearities ($\calF^{\rm m}$, $\calF^{\rm u}$) are ignored and we assume ($\rho_n(t), u_n(t) \propto e^{- i \om_n t}$), one finds the dispersion relation $(\om_n - n \thickbar u)^2 = n^2 (1 + \eps^2 n^2)$ familiar from \S \ref{s:dispersive-sound-waves}. To deal with the fully nonlinear evolution given by (\ref{e:cont-vel-baro-fourier}), we use a semi-implicit numerical scheme outlined in Appendix \ref{a:numerical-scheme}.

%----------
\subsection{Numerical results: Avoidance of gradient catastrophe, solitons and recurrence}
\label{s:numerical-results}
%----------

The above numerical scheme for $\g = 2$ is implemented by truncating the Fourier series after $n_{\rm max} = 16$ modes. The evolution is done for 750 time steps $(0 \leq \hat t \leq 15)$, each of size $\D = 0.02$, starting with the nondimensional ICs
	\beqs
	\hat \rho(\hat x,0) &=& 1 + \del \tl\rho(\hat x,0) \quad \text{and} \quad \hat u(\hat x,0) = 0 \quad \text{where} \cr
	\tl \rho(\hat x,0) &=& \sin \hat x \quad \text{or} \quad \cos 2 \hat x \quad\text{and} \quad \del = 0.1.
	\eeqs
We find that at early times, where $\hat u$ has a negative slope, its gradient increases and decreases where the slope is positive. Without the regularization $(\eps = 0)$, the higher Fourier modes can then get activated and the velocity and density profiles become highly oscillatory with steep gradients. Moreover, amplitudes begin to grow and the code eventually ceases to conserve energy and momentum.

\begin{figure}
\begin{center} 
 \includegraphics[width = 8.7cm]{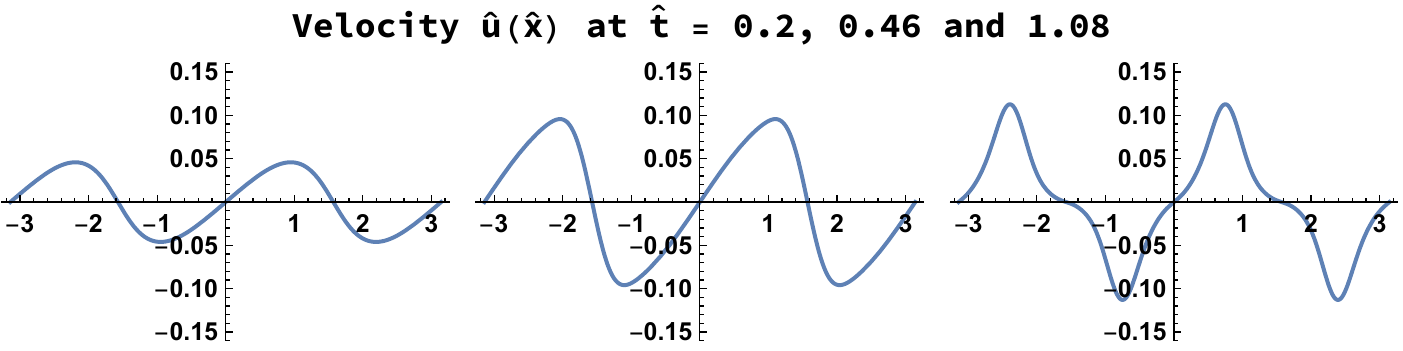} \caption{Evolution of velocity for IC $\hat \rho = 1 + 0.1 \cos{2 \hat x}$ and $\hat u = 0$ showing how the gradient catastrophe is averted through the formation of a pair of solitary waves in the velocity profile.}
 \label{f:u-reg-soliton-pair-forms}
 \end{center}
\end{figure}

\begin{figure}
  \begin{subfigure}[t]{5cm}
    \includegraphics[width=5cm]{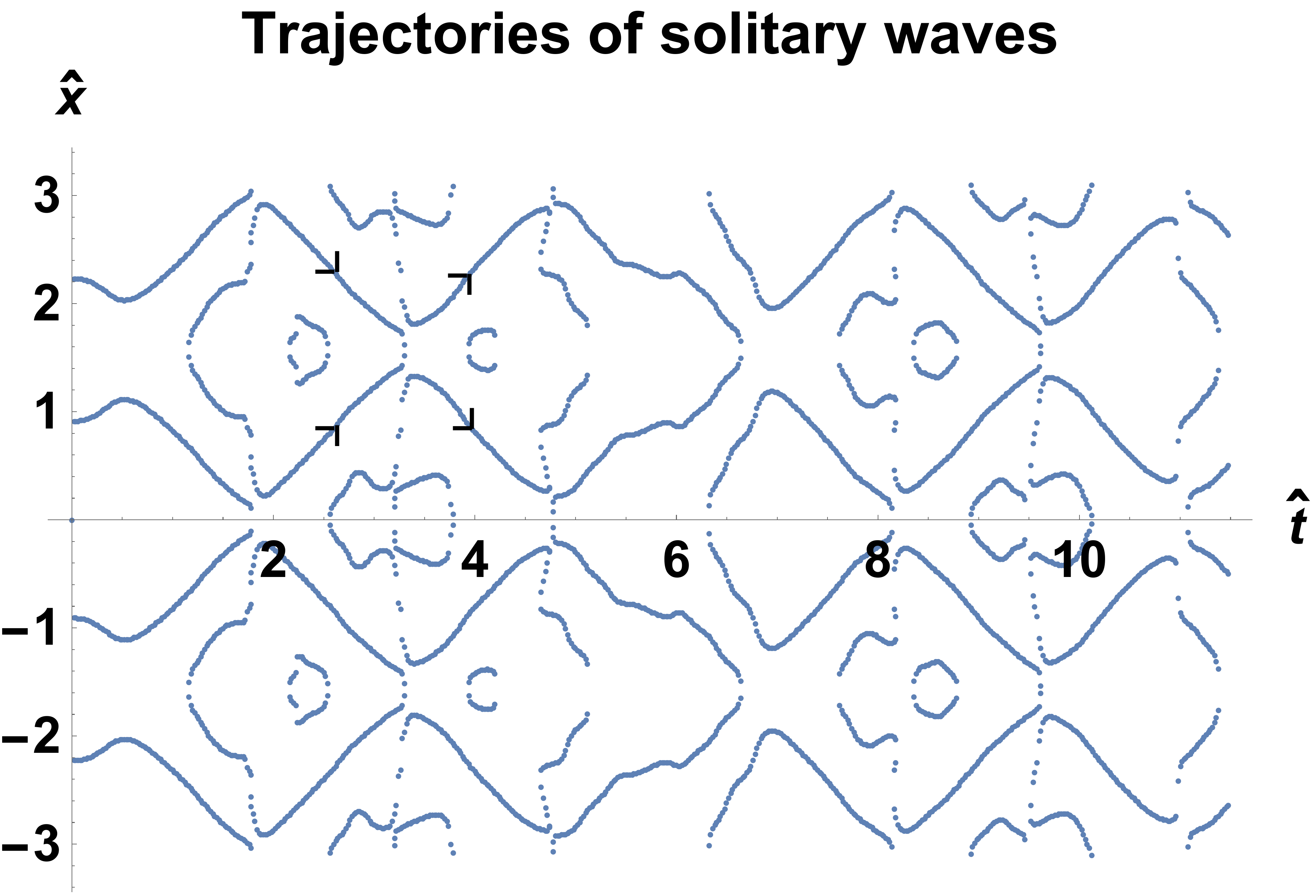}
    \caption{}
  \end{subfigure}
  \begin{subfigure}[t]{5cm}
    \includegraphics[width=5cm]{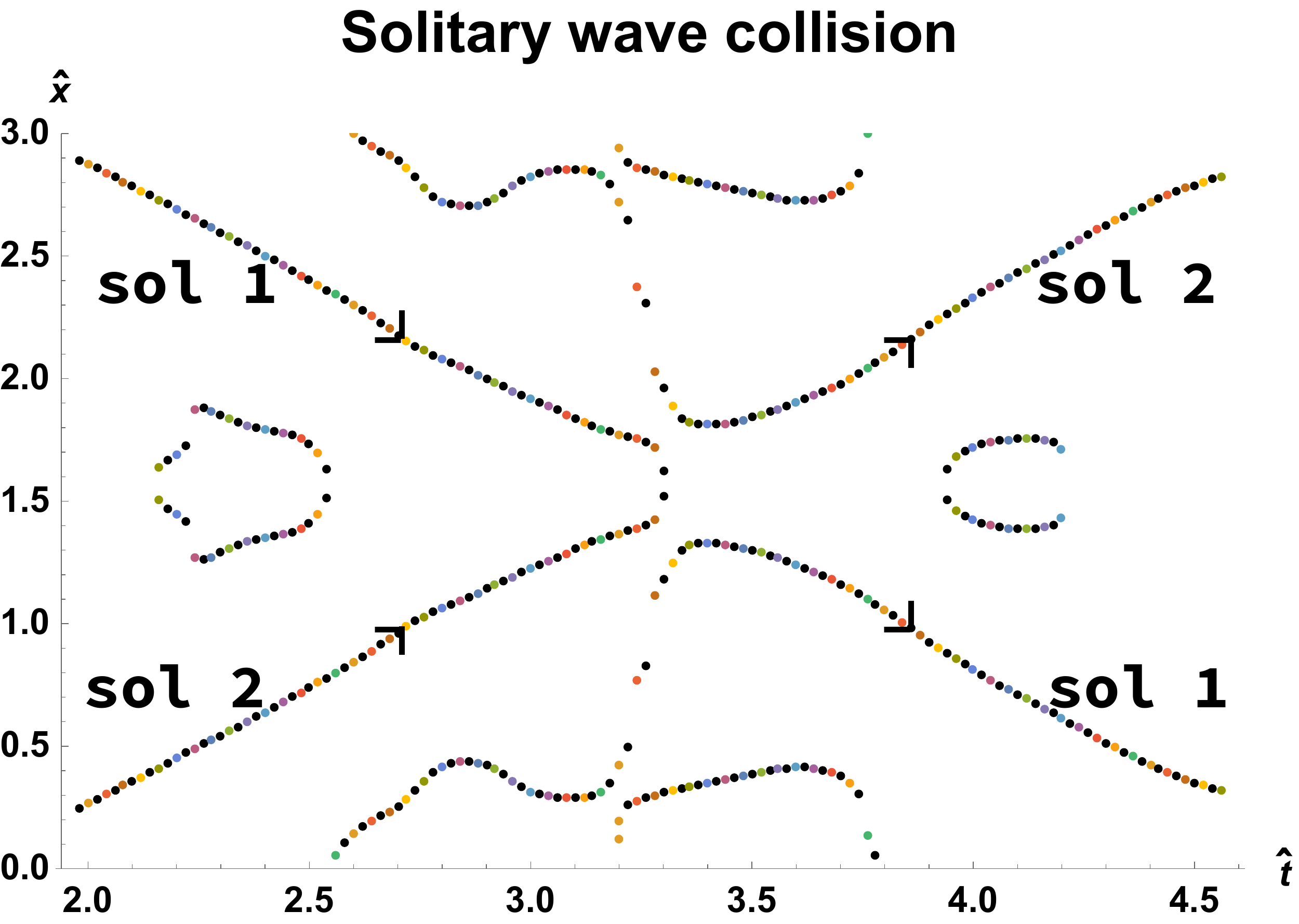}
    \caption{}
  \end{subfigure}
\caption{\label{f:solitary-wave-coll-cos2x} (a) Space-time trajectories of the `centers' of `solitary waves' in the velocity profile (for initial condition $\hat \rho = 1 + 0.1 \cos{2 \hat x}$ and $\hat u = 0$) showing several collisions. The locations of the `centers' are determined by finding the maxima/minima of $\hat u$ at each instant of time. (b) Close-up of one collision of solitary waves (sol 1 and sol 2), showing approximate asymptotic straight-line trajectories and phase shifts. The figures include the trajectories of the crests/troughs of small ripples that typically arise and go away in pairs but do not qualify as solitary waves.}
\end{figure}

By contrast, in the presence of the regularization (say $\eps = 0.2$), we find that the above gradient catastrophe is averted and energy is conserved (to about 3 parts in 1000) while momentum is conserved to machine precision. In fact, we find that the real and imaginary parts  $Q_r$ and $Q_i$ of the next conserved quantity (\ref{e:Qr-Qi}) are also conserved. Interestingly, when $u$ develops a steep negative gradient, a pair of solitary waves emerge at the top (wave of elevation) and bottom (wave of depression) of the $u$ profile and the gradient catastrophe is avoided (see Fig.~\ref{f:u-reg-soliton-pair-forms}). This mechanism by which the incipient shock-like discontinuity is regularized is to be contrasted with KdV, where an entire train of solitary waves can form \cite{kruskal-zabusky}. However, as with KdV, our solitary waves can suffer a head-on collision and pass through each other. After the collision, they re-emerge with roughly similar shapes and a phase shift. Fig. \ref{f:solitary-wave-coll-cos2x} shows the space-time trajectories of the centers of several of these solitary waves, showing their collisions. 

\begin{figure*}
 \includegraphics[width = 4cm]{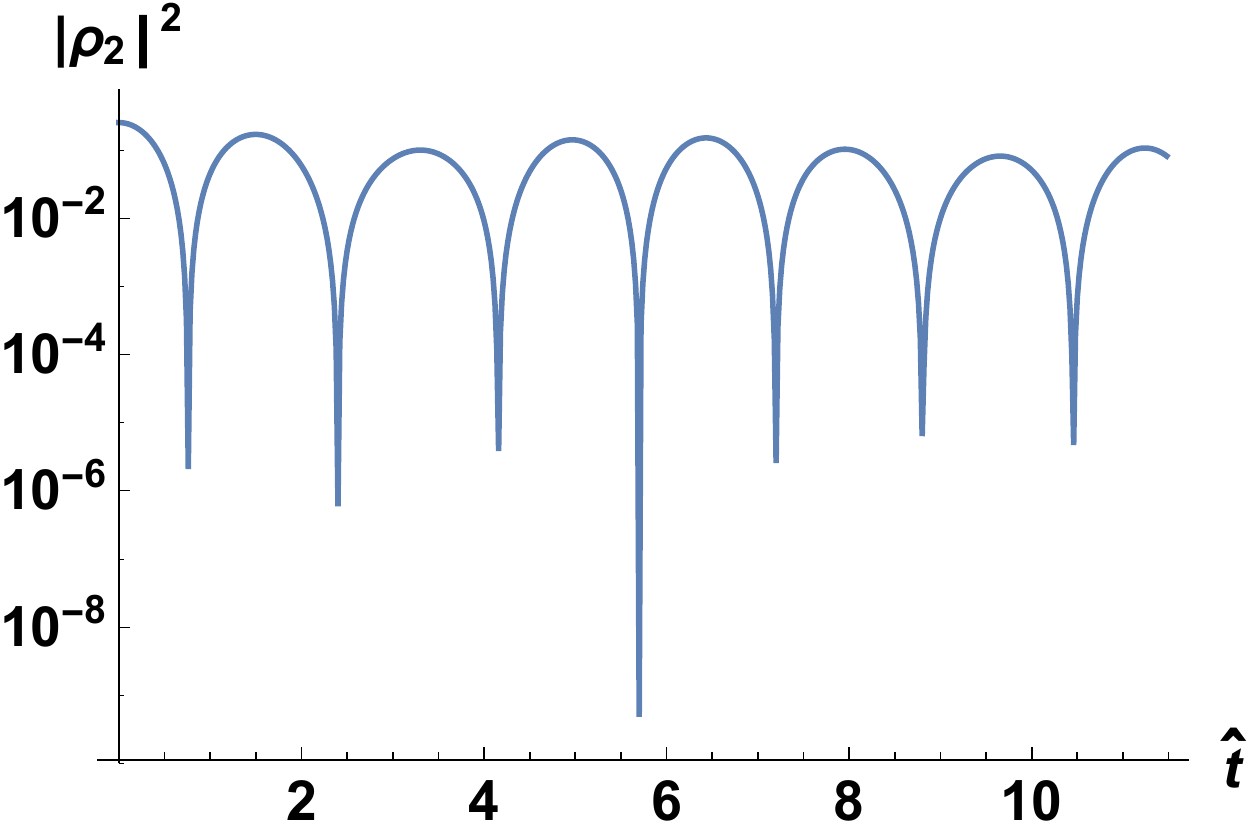}\qquad 
\includegraphics[width = 4cm]{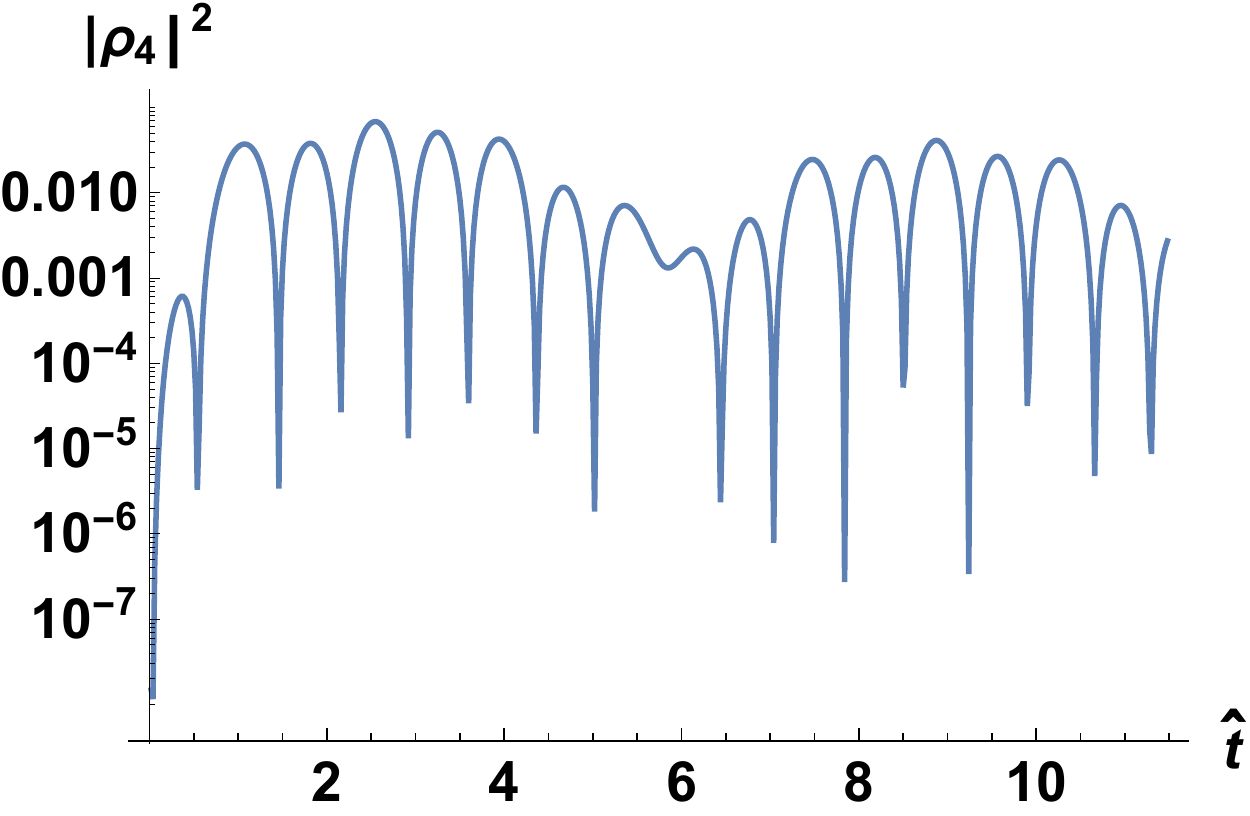}\qquad
 \includegraphics[width = 4cm]{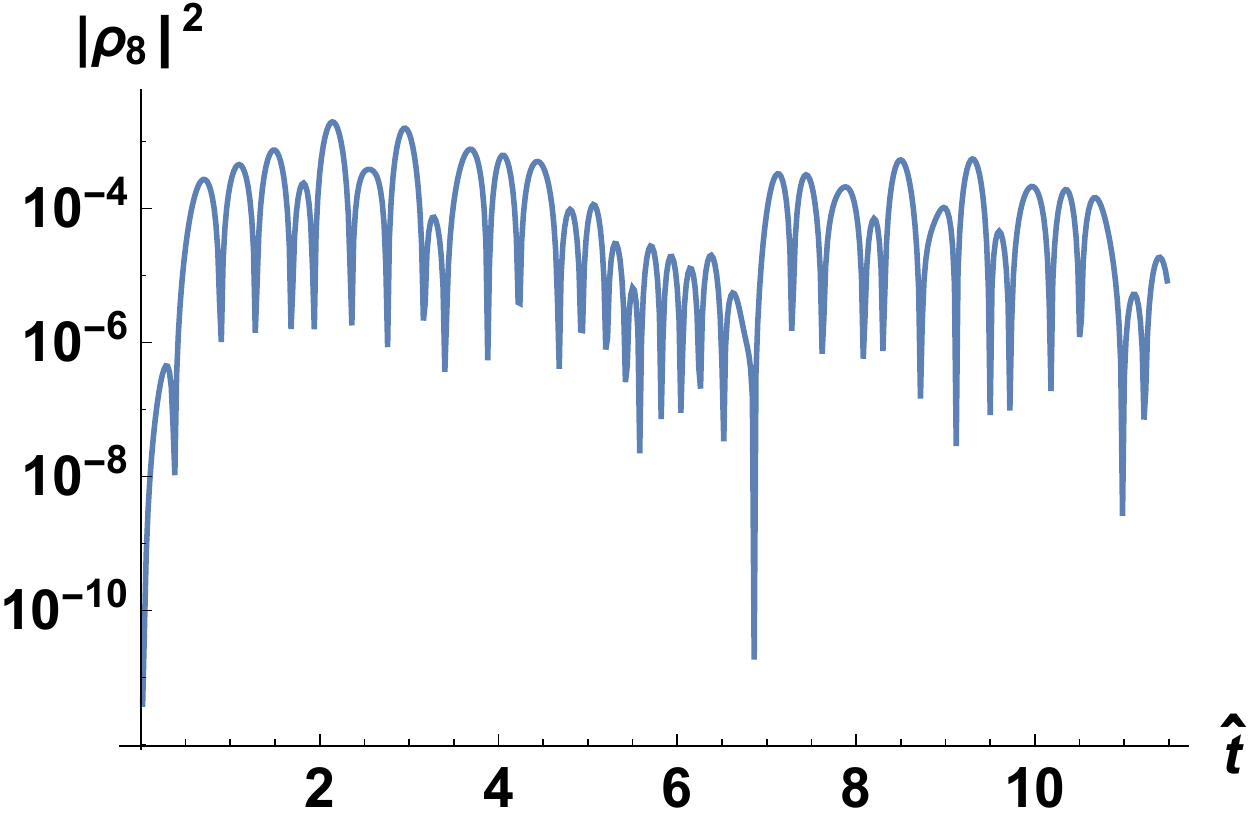}
 \includegraphics[width = 4cm]{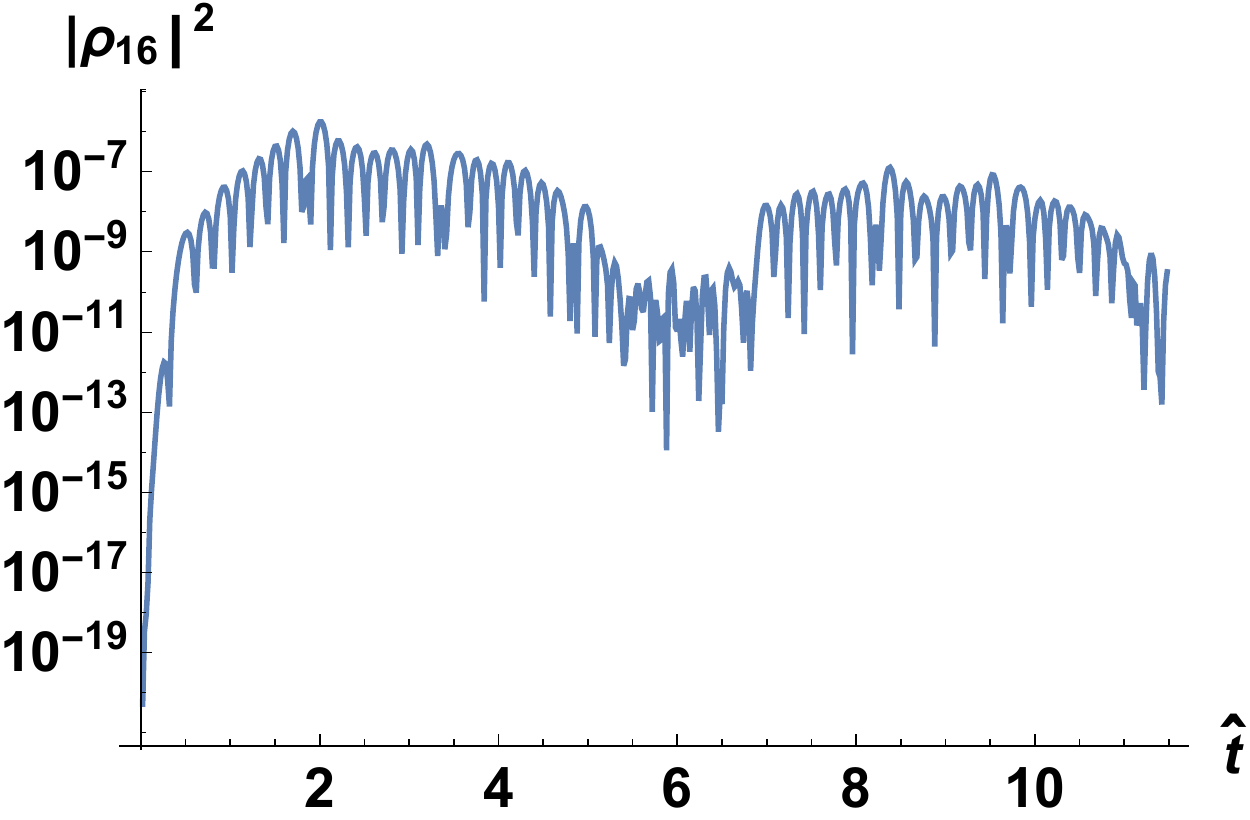}
 \caption{Time evolution of Fourier modes $|\rho_n|^2$ for $n = 2,4,8,16$ for IC $\tl \rho = \cos{2 \hat x}$ and $\tl u = 0$ showing that the higher modes remain uniformly small compared to the first few, justifying truncation of Fourier series.}
 \label{f:u-cos2x-four-modes-growth}
\end{figure*}

\begin{figure*}	
\begin{center}
	\begin{subfigure}[t]{4.3cm}
		\includegraphics[width=4.3cm]{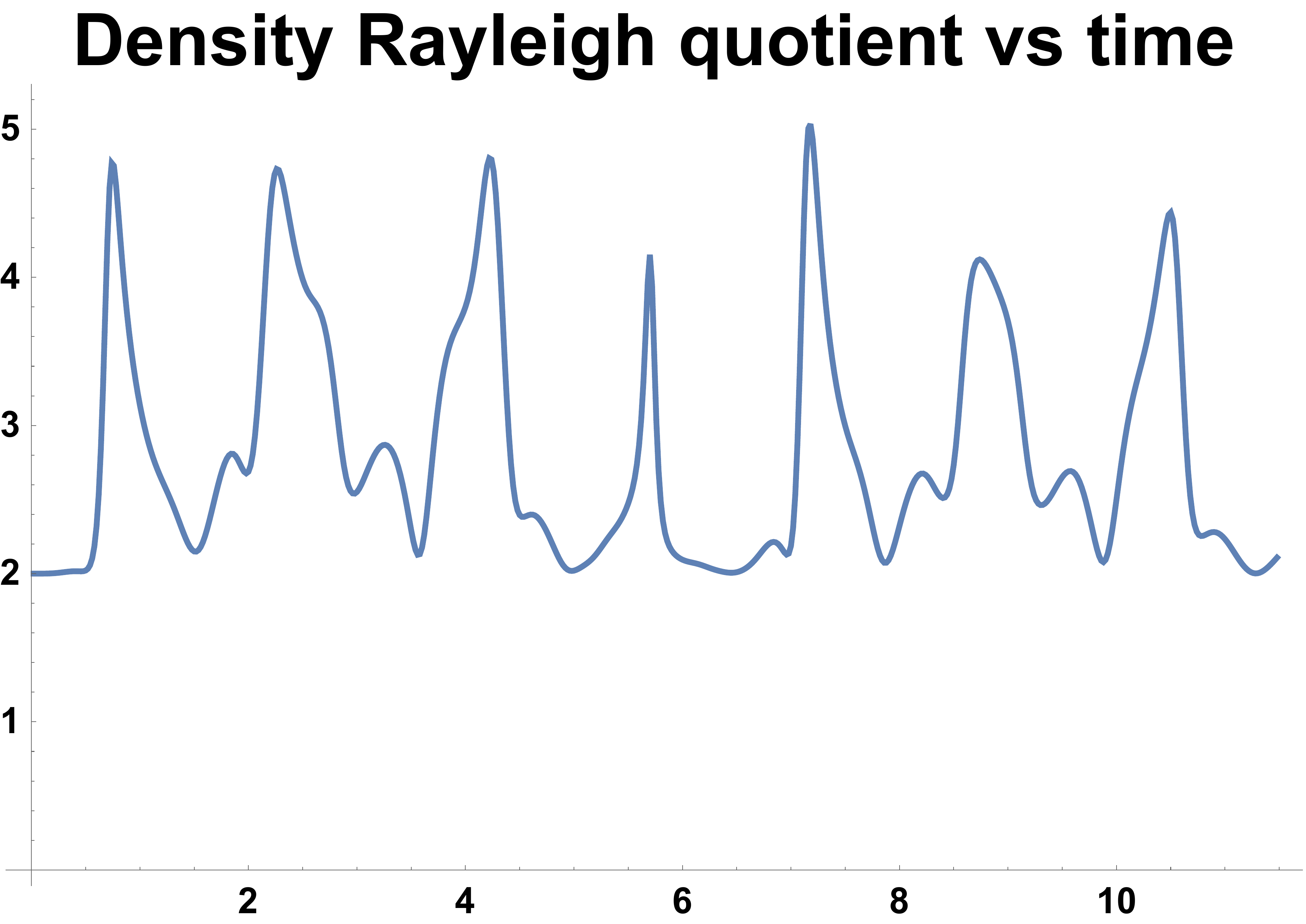}
		\caption{}
	\label{f:Rayleigh-quotient}
	\end{subfigure}
\qquad \qquad
	\begin{subfigure}[t]{5cm}
		\includegraphics[width=5cm]{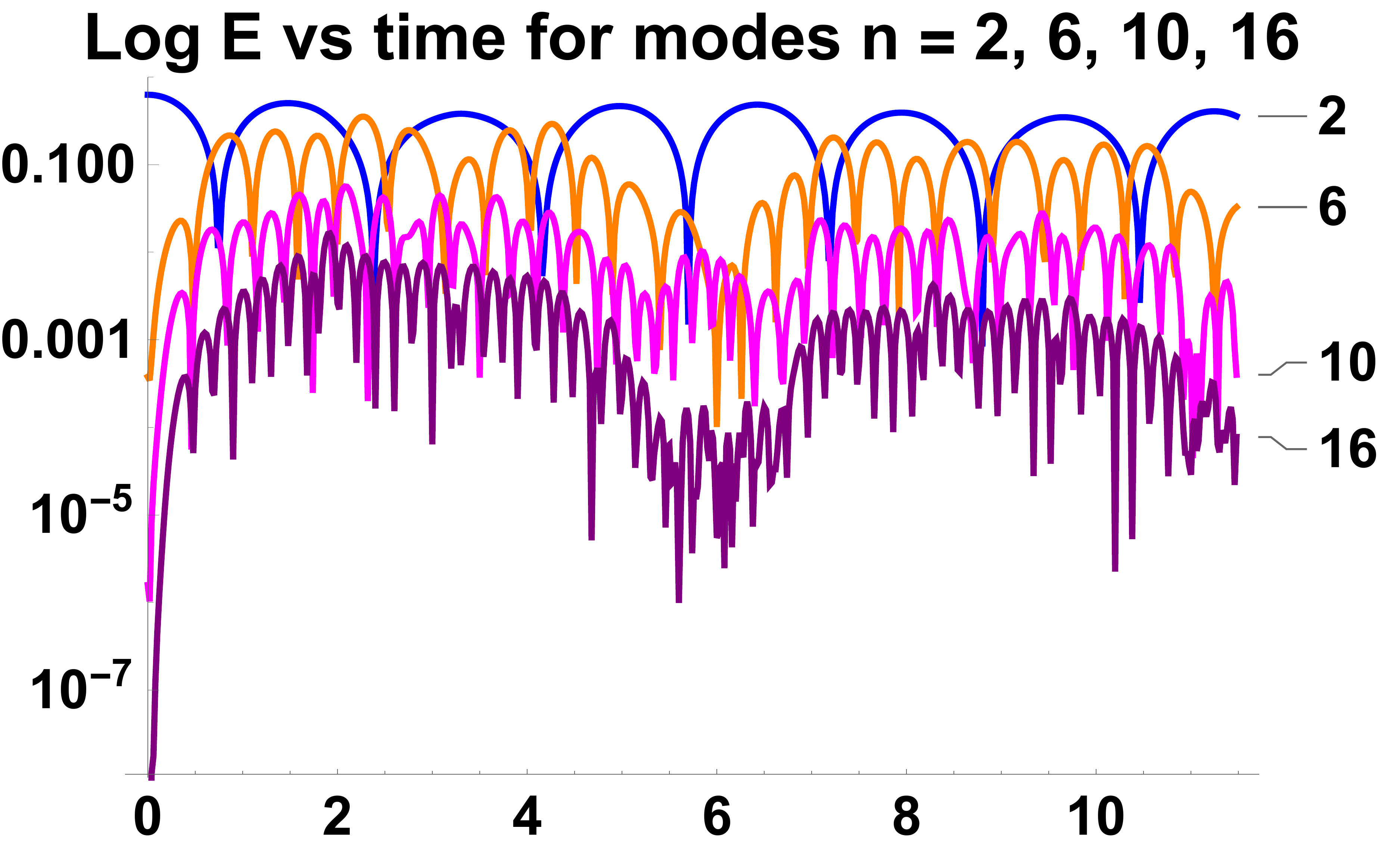}
		\caption{}
		\label{f:Four-time-evol-period}	
	\end{subfigure}
	\qquad \qquad 
	\begin{subfigure}[t]{5cm}
		\includegraphics[width=5cm]{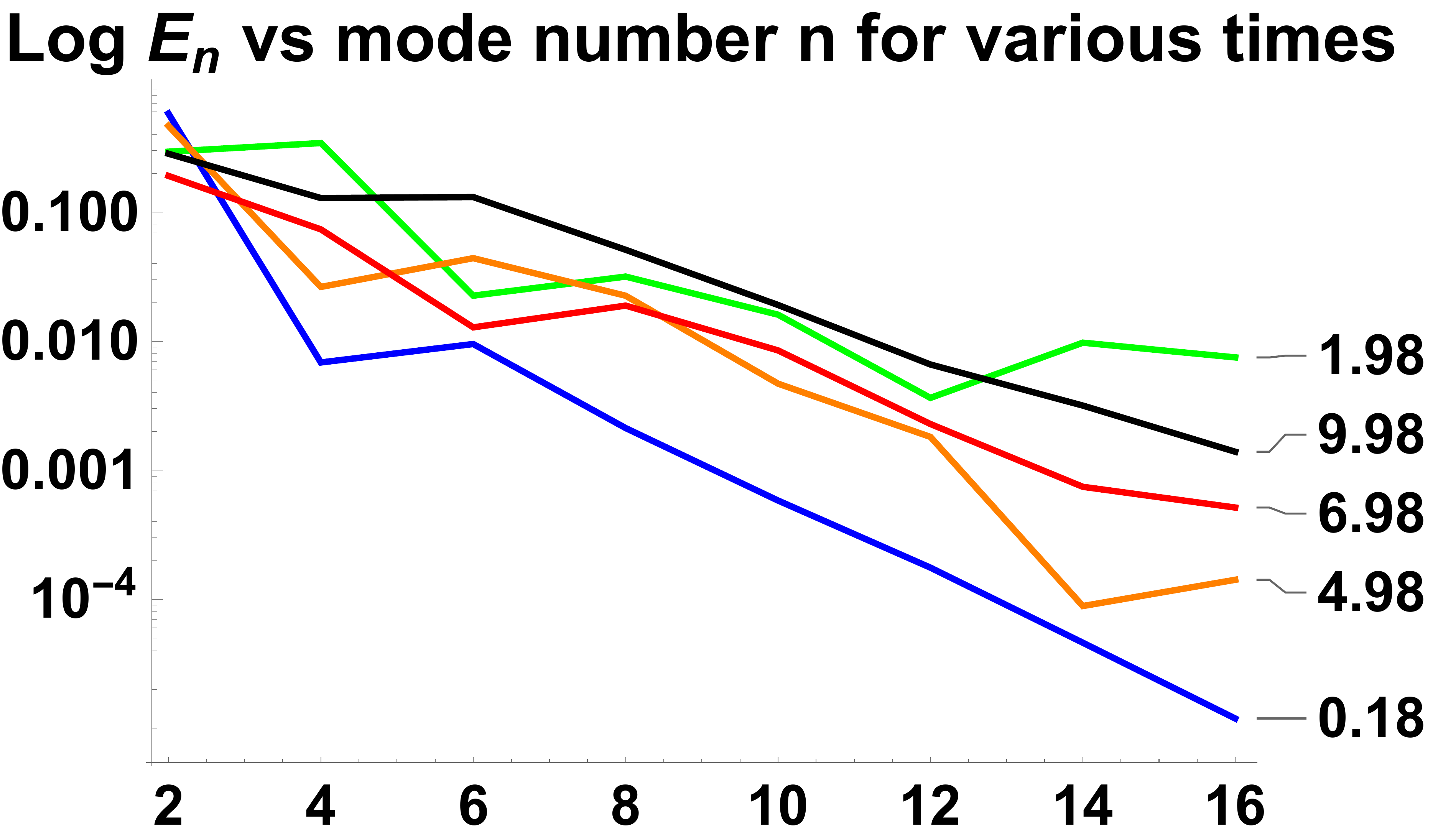}
		\caption{}
		\label{f:En-vs-n}	
	\end{subfigure}
\end{center}
	\caption{(a) Rayleigh quotient displays bounded oscillations indicating only a few modes are active. (b) Time evolution of $\log{E_n}$ for modes $n = 2,6,10,16$. The higher modes remain small with each one showing approximately periodic oscillations. (c) $\log{E_n}$ vs $n$ for a few values of $\hat t$ showing exponential drop with $n$. In all cases, the IC was $\tl \rho =\cos{2 \hat x}$ and $\tl u =0$.}
	\label{f:energy-spectral-distrib-plot}
\end{figure*}

We also find that higher modes $u_n$ and $\rho_n$ ($n \gtrsim 9$) grow from zero but soon saturate and remain a few orders of magnitude below the first few modes (see Fig.~\ref{f:u-cos2x-four-modes-growth}). This justifies truncating the Fourier series at $n_{\rm max} = 16$. The modes also display an approximate periodicity in time. This suggests recurrent motion\cite{thyagaraja-1979,thyagaraja-1983,SIdV}. This behavior is also captured in Fig.~\ref{f:Rayleigh-quotient} where we plot the Rayleigh quotient or mean square mode number
	\beq
	R =  \frac{\int |\hat \rho_{\hat x}|^2 d\hat x}{\int |\hat \rho|^2 d\hat x} = \frac{\sum_n n^2 |\rho_n|^2}{\sum_n |\rho_n|^2},
	\eeq
as a function of time. It is found to fluctuate between bounded limits indicating that effectively only a finite number of modes participate in the dynamics. Another interesting statistic is the spectral distribution of energy $(E_n)$ and its dependence on time. Fig.~\ref{f:Four-time-evol-period}, shows the time evolution of $\log{E_n}$ for $n = 2,6,10,16$ for the IC $\tl \rho = \cos{2\hat x}$ and $\tl u = 0$ and demonstrates that the energy in the higher modes remains small. Moreover, each mode $\rho_n$ and $u_n$ oscillates between an upper and lower bound and shows approximate periodicity with differing periods. In Fig.~\ref{f:En-vs-n}, $\log{E_n}$ vs $n$ is plotted for a few values of $\hat t$. Unlike the power law decay $n^{-5/3}$ in the inertial range of fully developed turbulence, here we see that $E_n$ drops exponentially with $n$. In particular, there is no equipartition of energy among the modes.

%--------------
\section{Connection to nonlinear Schr\"odinger and generalizations}
\label{s:r-gas-to-nlse}
%--------------

Our numerical results indicate recurrence and soliton-like scattering in 1d isentropic R-gas dynamics for $\g = 2$, suggesting integrability. Remarkably, in this case, R-gas dynamics 
is transformable into the defocussing (repulsive) cubic nonlinear Schr\"odinger equation (NLSE). More generally, adiabatic R-gas dynamics may be viewed as a novel generalization of the NLSE. Indeed, let us write the velocity field as $\bfv = \grad \phi + \frac{\la \grad \mu}{\rho} + \frac{\al \grad s}{\rho}$ where $s$ is the specific entropy and $\al,\la$  and $\mu$ are Clebsch potentials \cite{zakharov-kuznetsov}. As in treatments of superfluidity \cite{gross,pitaevskii} and `quantum hydrodynamics' \cite{madelung}, if we define the Madelung transform, $\psi(\bfr,t) = \sqrt{\rho} \exp{(i  \phi(\bfr)/2 \sqrt{\beta_*})}$, then the R-gas dynamic energy density (\ref{e:3d-hamiltonian}) becomes:
	\beqs \small
	&& {\cal E} = 2 \beta_* |\grad \psi|^2 + 2 \frac{\sqrt{\beta_*}}{|\psi|^2}\left( \al \grad s + \la \grad \mu\right) \cdot \left( \frac{\psi^* \grad \psi - \psi \grad \psi^* }{2i} \right) \cr
	&& + \frac{\la \al}{|\psi|^2} \grad \mu \cdot \grad s + \frac{\al^2 (\grad s)^2 + \la^2 (\grad \mu)^2 }{2|\psi|^2} + \frac{\bar p e^{s/c_V}}{(\g-1)} \frac{|\psi|^{2\g}}{\bar \rho^\g}.
	\label{e:hamil-3d-psi-s-alpha}
	\eeqs
%	\beq
%	H = \int \left[ 2 \beta_* |\grad \psi|^2 + 2 \frac{\sqrt{\beta_*}\al \grad s}{|\psi|^2} \left( \frac{\psi^* \grad \psi - \psi \grad \psi^* }{2i} \right) + \half \frac{\al^2 (\grad s)^2}{|\psi|^2} + \frac{\bar p e^{s/c_V}}{(\g-1)} \frac{|\psi|^{2\g}}{\bar \rho^\g} \right] d\bfr.
%	\eeq
Here $\bar p$ and $\bar \rho$ are constant reference pressure and density. The transformation from $(\rho, \phi)$ to $(\psi, \psi^*)$ is canonical. The corresponding equations of motion for $\psi, \al, \la, \mu$ and $s$ may be obtained using the canonical bosonic PBs $\{ \psi(\bfr), \psi^*(\bfr') \} = - (i/2 \sqrt{\beta_*}) \del(\bfr - \bfr')$, $\{ \la(\bfr), \mu(\bfr') \} = \del(\bfr - \bfr')$ and $\{ \al(\bfr), s(\bfr') \} = \del(\bfr - \bfr')$.

Specializing to isentropic potential flow where $s = \bar s$ is constant, $p = K (\g - 1) \rho^\g$ (\ref{e:entropy-barotropic}) and $\bfv = \grad \phi$, (\ref{e:hamil-3d-psi-s-alpha}) simplifies to
	\beq
	H = \int \left[ 2 \beta_* |\grad \psi|^2 + K |\psi|^{2\g} \right] d\bfr.
	\eeq
Using the above PBs for $\psi$, one finds  that $\psi$ satisfies the {\it defocusing} nonlinear Schr\"odinger equation
	\beq
	i \sqrt{\beta_*} \psi_t = - \beta_* \grad^2 \psi + \half \g K |\psi|^{2(\g-1)} \psi.
	\label{e:nlse-3d-gen-gamma}
	\eeq
Interestingly, in 1d, the R-gas dynamic form of the {\it focusing} cubic $(\g = 2)$ NLSE had been obtained in the context of the Heisenberg magnetic chain \cite{lakshmanan-ruijgrok,turski}. However, as noted in \cite{turski}, the Heisenberg chain leads to negative pressure! Returning to (\ref{e:nlse-3d-gen-gamma}), we see that the real and imaginary parts of the NLSE correspond to the Bernoulli and continuity equations. The $\grad^2\psi$ term leads to the divergence of the mass flux, $\bfv^2$ and regularization terms in the isentropic R-gas dynamic equations
	\beq
	\rho_t + \grad \cdot (\rho \bfv) = 0 \quad \text{and} \quad
	\phi_t = - \g K \rho^{\g-1} - \frac{\bfv^2}{2} + 2 \beta_* \frac{\grad^2 \sqrt{\rho}}{\sqrt{\rho}}.
	\eeq
Evidently, $\beta_*$ plays the role of $\hbar^2$. The nonlinear term $(\g K/2) |\psi|^{2 (\g-1)} \psi$ corresponds to the isentropic pressure $p = (\g-1) K \rho^\g$ whose positivity implies we get the defocusing/repulsive NLSE. Thus, our regularization term $2 \beta_*(\grad^2 \sqrt{\rho})/\sqrt{\rho}$ is like a quantum correction to the classical isentropic pressure. For $\g = 2$ we get the cubic NLSE or Gross-Pitaevskii equation (without an external trapping potential). Note that 1d isentropic flow on the line with $\bfv = (u(x),0,0)$ is always potential flow: $u = \phi_x$. So the above transformation takes 1d isentropic R-gas dynamics (\ref{e:unsteady-barotropic-eqns}) to the defocusing 1d NLSE, which for $\g = 2$ and periodic BCs admits infinitely many conserved quantities in involution \cite{faddeev-takhtajan}. This explains our numerical observations of approximate phase shift scattering of solitary waves and recurrence.

%-------------------
\subsection{NLSE interpretation of steady R-gas dynamic cavitons and cnoidal waves}
\label{s:caviton-to-nlse}
%-------------------

It turns out that steady solutions of 1d isentropic R-gas dynamics (\S \ref{s:steady-trav-quadrature}) correspond to NLSE wavefunctions $\psi$ with harmonic time dependence. For $\g = 2$, our aerostatic caviton corresponds to the dark soliton of NLSE. More generally, aerostatic steady solutions correspond to $\psi$ of the form $\sqrt{\rho(x)} \exp(-i F^{\rm u} t/2\sqrt{\beta_*})$ where $F^{\rm u}$ is the constant velocity flux (\ref{e:barotropic-curr}). Finally, non-aerostatic cnoidal waves correspond to interesting asymptotically plane wave NLSE wavefunctions modulated by a periodic cnoidal amplitude. Here, we consider the cavitons and $\text{cn}$ waves in increasing order of complexity.

\noindent{\bf Aerostatic caviton:} The simplest caviton solution of \S \ref{s:exact-cavitons-cnoidal-g-2} is aerostatic: 
    \beqs
    && \rho(x) = \rho_\pt \tanh^2\left( \frac{x}{2\la_\pt} \right), \quad u(x)=0 \quad \text{and}  \cr
    && p(x) = K \rho^2 \quad \text{where} \quad K = \frac{c_\pt^2}{2\rho_\pt},
    \eeqs
Here, $\rho_\pt, \la_\pt$ and $c_\pt^2$ are positive constants. The corresponding specific enthalpy and velocity flux are 
    \beqs
    && h = 2 K \rho = c_\pt^2 \tanh^2\left( \frac{x}{2\la_\pt}\right) \quad \text{and} \cr
    && F^{\rm u} = \frac{u^2}{2} + h - 2\beta_*\frac{(\sqrt{\rho})_{xx}}{\sqrt{\rho}} = c_\pt^2 \quad \text{where} \quad \beta_* = \la_\pt^2 c_\pt^2.
    \eeqs
Thus, the Bernoulli equation $\phi_t + F^{\rm u} = 0$ (\ref{e:barotropic-curr}) is satisfied provided we take the velocity potential $\phi = -c_\pt^2\, t$ to be time-dependent. The resulting $\psi$ is the dark soliton solution of the defocusing NLSE (see \S 6.6 of \cite{ablowitz})
    \beq
    \psi = \sqrt{\rho} \;\exp\left( {\frac{i\phi}{2\sqrt{\beta_*}}} \right) =  \sqrt{\rho_\pt} \tanh\left(\frac{x}{2\la_\pt}\right) {\rm e}^{{-\frac{ic_\pt^2t}{2\sqrt{\beta_*}}} }.
    \eeq
\noindent{\bf Non-aerostatic caviton:} More generally, for a non-aerostatic caviton (for $0 < M_\pt < 1$ and $\rho_\pt, \la_\pt$ and $c_\pt^2$ positive constants)
    \beqs
    && \rho(x) = \rho_\pt \left( 1 - (1 - M_\pt^2) \sech^2\left( \sqrt{\frac{1 - M_\pt^2}{4}} \frac{x}{\la_\pt} \right) \right), \cr 
    && u(x) = \frac{c_\pt M_\pt \rho_\pt}{\rho(x)} \quad \text{and} \quad p(x) = \frac{c_\pt^2}{2\rho_\pt} \rho(x)^2.
    \eeqs
In this case, the constant velocity flux is $F^{\rm u} = c_\pt^2 (2 + M_\pt^2)/2$. The resulting velocity potential is
    \beqs
    \phi &=& - \half c_\pt^2 (2 + M_\pt^2)t + c_\pt  M_\pt x \cr
    && + 2 c_\pt \la_\pt \arctan\left( \frac{\sqrt{1 - M_\pt^2}}{M_\pt} \tanh \left( \frac{\sqrt{1 - M_\pt^2}}{2\la_\pt}x \right) \right).\quad 
    \eeqs
Thus, $\psi$ is asymptotically a plane wave with phase speed ${c_\pt (2 + M_\pt^2)/2M_\pt}$. This $\psi$ may be regarded as a high-frequency carrier wave modulated by a localized signal.
% {\Blue For a gray soliton $\rho$ is traveling, so this is not a gray soliton.}

{\fl \bf Aerostatic snoidal waves:} The simplest steady periodic solutions is the aerostatic snoidal wave (\ref{e:cnoidal-wave-gamma=2}):
    \beq
    u \equiv 0, \;\; \rho = \rho_\pt (1-2\ka_\pt)\; \text{sn}^2\left(\frac{x}{2\la_\pt}, 1-2\ka_\pt \right) \;\;\text{for} \;\; 0< \ka_\pt < \frac{1}{2}.
    \eeq
Here, $F^{\rm u} = c_\pt^2 (1 - \ka_\pt)$ and $\phi = - F^{\rm u} \,t$. The resulting $\psi$ is a snoidal wave with harmonic time dependence:
    \beq
    \psi = \sqrt{\rho_\pt(1-2\ka_\pt)} \; \text{sn}\left(\frac{x}{2\la_\pt}, 1-2\ka_\pt \right) \text{e}^{-\frac{ i c_\pt^2 (1 - \ka_\pt)\,t}{2\sqrt{\beta_*}}}.
    \eeq
\noindent{\bf Non-aerostatic cnoidal waves:} Finally, in the general case, we have from (\ref{e:cnoidal-wave-rho-til-triangle}), 
    \beqs
    && \rho(x) = \rho_\circ\left[ \tl\rho_+ - (\tl\rho_+ - \tl\rho_-) \,{\rm cn}^2\left( \frac{\sqrt{1 - \tl \rho_-}}{2} \frac{x}{\la_\circ}, \frac{\tl\rho_+ - \tl\rho_-}{1 - \tl\rho_-} \right) \right], \cr 
    && u(x) = \frac{c_\circ M_\circ \rho_\circ}{\rho(x)} \quad \text{and} \quad
    p(x) = \frac{c_\circ^2}{2\rho_\circ}\rho(x)^2
    \eeqs
where $0 < \ka_\circ < 1/2$ and $0 < M_\circ < 1 - \sqrt{2\ka_\circ}$ (lower triangular region of Fig.~\ref{f:kappa-M-K>0}). Here $\tl\rho_{\pm}(\ka_\circ, M_\circ)$ are as in (\ref{e:steady-rho-plus-minus}). Furthermore, $\rho_\circ, c_\circ^2$ and $\la_\circ$ are positive constants that set scales. As before, $\phi = - F^{\rm u} t + \int_0^x u(x') \,dx'$, where the constant velocity flux $F^{\rm u}$ depends on $c_\circ, \ka_\circ$ and $M_\circ$, but is independent of $\rho_\circ$ and $\la_\circ$. Though an explicit formula for $\phi(x,t)$ is not easily obtainable, it is evident that for large $x$, $\phi$ grows linearly in $x$ with a subleading oscillatory contribution. Thus, $\psi$ has a purely harmonic time-dependence $\exp \left( -i F^{\rm u} t/2\sqrt{\beta_*} \right)$, a periodic cnoidal modulus $|\psi| = \sqrt{\rho(x)}$ and an argument that asymptotically grows linearly in $x$. Thus, asymptotically, $\psi$ is a plane wave modulated by the periodic amplitude $\sqrt{\rho(x)}$.

%-------------------
\subsection{Conserved quantities and Rayleigh quotient of NLSE and R-gas dynamics}
\label{s:rayleigh-quotient-NLSE-conserved-quantities}
%-------------------

The cubic 1d NLSE admits an infinite tower of conserved quantities. The first three are
	\beqs
	&& N = \int |\psi|^2 \: dx, \quad P_{\rm NLSE} = \int \Im (\psi^* \psi_x) \: dx, \quad \text{and} \cr
	&& E_{\rm NLSE} = \int \left(|\psi_x|^2 + \frac{K}{2 \beta_*} |\psi|^4 \right) {\beta_*^{1/4}} dx.
	\eeqs
These correspond to the mass $M = N$, momentum $P = 2 \sqrt{\beta_*} P_{\rm NLSE}$ and energy $H = 2 \beta_*^{3/4} E_{\rm NLSE}$ of R-gas dynamics. The next conserved quantity of NLSE (with periodic BCs) is \cite{faddeev-takhtajan}
	\beq
	Q = \int_{-L}^L \left[ \sqrt{\beta_*} \psi^* \psi_{xxx} - \frac{K |\psi|^2}{2 \sqrt{\beta_*}} (\psi \psi^*_x + 4 \psi^* \psi_x) \right]dx.
	\eeq
$\Re Q$ and $\Im Q$  correspond to the following in R-gas dynamics:
	\beqs
	&& Q_r = \frac{3\sqrt{\beta_*}}{4} \int_{-L}^L \left( \frac{\rho_x^3}{2 \rho^2} - \frac{\rho_x \rho_{xx} }{\rho} \right) \: dx \quad \text{and} \cr
	&& Q_i = - \int_{-L}^L \left[ \frac{u_x \rho_x}{2} + \frac{3 u \rho_x^2}{8 \rho} + \frac{u^3 \rho}{8 \beta_*} + \frac{ 3 K u \rho^2}{4\beta_*} \right] dx.
	\label{e:Qr-Qi}
	\eeqs
In \S \ref{s:numerical-scheme} we use the conservation of $Q_r$ and $Q_i$ to test our numerical scheme. In Ref.~\cite{thyagaraja-1979}, it was shown that for the cubic 1d NLSE, the Rayleigh quotient or mean square mode number for periodic BCs,
    \beq
    R_{\rm NLSE} = \frac{\int_{-L}^L |\psi_x|^2\,dx}{\int_{-l}^l |\psi|^2\, dx},
    \eeq
is related to the number of active degrees of freedom and is bounded in both focussing and defocussing cases. This has a simple interpretation in R-gas dynamics, since
    \beq
    R_{\rm NLSE} = \frac{1}{2\beta_* M} \int_{-L}^L \left( \half \rho u^2 + \frac{\beta_*}{2} \frac{\rho_x^2}{\rho} \right)\, dx \leq \frac{H}{2\beta_* M}.
    \eeq
Consequently, 1d isentropic R-gas dynamic potential flows are recurrent in the sense discussed in \cite{thyagaraja-1979}.

%--------------
\subsection{Connection to vortex filament and Heisenberg chain equations}
\label{s:neg-pressure-vortex-filament}
%--------------

Intriguingly, the {\it negative pressure} $\g = 2$ isentropic R-gas dynamic equations in 1d for $\rho$ and $u$ (\ref{e:unsteady-barotropic-eqns}) are equivalent to the vortex filament equation $\dot \xi = G\, \xi' \times \xi''$ where $G$ is a constant. Here the curve $\xi(x,t)$ represents a vortex filament with tangent vector $\xi'$. The equivalence is most transparent when the vortex filament equation is expressed in the Frenet-Serret frame, with curvature $\ka = \sqrt{\rho}$ and torsion $\tau = u$ \cite{lakshmanan-ruijgrok}. The FS frame ($\xi', n, b$) is an orthonormal basis along the filament, where $n = \xi''/\ka$ is the unit normal and $b = \xi'\times n$ the binormal. The FS equations describe how the frame changes along the filament.
	\beq
	\xi'' = \ka n, \quad n' = -\ka \xi' + \tau b \quad \text{and} \quad b' = -\tau n
	\eeq
We may use the vortex-filament and FS equations to find evolution equations for the FS frame: 
	\beqs
	&& \frac{\dot \xi'}{G} = - \ka \tau n + \ka' b, \quad 
	\frac{\dot n}{G} = \ka \tau \xi' + \left( \frac{\ka''}{\ka} - \tau^2 \right) b \cr
	&& - \ov{\ka}\left( 2\ka' \tau + \ka \tau' + \frac{\dot \ka}{G} \right) n \cr
	&& \text{and} \quad 
	\frac{\dot b}{G} = - \left(\ka' \xi' +  \left( \frac{\ka''}{\ka} - \tau^2 \right) n\right).
	\label{e:evol-of-fs-frame}
	\eeqs 
Eqn.~(\ref{e:evol-of-fs-frame}) leads to evolution equations for $\ka$ (using $\dot n \cdot n = 0$ as $n\cdot n = 1$) and $\tau = n'\cdot b$:
	\beq
	\dot\ka = -G \left( 2\ka'\tau + \ka\tau' \right) \quad \text{and}\quad \dot \tau = G \left( \frac{\ka^2}{2} - \tau^2 + \frac{\ka''}{\ka}  \right)'.
	\label{e:ka-tau-evol}
	\eeq
Taking $G = 1/2$, $\ka = \sqrt{\rho}$ and $\tau = u$, (\ref{e:ka-tau-evol}) reduces to the continuity and velocity equations for $\g = 2$ isentropic R-gas dynamics with $\beta_* = 1/2$, but with a negative $p = -\rho^2/8$ \cite{turski}. Furthermore, it is well-known \cite{sgrajeev} that the vortex filament equation is related to the Heisenberg magnetic chain equation $\dot S = G \,S \times S''$ with $S = \xi'$. An open question is to give a geometric or magnetic chain interpretation of the {\it positive pressure} R-gas dynamic equations as well as those for general $\g$. It is noteworthy that negative pressures (relative to atmospheric pressure) can arise in real flows, for instance in the presence of strong currents \cite{ali-kalisch-2}.

%--------------
\section{Discussion}
\label{s:discussion}
%--------------

It is a significant feature of our attempt to conservatively regularize singularities in gas dynamics that it has led us (in the case of isentropic potential flow) to the {\it defocusing} NLSE. Heuristically, the defocusing interaction tends to amplify linear dispersive effects and thereby prevent blowups. For $\gamma = 2$, this connection to the cubic NLSE should provide powerful tools [including the inverse scattering transform in 1d (see \S 9.10 of \cite{ablowitz}, \cite{faddeev-takhtajan} and \cite{novikov})] to deal with the initial-boundary value problems in various dimensions as well as alternative numerical schemes. Moreover, the bound on the NLSE Rayleigh quotient obtained in \cite{thyagaraja-1979} (see also \S\ref{s:rayleigh-quotient-NLSE-conserved-quantities}) generalizes to 2d and 3d as well as to nonlinearities other than cubic ($\gamma \ne 2$). This would have implications for recurrence in more general R-gas dynamic isentropic potential flows, even in the absence of integrability. The techniques of dispersive shock wave theory \cite{whitham,El-Hoefer,miller} could provide additional tools to address R-gas dynamic flows.

In \cite{tao-nonlinear-dispersive,tao-global-behav}, a classification of semilinear PDEs [perturbations of linear equations by lower order nonlinear terms] into subcritical, critical and supercritical, based on conserved quantities (mass, energy etc.), scaling symmetries and regularity of initial data is described. Though the equations of R-gas dynamics given in \S \ref{s:3d-hamil-form-R-gas-dyn} are not semilinear, in the special case of isentropic potential flow, the transformation to NLSE makes them semilinear. It is thus interesting to examine the implications of this classification for R-gas dynamics. For example, the critical scaling regularity \cite{tao-nonlinear-dispersive} of NLSE in $d$ spatial dimensions is $s_c = d/2 - 1/(\g-1)$. Thus, if the initial data is such that the number of particles and energy are finite (so that $\psi, \grad \psi \in L^2$ and $\psi \in H^1$), then according to the scaling heuristic, the NLSE is subcritical for any $\g > 1$ in 1d and 2d and also for $1 < \g < 3$ in 3d.

In \S\ref{s:patched-shock-weak-sol} we argued that 1d R-gas dynamics does not admit any smooth or continuous shock-like steady solutions. In fact, we found that if we try to patch half a caviton at its trough density with a constant state, then the mass, momentum and energy fluxes in the pre-shock region cannot all match their values in the post-shock region, so that the Rankine-Hugoniot conditions are violated. We conjecture that this absence of steady shock-like solutions is a general feature of a wide class of conservatively regularized gas dynamic models. Loosely, this is like d'Alembert's `theorem' that continuous solutions of Euler's equations cannot ever lead to drag, although possibly to lift. On the other hand, inclusion of viscosity {\it does} permit drag as well as steady shock-like solutions as in the Burgers equation \cite{whitham, ablowitz}. Allowing for non-steady solutions, we find that in R-gas dynamics, the gradient catastrophe is averted through the formation of a pair of solitary waves (see \S\ref{s:numerical-results}). It would be interesting to see if this mechanism is observed in any physical system, say, one where dissipative effects are small as in nonlinear optics, weak shocks, cold atomic gases or superfluids. For further discussion on steepening gradients and criteria for detecting wave breaking in dispersive hydrodynamics, see \cite{hatland-kalisch,wilkinson-banner}.

Though we have formulated R-gas dynamics in 3d, our analytic and numerical solutions have been restricted to 1d. It would be interesting to extend this work to higher dimensional problems such as oblique shocks and the Sedov-Taylor spherical blast wave problem. The mechanical and thermodynamic stability of our traveling caviton and periodic wave solutions is also of interest. One also wishes to examine whether the capillarity energy considered here arises from kinetic theory in a suitable scaling limit of small Knudsen number as for the Korteweg equation \cite{gorban-karlin,huang-wang-wang-yang}. Finally, our Hamiltonian and Lagrangian formulations of R-gas dynamics can be used as starting points in formulating the quantum theory. The transformation to NLSE provides another approach to quantization for isentropic potential flow especially when $\g = 2$.
Though we have focused on the conservatively regularized model, a more complete and realistic treatment would have to include viscous dissipation just as in the KdV-Burgers equation.

Our attempt to generalize KdV to include the adiabatic dynamics of density, velocity and pressure has led to a interesting link between KdV and NLSE that is quite different from the well known ones (see E.g. \cite{ablowitz}). In fact, we may view R-gas dynamics as a natural generalization of both. While it extends the KdV idea of a minimal conservative dispersive regularization to adiabatic gas dynamics in any dimension and shares with it the cnoidal and $\sech^2$ solutions, it also reduces to the defocusing cubic NLSE for isentropic potential flow of a gas with adiabatic index two. Thus, the cubic 1d NLSE is a very special member of a larger class of R-gas dynamic equations that make sense in any dimension and for nonlinearities other than cubic while also allowing for adiabatically evolving entropy and vorticity distributions.

% The sech$^2$ and cnoidal waves familiar from KdV are solutions of R-G dyn and so they are also solutions of NLSE.

\begin{acknowledgements}

AT gratefully acknowledges the support and hospitality of CMI. We thank anonymous referees for suggesting references and improvements. This work was supported in part by the Infosys Foundation, J N Tata Trust and the Science and Engineering Research Board, Govt. of India via grants CRG/2018/002040 and MTR/2018/000734.
 
\end{acknowledgements}

\appendix

%---------------
\section{Vector field and phase portrait for steady solutions}
\label{a:vect-fld-phase-portrait}
%---------------

With $x$ and $\rho$ viewed as time and position, steady, isentropic density profiles must satisfy the Newtonian ODE (\ref{e:steady-reg-gas-eqn-rho}):
    \beqs
    && \beta_* \rho_{xx} = - V'(\rho) + \frac{(\g + 1) \beta_*}{2} \frac{\rho_x^2}{\rho} \quad \text{for} \quad \rho > 0 \quad \text{where} \cr
    && V(\rho) =  \g F^{\rm p} \rho - \frac{ (\g -1)F^{\rm u} \rho^2}{2}  - \frac{(\g + 1) ( F^{\rm m})^2}{2}\log \rho.
    \label{e:steady-newtons-law}
    \eeqs 
Ignoring the velocity-dependent force, two cases arise: (a) for $ F^{\rm u} < 0$, $V$ has only bound states, corresponding to periodic $\rho$ and (b) for $ F^{\rm u} > 0$, there are periodic waves and a caviton, provided $V$ has a local minimum [this happens when $F^{\rm p}$ and $\D =  (F^{\rm p})^2\g^2 - 2(\g^2 - 1) F^{\rm u}  (F^{\rm m})^2$ are both positive]. The velocity-dependent force $\propto \rho_x^2/\rho$ is reminiscent of the drag force $\propto -|\rho_x| \rho_x$ on a body at high Reynolds number. However, while air drag tends to decrease the speed, this force is positive and tends to increase the velocity $\rho_x$. Moreover, though the naive `energy' $\beta_*\rho_x^2/2 + V$ is {\it not} conserved, Eqn.~(\ref{e:steady-newtons-law}) is non-dissipative and in fact Hamiltonian (see Appendix \ref{a:hamil-lagr-for-steady-sol}). Since the velocity-dependent force tends to push the particle outwards, it cannot convert a scattering state into a bound state. We also find that the qualitative nature of the phase portraits is not significantly altered by this force.

\begin{figure*}	
	\begin{center}
			\begin{subfigure}[t]{4cm}
		\includegraphics[width=4cm]{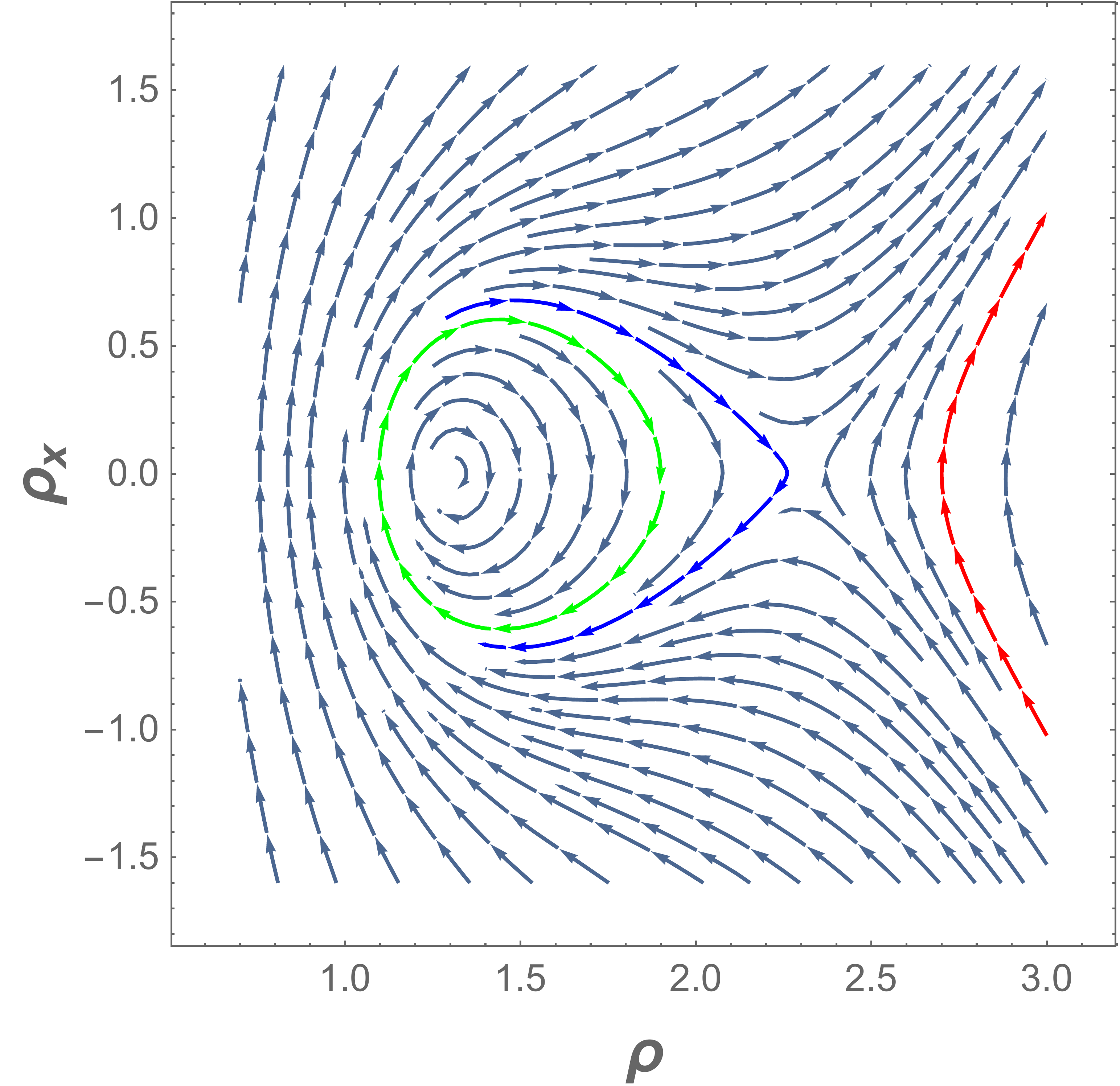}
		\caption{}
		\label{f:phase-por-caviton}	
	\end{subfigure}
	\quad
	\begin{subfigure}[t]{4cm}
		\includegraphics[width=4cm]{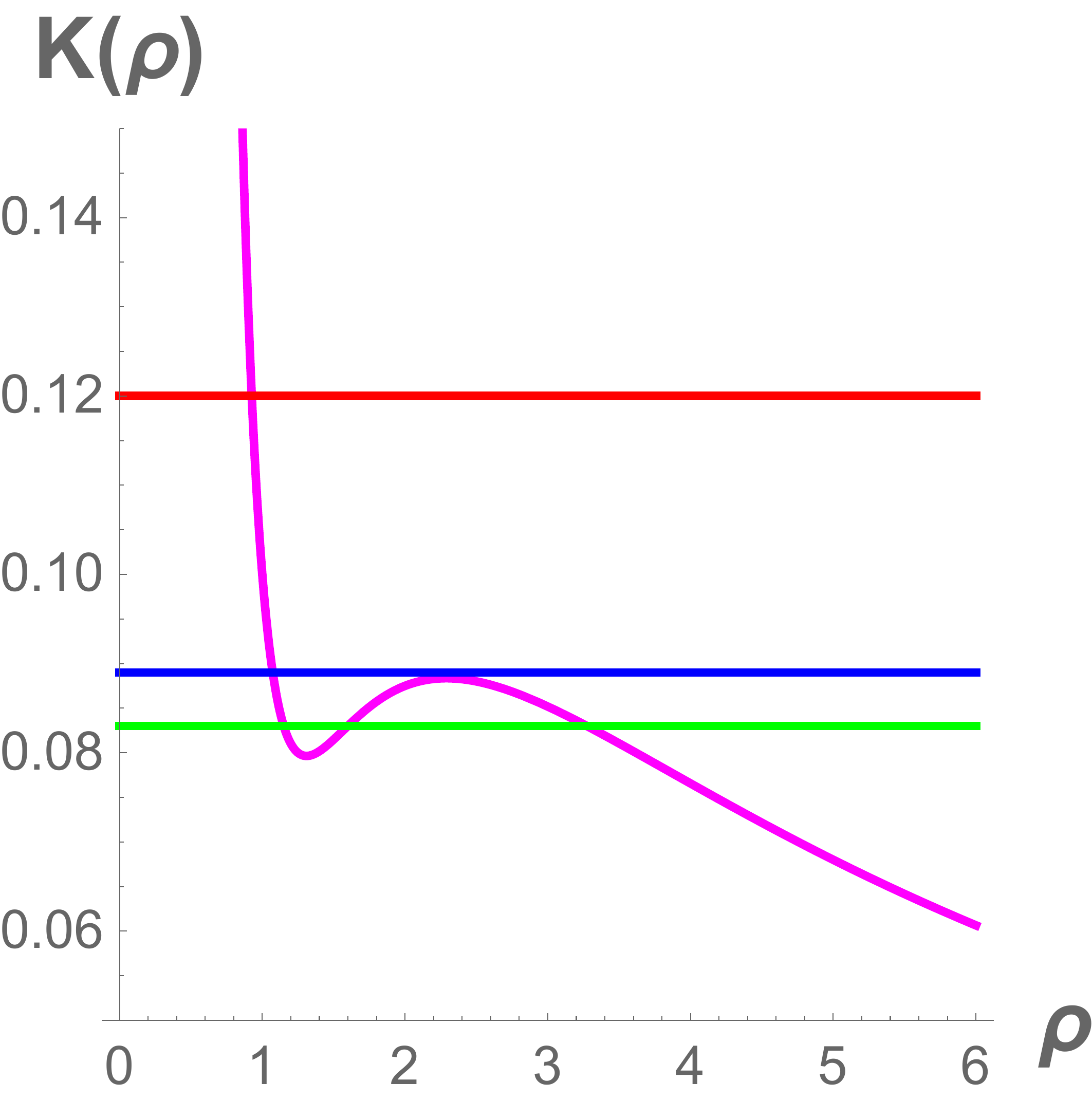}
		\caption{}
	\end{subfigure} 
	\quad
	\begin{subfigure}[t]{4cm}
		\includegraphics[width=4cm]{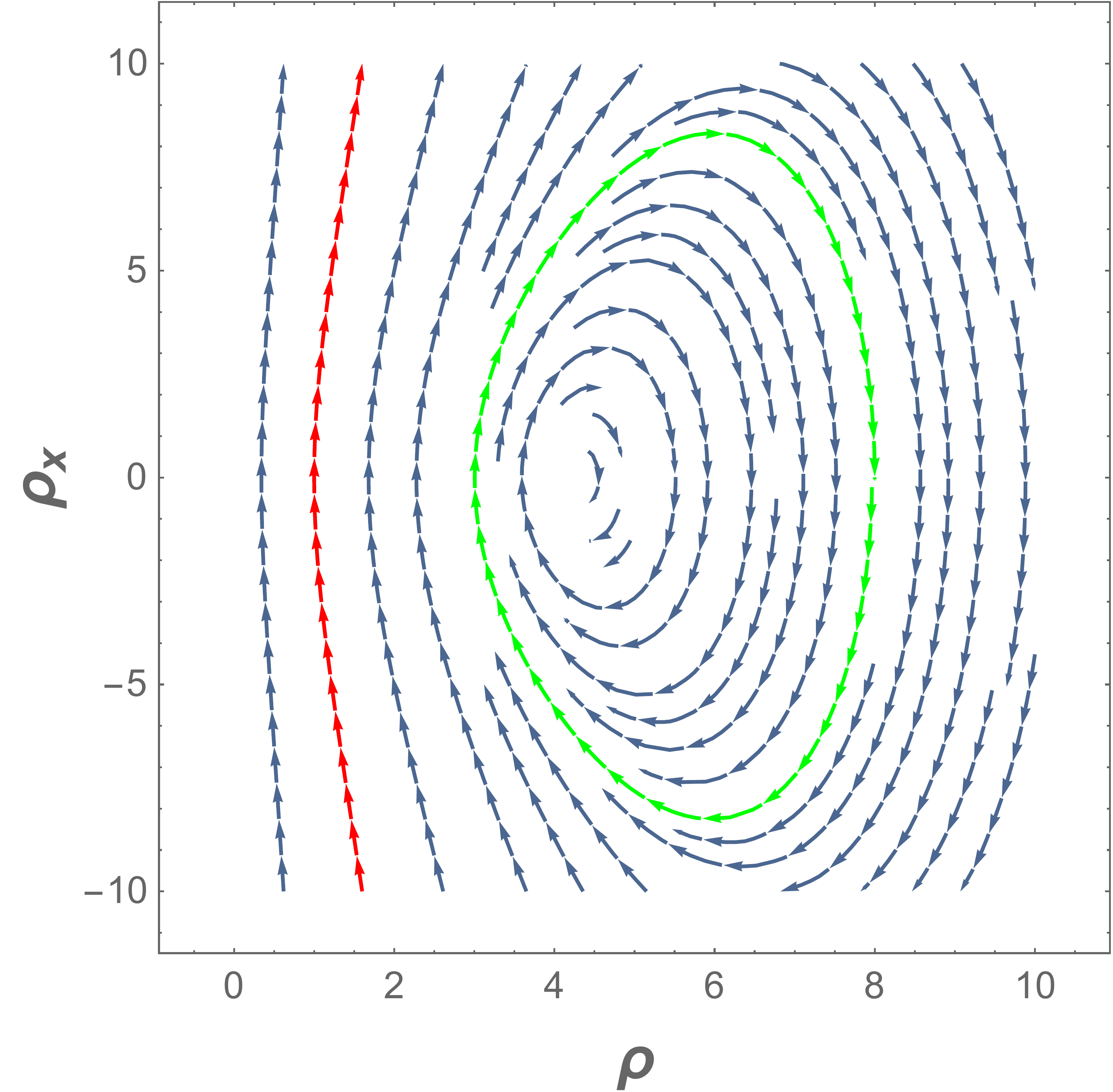}
		\caption{}
	\end{subfigure} 
	\quad
	\begin{subfigure}[t]{4cm}
		\includegraphics[width=4cm]{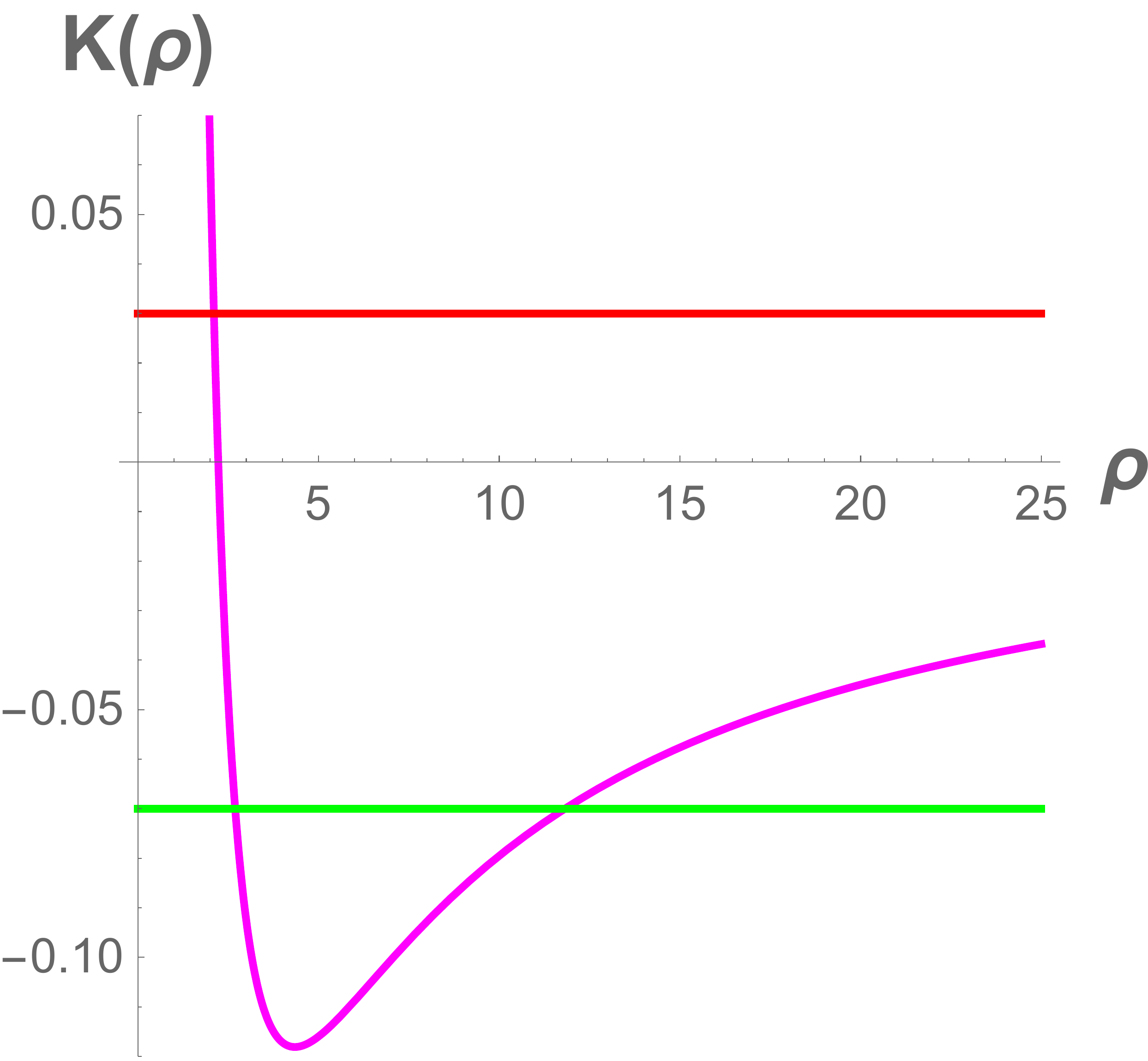}
		\caption{}
	\end{subfigure}
	\end{center}
	\caption{(a) and (b): The vector field $W$ on the $\rho$-$\rho_x$ phase plane for $\g =2$ and $\beta_* = 0.1$ and the entropy constant $K(\rho_\pt)$ (\ref{e:K-at-rhostar}) (which labels trajectories) for $F^{\rm m} = 1$, $F^{\rm p} = 0.9$ and $F^{\rm u} = 0.5$. As $K$ is decreased from infinity, we encounter unbounded solutions followed by a bounded caviton separatrix emanating from the X-point. The caviton encircles periodic orbits around the O-point. The X and O points are the only constant solutions. (c) and (d): The vector field $W$ and corresponding $K$ for $F^{\rm m} = 1$, $F^{\rm p} = - 2$ and $F^{\rm u} = -1$. Scattering states for $K > 0$ are followed by periodic orbits around the O-point with $K < 0$ and an infinite caviton separatrix at $K = 0$. Solutions with $K <0$ have negative pressure.}
	\label{f:vec-fld}
\end{figure*}

%-------

Introducing $\eta = \rho_x$, (\ref{e:steady-newtons-law}) defines a vector field $W$ on the right half $\rho$-$\rho_x$ phase plane:
    \beq
    \colvec{2}{\rho}{\eta}_x = W \equiv \colvec{2}{\eta}{- \frac{V'(\rho)}{\beta_*}+ \frac{(\g + 1)}{2} \frac{\eta^2}{\rho}}.
    \label{e:vect-field-W}
    \eeq
Though $\eta^2/\rho$ is singular along the $\rho = 0$ axis, it is `shielded' by the repulsive logarithmic potential in $V$. Bounded integral curves of $W$ correspond to bounded steady densities. Fixed points (FPs) of $W$ correspond to constant density solutions. They are located at $(\rho_{p,m}, 0)$ where $\rho_{p,m}$ are the extrema of $V$:
    \beq
    \rho_{p,m} = \frac{ \g F^{\rm p} \pm \sqrt{\D}}{2 (\g-1)  F^{\rm u}}.
    \label{e:rho-pm-fixedpts-steady}
    \eeq 
    
\begin{table*}
    \parbox{.45\linewidth}{
    \centering
    \caption*{\bf (a) Non-aerostatic steady solutions \vspace{-.2cm}}
    \begin{tabular}{|c|c|c|c|}
    \hline
        $ F^{\rm p}$ &   $F^{\rm u}$ & Fixed point & Bounded solutions  \\
        \hline
        $+$ & $+$ & O and X point & periodic, caviton \\
        \hline
        $+$ & $-$ & O point & periodic  $(K < 0)$\\ 
        \hline
        $-$ & $-$  & O point & periodic  $(K < 0)$ \\
        \hline
        $-$ & $+$  & none & none \\
        \hline
    \end{tabular}
    }
    \hfill
    \parbox{.45\linewidth}{
    \centering
    \caption*{\bf (b) Aerostatic steady solutions \vspace{-.3cm}}
    \begin{tabular}{|c|c|c|c|}
        \hline
        $ F^{\rm p}$ &  $F^{\rm u}$ & Fixed point & Bounded solutions  \\
        \hline
        $+$ & $+$ & X point & periodic, caviton \\
        \hline
        $-$ & $-$  & O point & periodic $(K < 0)$\\
        \hline
        $+$ & $-$ & none & periodic $(K < 0)$\\ 
        \hline
        $-$ & $+$  & none & none \\
        \hline
        0 & + & none & none \\
        \hline
        0 & $-$ & none & elevatons $(K < 0)$ \\
     \hline
     0 & 0 & none & none \\
     \hline
    \end{tabular}
    }
    \caption{Nature of fixed points and bounded solutions on $\rho$-$\rho_x$ half-plane. (a) General non-aerostatic case, when $ F^{\rm m}, F^{\rm e} \ne 0$ and $\D > 0$. (b) Aerostatic limit $(u \equiv 0,  F^{\rm m} =  F^{\rm e} = 0)$. $K < 0$ corresponds to solutions with negative pressure.}
    \label{t:nonaero-aero-steady}
\end{table*}

There may be two, one or no FPs in the physical region $\rho > 0$. We are interested in the cases where there is at least one FP in the physical region, as otherwise $\rho$ is unbounded. This requires $\D > 0$. Assuming this is the case and also assuming that the flow is not aerostatic $( F^{\rm m} \neq 0$ or $u \not \equiv 0)$, we find that there are two physical FPs if $ F^{\rm p}$ and $F^{\rm u}$ are both positive, one fixed point if $F^{\rm u} < 0$ and none otherwise. The character of these FPs may be found by linearizing $W$ around them. Writing $\rho = \rho_{p,m} + \del\rho$ and $\eta = 0 + \del\eta$, we get
    \beq
	\DD{}{x} \colvec{2}{\del \rho}{\del \eta} = A \colvec{2}{\del \rho}{\del \eta} \quad \text{where} \quad A = \colvec{2}{0 & 1}{- V''(\rho_{p,m})/\beta_* & 0}. 
    \eeq
The eigenvalues of the coefficient matrix $A$ are
    \beq
	\la = \pm \sqrt{\frac{- V''(\rho_{p,m})}{\beta_*}} \quad \text{where} \quad
	V''(\rho_{p,m}) = \mp \frac{\sqrt{\D}}{\rho_{p,m}}.
	\label{e:Vprpr-formula}
    \eeq
Thus, the physical FPs of $W$ must either be X or O points (saddles or centers in the linear approximation) according as $V'' < 0$ (real eigenvalues) or $V'' > 0$ (imaginary eigenvalues). The Hartman-Grobman Theorem guarantees that the linear saddles remain saddles even upon including nonlinearities. Moreover, the linear O-point at $(\rho_m,0)$ is always a true O-point since we may verify that $(\rho_m,0)$ is a minimum of the conserved quantity $K$ (\ref{e:rhox-vs-rho-diff-eqn}). Thus, as summarized in Table \ref{t:nonaero-aero-steady}(a), there are two types of phase portraits leading to bounded solutions $\rho(x)$: (i) if $ F^{\rm p}$ and $ F^{\rm u}$ are both positive, then $W$ has an O-point at $(\rho_m, 0)$ and an X-point at $(\rho_p, 0)$ to its right and (ii) if $ F^{\rm u}<0$, $W$ has only one physical fixed point, an O-point at $(\rho_m, 0)$. As shown in Fig.~\ref{f:vec-fld} (a,b), in case (i) we have two types of bounded solutions: periodic waves corresponding to closed curves around the O-point $(\rho_m,0)$ and a solitary wave corresponding to the separatrix orbit that begins and ends at $(\rho_p,0)$ and encircles the O-point. Since $\rho_p > \rho_m$, a solitary wave must be a caviton. In case (ii) the only bounded solutions $\rho(x)$ are periodic waves corresponding to closed curves encircling the O-point $(\rho_m,0)$ as shown in Fig.~\ref{f:vec-fld} (c,d).  

%-----------

{\fl \bf Isentropic aerostatic steady solutions:} In the aerostatic limit $(u \equiv 0)$, both the fluxes $F^{\rm e}$ and $F^{\rm m}$ vanish, though their ratio $F^{\rm e}/F^{\rm m} = F^{\rm u}$ is finite. Eqn. (\ref{e:steady-reg-gas-eqn-rho}) for steady solutions becomes
    \beqs
	&& \beta_* \rho_{xx} = - V'(\rho) + \frac{(\g + 1) \beta_*}{2} \frac{\rho_x^2}{\rho} \quad \text{where} \cr
	&& V(\rho) = \g  F^{\rm p}\rho - (\g - 1) \frac{ F^{\rm u}}{2} \rho^2.
    \eeqs
In this limit, the small-$\rho$ logarithmic barrier in $V$ (\ref{e:steady-newtons-law}) is absent, and the singularity along $\rho = 0$ becomes `naked'. One of the FPs in Eqn.~(\ref{e:rho-pm-fixedpts-steady}) tends to $(0,0)$, while the other one tends to $(\g  F^{\rm p}/(\g - 1) F^{\rm u}, 0)$. Table \ref{t:nonaero-aero-steady}(b) summarizes the nature of physical fixed points and bounded solutions for various possible signs of $ F^{\rm p}$ and $ F^{\rm u}$. Interestingly, for $ F^{\rm p} = 0$ and $ F^{\rm u} < 0$, there is a family of solitary waves of {\it elevation}, though with negative pressure.

%---------------
\section{Canonical formalism for steady solutions}
\label{a:hamil-lagr-for-steady-sol}
%---------------

The equation for steady solutions (\ref{e:steady-reg-gas-eqn-rho}) describes a mechanical system with 1 degree of freedom and conserved quantity $K(\rho,\rho_x)$ (\ref{e:rhox-vs-rho-diff-eqn}). Here we give a canonical formulation for (\ref{e:steady-reg-gas-eqn-rho}) by taking $K$ to be the Hamiltonian. We seek a suitable PB $\{\rho, \rho_x \}$ so that Hamilton's equations $\rho_x = \{\rho, K \}$ and $\rho_{xx} = \{\rho_x, K \} $ reproduce (\ref{e:steady-reg-gas-eqn-rho}). The former gives
	\beqs
	&& \{\rho, K\} = \left\{ \rho, \half  \frac{\beta_* \rho_x^2}{\rho^{\g +1}} \right\} = \rho_x  \quad \text{or} \quad  \frac{\beta_* \rho_x}{\rho^{\g + 1}} \{\rho, \rho_x \} = \rho_x \cr
	&& \imply \quad  \{\rho, \rho_x \} = \frac{\rho^{\g + 1}}{\beta_*}.
	\eeqs
Using this PB, $\rho_{xx} = \{\rho_x , K \}$ reproduces  (\ref{e:steady-reg-gas-eqn-rho}). This PB is not canonical, but if we define $\varpi = \beta_* \rho_x/\rho^{\g + 1}$ then $\{ \rho , \varpi \} = 1$ so that $\varpi$ is the momentum conjugate to $\rho$. The corresponding Hamiltonian is
	\beq
	K(\rho,\varpi) = \half  \frac{\rho^{\g + 1}}{\beta_*} \varpi^2 + U(\rho),
	\eeq
with $U$ as in (\ref{e:rhox-vs-rho-diff-eqn}). The `mass' factor in the kinetic term is `position' $(\rho)$ dependent. In terms of the contravariant `mass metric' $m^{-1}(\rho) = {\rho^{\g + 1}}/{\beta_*}$, $K = (1/2) m^{-1} \varpi^2 + U$. The corresponding Lagrangian is
	\beq
	L = \text{ext}_\varpi ( \varpi \rho_x  - K) =  \half \frac{\beta_*}{\rho^{\g + 1}} \rho_x^2 - U  = \half m(\rho) \rho_x^2 - U.
	\eeq
The Euler-Lagrange equation reduces to (\ref{e:steady-reg-gas-eqn-rho}). Thus, we have Hamiltonian and Lagrangian formulations for both the full R-gas dynamic field equations and their reduction to the space of steady solutions.

%----------
\section{Parabolic embedding and Lagrange-Jacobi identity for steady flow}
\label{a:parabolic-embed-LJ-id}
%----------

For $\g \ne 2$ the quadrature in (\ref{e:quadrature-steady-rho}) cannot be done using elliptic functions, but could be done numerically. An alternative approach is to take a linear combination of the equations in (\ref{e:rhoxx-steady-eqn-two-versions}) to obtain a form of the steady equation for $\rho$ without the velocity dependent term but with a generally non-integral power of $\rho$:
	\beq
	\beta_* \rho_{xx} = (\g + 1) K \rho^\g - 2 F^{\rm u} \rho +  F^{\rm p}.
	\label{e:steady-eqn-rhoxx-no-vel-dep-term}
	\eeq
If we introduce a pseudo-time $\tau$, then steady solutions can be obtained via a parabolic embedding in a nonlinear heat equation with a source:
	\beq
	\rho_{\tau}-\beta_{*}\rho_{xx}
 	= - (\g + 1) K \rho^\g + 2 F^{\rm u} \rho - F^{\rm p}.
	\eeq
By prescribing suitable BCs and starting with an arbitrary initial condition, the solution of this PDE should relax to the stable solutions of (\ref{e:steady-eqn-rhoxx-no-vel-dep-term}).

Eqn. (\ref{e:steady-eqn-rhoxx-no-vel-dep-term}) may also be used to derive additional virial/Lagrange-Jacobi-type identities by multiplying it by $\rho$ and using (\ref{e:rhox-vs-rho-diff-eqn}) and repeating the process. The first two such identities are 
	\beqs
	&& \frac{\beta_*}{2} (\rho^2)_{xx} = (\g+3) K \rho^{\g+1} - 4 F^{\rm u} \rho^2 + 3 F^{\rm p} \rho - (F^{\rm m})^2 \; \text{and}\cr
	&& \frac{\beta_*}{3} (\rho^3)_{xx} = (\g+5) K \rho^{\g+2} - 6 F^{\rm u} \rho^3 + 5 F^{\rm p} \rho^2 - 2 (F^{\rm m})^2 \rho. \quad \;\;\;\:
	\label{e:lagrange-jacobi-idnetities}
	\eeqs 
Integrating (\ref{e:steady-eqn-rhoxx-no-vel-dep-term}) and (\ref{e:lagrange-jacobi-idnetities}) with periodic BCs we get a hierarchy of integral invariants for steady soultions 
	\beqs
	&& \int_{-L}^L \left[ (\g + 1) K \rho^\g - 2 F^{\rm u} \rho +  F^{\rm p} \right] dx \cr
	&& = \int_{-L}^L \left[ (\g+3) K \rho^{\g+1} - 4 F^{\rm u} \rho^2 + 3 F^{\rm p} \rho - (F^{\rm m})^2 \right] dx \cr
	&& = \int_{-L}^L \left[ (\g+5) K \rho^{\g+2} - 6 F^{\rm u} \rho^3 + 5 F^{\rm p} \rho^2 - 2 (F^{\rm m})^2 \rho \right] dx \cr
	&& = 0 \quad \text{etc.}  
	\eeqs
These integral identities can provide valuable checks on any numerics used to obtain steady solutions.

%------------
\section{Semi-implicit spectral scheme for time evolution}
\label{a:numerical-scheme}
%------------

Here, we describe the scheme used to solve the IVP for the 1d $\g = 2$ isentropic R-gas dynamic equations (\ref{e:cont-vel-baro-fourier}). To include the effects of the nonlinear terms in (\ref{e:cont-vel-baro-fourier}), we discretize time $\hat t = j \D$, $j = 0,1,2, \ldots$, denote the Fourier modes $\rho_n (j \D) = \rho_n^j$ etc. and use a centered difference scheme for the linear part:
	\beqs
	&& \rho_n^{j+1} = \rho_n^j - \frac{i n \D}{2} \left( \thickbar u (\rho_n^{j} + \rho_n^{j+1}) + u_n^j + u_n^{j+1} \right) - \D \del (\calF^{\rm m})_n^{j} \cr
	&& u_n^{j+1} = u_n^j - \frac{i n \D}{2} \left( \thickbar u (u_n^j + u_n^{j+1}) + (1 + \eps^2 n^2) (\rho_n^j + \rho_n^{j+1}) \right) \cr
	&& \qquad \quad - \D \del (\calF^{\rm u})_n^{j}.
	\eeqs
For simplicity, the nonlinear terms are treated explicitly; treating the linear terms explicitly leads to numerical instabilities. In matrix form, the equations read
	\beqs
	&& A \colvec{2}{\rho_n^{j+1}}{u_n^{j+1}}
	= B \colvec{2}{\rho_n^{j}}{u_n^{j}} 
	- \D \del \colvec{2}{(\calF^{\rm m})_n^{j}}{(\calF^{\rm u})_n^{j}} \quad 
	\text{where} \cr
	&& A = I + \frac{i n \D}{2} \colvec{2}{ \thickbar u & 1}{p_n^2 & \thickbar u} \quad \text{and} \quad
	B =  I - \frac{i n \D}{2} \colvec{2}{ \thickbar u & 1}{p_n^2 & \thickbar u}. \quad 
	\eeqs
with $p_n^2 = (1 + \eps^2 n^2)$. $B$ is related to $A$ via conjugation or by $\D \to - \D$. Thus, the variables at the $j+1^{\rm st}$ time step are
	\beqs
	 && \colvec{2}{\rho_n^{j+1}}{u_n^{j+1}}
	= U \colvec{2}{\rho_n^{j}}{u_n^{j}} 
	- \D \del A^{-1} \colvec{2}{(\calF^{\rm u})_n^{j}}{(\calF^{\rm u})_n^{j}} \quad \text{where}  \cr
	&& U = A^{-1} B = \ov{\det A} \left[ I + n \D \colvec{2}{ \frac{\D}{4} n (\thickbar u^2 - p_n^2 ) & -i}{-i p_n^2 &  \frac{\D}{4} n(\thickbar u^2 - p_n^2 )} \right], 
	\cr
	&& (\det A) \, A^{-1} = I +  \frac{i n \D}{2} \colvec{2}{\thickbar u & - 1}{- p_n^2 &\thickbar u } \quad \text{and} \cr
	&&\det A = \left(1 + \frac{i n \thickbar u \D}{2} \right)^2 + \frac{n^2 \D^2 p_n^2}{4}.
	\eeqs
$U$ is unitary with respect to the inner product $\left\langle (\rho, u), (\tl \rho, \tl u)  \right\rangle = p_n^2 \rho^* \tl \rho + u^* \tl u$, which ensures that the linear evolution conserves $H_1$ of (\ref{e:sound-waves-cons-qtys}). An advantage of this scheme is that conservation of $\int \rho \, dx $ and $\int u \, dx$ are automatically satisfied to round-off accuracy. We also find that $Q_r$ and $Q_i$ (\ref{e:Qr-Qi}) are quite accurately conserved in our numerical evolution for the ICs in \S \ref{s:numerical-results}. Moreover, since $\calF^{\rm m}$ (\ref{e:cont-linear-nonlinear-split}) and $\calF^{\rm u}$ (\ref{e:vel-linear-nonlinear-split}) are divergences, their Fourier coefficients can be calculated by integration by parts without any differentiation.

\scriptsize

%-----------------------

\end{document}